\documentclass[a4paper,oneside,final,notitlepage,onecolumn,11pt,english]{article}

\pdfoutput=1

\usepackage{ifpdf,epsfig,array,amsmath,amssymb,psfrag,graphicx}
\usepackage{verbatim}

\usepackage[normalem]{ulem}
\usepackage{color}
\definecolor{blue}{rgb}{0,0,1}
\definecolor{red}{rgb}{1,0,0}

\newcommand{\inserted}[1]{\textcolor{black}{#1}}

\usepackage{psfrag}
\psfrag{EBTE0}{\footnotesize $E^{BT}_{tot}/E_0$}
\psfrag{EDDE0}{\footnotesize $E^{DD}_{tot}/E_0$}
\psfrag{deltaE}{\footnotesize $\delta E$}

\begin{document}

\title{Gravitational waves from spinning eccentric binaries\thanks{This paper is dedicated to the memory of our colleague and friend P\'{e}ter Csizmadia a young physicist, computer expert and one of the best Hungarian mountaineers who disappeared in China's Sichuan near the Ren Zhong Feng peak of the Himalayas October 23, 2009. We started to develop CBwaves jointly with P\'eter a couple of moth before he left for China.}}
\author{\small
\fbox{P\'{e}ter Csizmadia}, Gergely Debreczeni\thanks{email: debreczeni.gergely@wigner.mta.hu},\, Istv\'{a}n R\'{a}cz\thanks{email: racz.istvan@wigner.mta.hu }\,\, and M\'{a}ty\'{a}s Vas\'{u}th\thanks{email: vasuth.matyas@wigner.mta.hu}
\\ 
\small WIGNER RCP, RMKI \\
\small  H-1121 Budapest, Konkoly Thege Mikl\'os \'ut 29-33.\\
\small Hungary
}
\maketitle

\begin{abstract}
This paper is to introduce a new software called CBwaves which
provides a fast and accurate computational tool to determine the
gravitational waveforms yielded by generic spinning binaries of
neutron stars and/or black holes on eccentric orbits. This is done
within the post-Newtonian (PN) framework by integrating the
equations of motion and the spin precession equations while the
radiation field is determined by a simultaneous evaluation of the
analytic waveforms. In applying CBwaves various physically
interesting scenarios have been investigated. In particular, we have \inserted{studied the appropriateness of the adiabatic approximation, and}
justified that the energy balance relation is indeed insensitive to
the specific form of the applied radiation reaction term. By
studying eccentric binary systems it is demonstrated that circular
template banks are very ineffective in identifying binaries even if
they possess tiny residual orbital eccentricity.  In addition, by
investigating the validity of the energy balance relation we show
that, on contrary to the general expectations,  the post-Newtonian
approximation should not be applied once the post-Newtonian
parameter gets beyond \inserted{the} critical value $\sim\inserted{ 0.08}-0.1$. Finally, by
studying the early phase of the gravitational waves emitted by
strongly eccentric binary systems---which could be formed e.g.\,in
various many-body interactions in the galactic halo---we have found
that they possess very specific characteristics which may be used to
identify these type of binary systems.
\end{abstract}


\maketitle

\section{Introduction}

The advanced versions of our current ground based interferometric
gravitational wave observatories such as LIGO \cite{advligo} and
Virgo \cite{advvirgo} are \inserted{expected} to do the first direct detection
soon after their restart in 2015. It is also expected that these
detectors observe yearly tens of gravitational wave \inserted{(GW)} signals emitted
during the final inspiral and coalescence of compact binaries
composed by neutron stars and low mass black holes. The
identification of these type of sources is attempted to be done by
making use of the matched filtering technique \cite{matchedfilter}
where templates are deduced by using various type of theoretical
assumptions. Among the physical quantities characterizing the
waveforms and the evolution of binaries the time dependence of the
wave amplitude and the orbital phase are of critical importance. In
determining the amplitude and the phase, along with several other
astrophysical properties of the sources, we need to apply templates
which are sufficiently accurate and cover the largest possible
parameter domain associated with the involved binaries.

\bigskip

The main purpose of the present paper is to introduce a new
computational tool, called \textit{CBwaves}, by the help of which
the construction of gravitational wave templates for the generic
inspiral of compact binaries can be done in a fast and accurate way.
\inserted{In developing it} our principal aim was to \inserted{make an accurate and fast software available for public use that is capable to} follow \inserted{the evolution of} generic configurations of spinning
and eccentric binaries with arbitrary orientation of spins and
arbitrary value of the eccentricity.

In carrying out this program the post-Newtonian framework
\cite{Blanchet06} has been applied---by making use of the analytic
setup \inserted{ which is a synthesis of the developments in}  {\color{black}\cite{Kidder,MoraWill,IyerWill,WillPNSO,WangWill,GopaIyer,Arun08,FBB,FBB2}}---, i.e.\,by integrating the
\inserted{3}.5PN accurate equations of motion and the spin precession equations
of the orbiting bodies while the radiation field is determined by a
simultaneous evaluation of the analytic waveforms which involves \inserted{all the} contributions \inserted{that have been worked out for generic eccentric orbits} up to \inserted{2}PN order. The equations of motion are
integrated by a fourth order Runge-Kutta method numerically. The
most important input parameters are the initial separation, the
masses, the spins, along with their orientations, of the involved
bodies and the initial eccentricity of the orbit. The waveforms are
calculated in time domain and they can also be determined in
frequency domain by using the implemented FFT. Moreover, CBwaves
does also provide the expansion of the radiation field in $s=-2$
spin weighted spherical harmonics. \inserted{The open source code of the current version of CBwaves may be downloaded from \cite{cbwavesdownload}.}

\bigskip

In the post-Newtonian approach various order of corrections are
added to the Newtonian motion where the fundamental scale of the
corrections are determined by the post-Newtonian parameter
$\epsilon\sim(v/c)^2\sim Gm/(rc^2)$, where $m$, $v$ and $r$ are the
total mass, orbital velocity and separation of the binary system.
Within this framework the damping terms are considered to be
responsible for the change of the motion of the sources in
consequence of the radiation of gravitational waves to infinity.
Gravitational radiation reaction---sometimes called to be Newtonian
radiation reaction---appears first at 2.5PN order, i.e., this
correction is of the order $\epsilon^{\frac52}$. It is known for
long that the relative acceleration term appearing in the radiation
reaction expressions is not unique. It is however argued by various
authors (see, e.g.\,\cite{IyerWill}) that the energy balance
relation has to be insensitive to the specific form of the applied
radiation reaction term. By making use of our CBwaves code the
effect of the two standard radiation reaction terms could be
compared. We have found that the energy balance relations are indeed
insensitive to the specific choices of the parameters in the
radiation reaction terms provided that a suitable coordinate
transformation is applied while switching between gauge
representations.

\bigskip

\inserted{It is expected that whenever the time scales of both the precession and shrinkage of the orbits of the investigated binaries are long, when they are compared to the orbital period, the adiabatic approximation is appropriate. This, in the particular case of quasi-circular inspiral orbits, means that they are expected to be correctly  approximated by nearly circular ones with a slowly shrinking radius (see for e.g. Section IV in \cite{Kidder}).
We investigated the validity of the adiabatic approximation by monitoring the rates of inspiral of the adiabatic approximation and of a corresponding time evolution yielded by CBwaves. \color{black}{It was found that the adiabatic approximation is pretty reliable as it produces, at 3.5PN level, less then $2\%$ faster decrease of the rate of inspiral than the corresponding time evolution. }}

\bigskip

\inserted{It was shown in \cite{yunes}---see the
appendix \cite{yunes} for detailed investigations--- }that a large fraction of
binaries emitting gravitational wave signals detectable by our \inserted{planned space based} detectors are expected to have orbits with non-negligible
eccentricity. Immediate examples are black hole binaries which may
be formed by tidal capture in globular clusters or galactic nuclei
\cite{wen,kocsis}. Although a circularization of these orbits will
happen by radiation reaction as the circularization process is slow
a tiny residual eccentricity may remain\inserted{ until the emitted wave gets to be detectable}. \inserted{Since the existence of binary systems with a slight residual eccentricity and with the possibility that they may be detected by our advanced or third generation GW detectors cannot be excluded} it is important
to \inserted{ to determine} in what extent such a residual eccentricity may affect
the detection performance of matched-filtering. \inserted{This type of investigation, based on the sensitivity of the initial LIGO detector, was done in \cite{exc}. By carrying out the corresponding investigations, based on the sensitivity of the advanced Virgo detector  \color{black}we shall further strengthen the conclusion} that circular template banks are very
ineffective in identifying\inserted{ these type of} binaries. It is also important to be emphasized that by using
templates based on the circular motion of binaries exclusively\inserted{ the} signal-to-noise-ratio (SNR)\inserted{ may be decreased significantly}, which in turn downgrades the performance of our\inserted{ next generation} detectors.

\medskip

The expectations concerning the applicability of the post-Newtonian
expansion are based on the assumption that the time scales of
precession and shrinkage are both long compared to the orbital
period until the very late stage of the evolution. On contrary to
this claim there are more and more evidences showing that the
post-Newtonian approximation should not taken seriously once the
post-Newtonian parameter reaches the parameter domain $\epsilon\sim
0.08-0.1$. By investigating the validity of the energy balance
relation we show that indeed it gets to be violated as soon as the
post-Newtonian parameter gets to be close to the critical upper
bound $0.1$. Note that these findings are in accordance with the
claims of \cite{janna} where the relative significance of the higher
order contributions was monitored. What is really unfavorable is
that this sort of loss of accuracy is getting to be more and more significant right
before reaching the frequency ranges of our current\inserted{ly upgraded} ground based GW
detectors.

\medskip

It is \inserted{shown in \cite{kocsis}} that in consequence of many-body interactions
strongly eccentric black hole binaries  are formed in the halo of
the galactic supermassive black hole. The early phase of
the gravitational waves emitted by these type of strongly eccentric
binary systems possesses burst type character. It is true that the
amplitude and the frequency of the gravitational waves emitted by
such systems change significantly during the inspiral due to the
circularization effect. Nevertheless, we have found that the
frequency-domain waveforms of the early phase of highly eccentric
binary systems possess very specific characteristics which may
suffice to determine the physical parameters of the system.

\medskip

This paper is organized as follows. In Section \ref{motion} some of the basics of the\inserted{ applied} analytic setup
will be summarized.  In Section \ref{CBwaves} a short introduction
of the CBwaves software is given. Section \ref{results} is to
present our main results. Subsection \ref{non-spin} deals with
non-spinning circular waveforms, while in  subsection \ref{rad-reac}
the gauge dependence of the radiation reaction is examined. A short
discussion in subsection \ref{dom-val} on the range of applicability
of the PN approximation is followed by the investigation of
eccentric motions in subsection \ref{ecc-motion} introducing the
found universalities in the evolution of eccentricity, along with
our justification of the loss of SNR whenever tiny residual
eccentricities are neglected by applying circular waveforms
exclusively. Subsection \ref{spinning-binaries} is to introduce our
finding regarding spinning and eccentric binaries while section
\ref{Summary} contains our final remarks.

\section{The motion and radiation of the binary system}\label{motion}

As mentioned above in detecting low mass black hole or neutron star
binaries we need to have template banks built up by sufficiently
accurate waveforms which cover the largest possible parameter domain
associated with the involved binaries. In determining the motion of
the bodies and the yielded waveforms the analytic setup \inserted{ which is a synthesis of the developments in}  {\color{black}\cite{Kidder,MoraWill,IyerWill,WillPNSO,WangWill,GopaIyer,Arun08,FBB,FBB2}} is applied.

In the post-Newtonian formalism \cite{Blanchet06} the spacetime is
assumed to be split into the near and wave zones. The field
equations for the perturbed Minkowski metric is solved in both
regions. In the near zone the energy-momentum tensor is nonzero and
retardation is negligible (post-Newtonian expansion), while in the
wave zone the vacuum Einstein equation are solved (post-Minkowskian
expansion). In the overlap of these regions the solutions are
matched to each other. As a result of the process the radiation
field far from the source is expressed in terms of integrals over
the source (the source multipole moments). For the special case of
compact binary systems the source integrals are evaluated with the
point mass assumption and this hypothesis leads to various
regularization issues and ambiguities already at 3PN, see e.g.
\cite{Blanchet04,DJS01}. \inserted{Nevertheless, } the analytic
setup of the present work was chosen such that the motion of the
binary is taken into account only up to \inserted{3}.5PN order while the
radiation field up to \inserted{2}PN order.

In harmonic coordinates the radiation field $h_{ij}$ far from the
source is decomposed as \cite{Kidder}
\begin{equation}
h_{ij}=\frac{2G\mu }{c^{4}D}\left(
Q_{ij}+P^{0.5}Q_{ij}+PQ_{ij}+PQ_{ij}^{SO}+P^{1.5}Q_{ij}+P^{1.5}Q^{SO}_{ij}
\inserted{ + P^{2}Q_{ij}+P^{2}Q_{ij}^{SS}} \right), \label{Wform}
\end{equation}%
where $D$ is the distance to the source and
$\mu=m_{1}m_{2}/(m_{1}+m_{2})$ is the reduced mass of the system. We
have collected the relevant terms up to 1.5PN order. $Q_{ij}$ is the
quadrupole (or Newtonian) term, $P^{0.5}Q_{ij}$, $PQ_{ij}$ \inserted{,} $P^{1.5}Q_{ij}$\inserted{ and $P^{2}Q_{ij}$
\cite{WW}} are higher order relativistic corrections, $PQ^{SO}_{ij}$
\inserted{,} $P^{1.5}Q^{SO}_{ij}$\inserted{ and
$P^{2}Q_{ij}^{SS}$} are spin-orbit \inserted{and spin-spin} terms.
Note that the 1.5PN \inserted{and 2PN} order
contributions to the waveform due to wave tails---which depend on
the past history of the binary
\cite{W93,Blanchet08}---\inserted{are} neglected.
The detailed expressions of the \inserted{ involved} contributions are given in \cite{Kidder,WW}, and for \inserted{ the sake of completeness}
they are also summarized in Appendix A.

For a plane wave traveling in the direction $\mathbf{\hat{N}}$,
which is a unite spatial vector pointing from the center of mass of
the source to the observer, the transverse-traceless (TT) part of
the radiation field is given as \cite{Maggiore}
\begin{equation}
h_{ij}^{TT}=\Lambda _{ij,kl}\,h_{kl}\,,  \label{TTproj}
\end{equation}
where
\begin{equation}
\Lambda _{ij,kl}(\mathbf{\hat{N}})=P_{ik}P_{jl}-\frac{1}{2}P_{ij}P_{kl},\ \ \ {\rm and}\ \ \
P_{ij}(\mathbf{\hat{N}})=\delta _{ij}-N_{i}N_{j}\,.  \label{TTproj2}
\end{equation}

\begin{figure}[htb]
\begin{center}
\includegraphics[width=9cm]{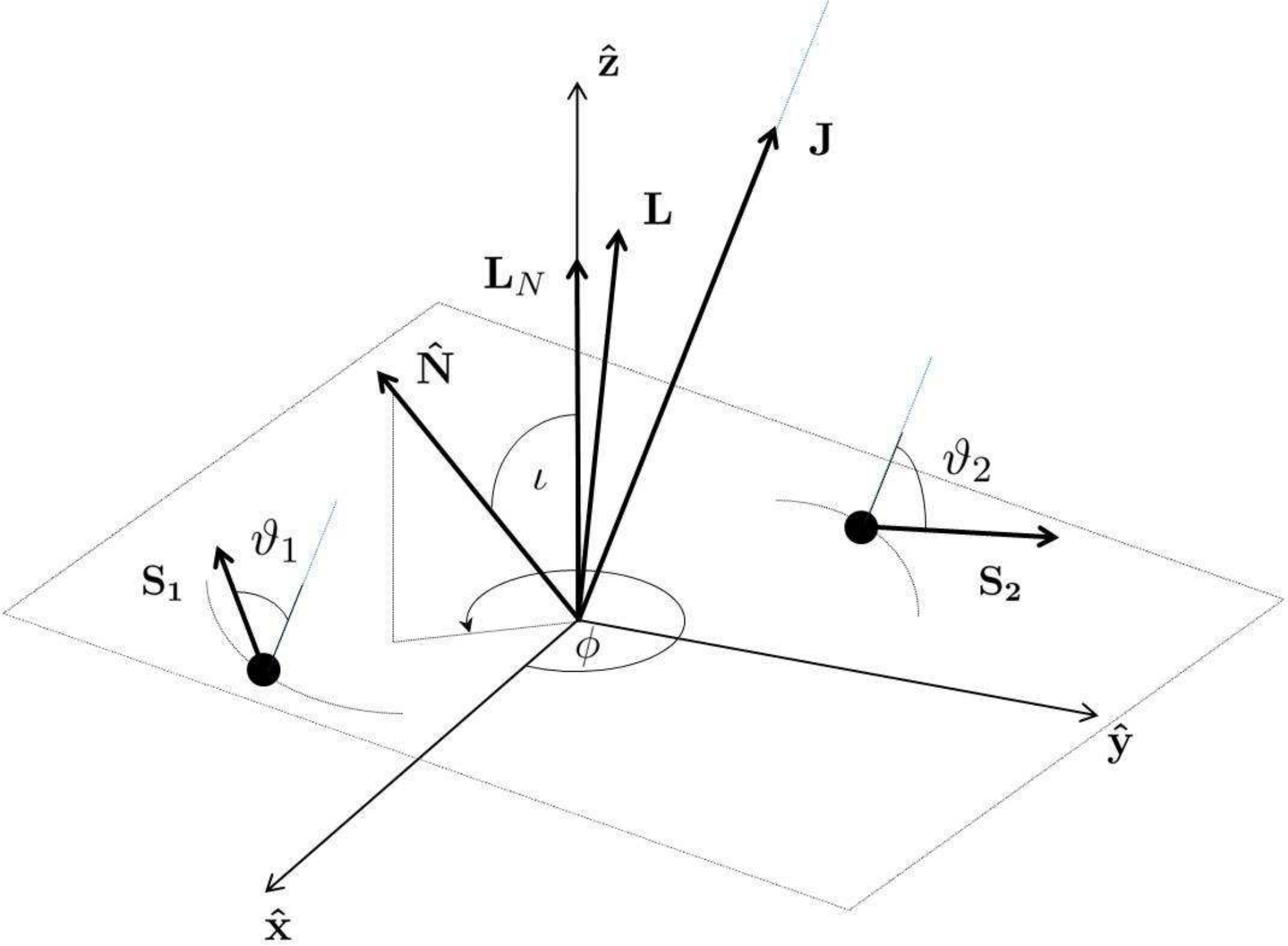}
\end{center}
\caption{\footnotesize The relative orientation of the main vectors
characterizing the binary with respect to the source frame
$(\mathbf{\hat{x}},\mathbf{\hat{y}},\mathbf{\hat{z}})$, along with
the polar angles $\iota $ and $\phi $ of $\mathbf{\hat{N}}$, are
shown. This figure also indicates the relative orientation of the
total and the orbital angular momentum vectors, $\mathbf{J}$ and
$\mathbf{L}$, and that of the individual spin vectors $\mathbf{S}_1$
and $\mathbf{S}_2$ with respect to $\mathbf{J}$ by indicating the
angles $\vartheta_i=\cos^{-1}(\mathbf{J} \cdot
\mathbf{S}_i/\|\mathbf{J}\|\|\mathbf{S}_i\|)$, where $i$ takes the
values $1,2$.} \label{SourceFrame}
\end{figure}

Following \cite{LSCCommon} an orthonormal triad, called the
radiation frame, is chosen as
\begin{eqnarray}
\mathbf{\hat{N}} &=&\mathbf{(}\sin \iota \cos \phi ,\sin \iota \sin
\phi,\cos \iota \mathbf{),} \label{trafo1} \\
\mathbf{\hat{p}} &=&(\cos \iota \cos \phi ,\cos \iota \sin \phi
,-\sin \iota),  \label{trafo2} \\
\mathbf{\hat{q}} &=&(-\sin \phi ,\cos \phi ,0) \label{trafo3}
\end{eqnarray}%
where the polar angles $\iota $ and $\phi $, determining the
relative orientation of the radiation frame with respect to the
source frame $(\mathbf{\hat{x}},\mathbf{\hat{y}},\mathbf{\hat{z}})$
as indicated on Fig.\,\ref{SourceFrame}. The basis
vectors of the source frame $\mathbf{\hat{x}}$ and $\mathbf{\hat{y}}$ are supposed to span the initial orbital
plane such that $\mathbf{\hat{x}}$ and $\mathbf{\hat{z}}$ are parallel to the separation vector $\mathbf{r}=\mathbf{x_{1}}-\mathbf{x_{2}}$ and the Newtonian part of the angular momentum $\mathbf{L}_{N}=\mu \mathbf{r}\times\mathbf{\dot r}$, both at the beginning of the orbital evolution, respectively. According to the particular relations given by Eqs.\,(\ref{trafo1})-(\ref{trafo3}) the source and radiation frames can be transformed into each other simply by two consecutive rotations. Rotating first the radiation frame around the $\mathbf{\hat{z}}$ axis by angle $-\phi$ which is followed by a rotation around the $\mathbf{\hat{y}}$ axis by the angle $-\iota$.

\begin{figure}[htb]
\begin{center}
\includegraphics[width=7cm]{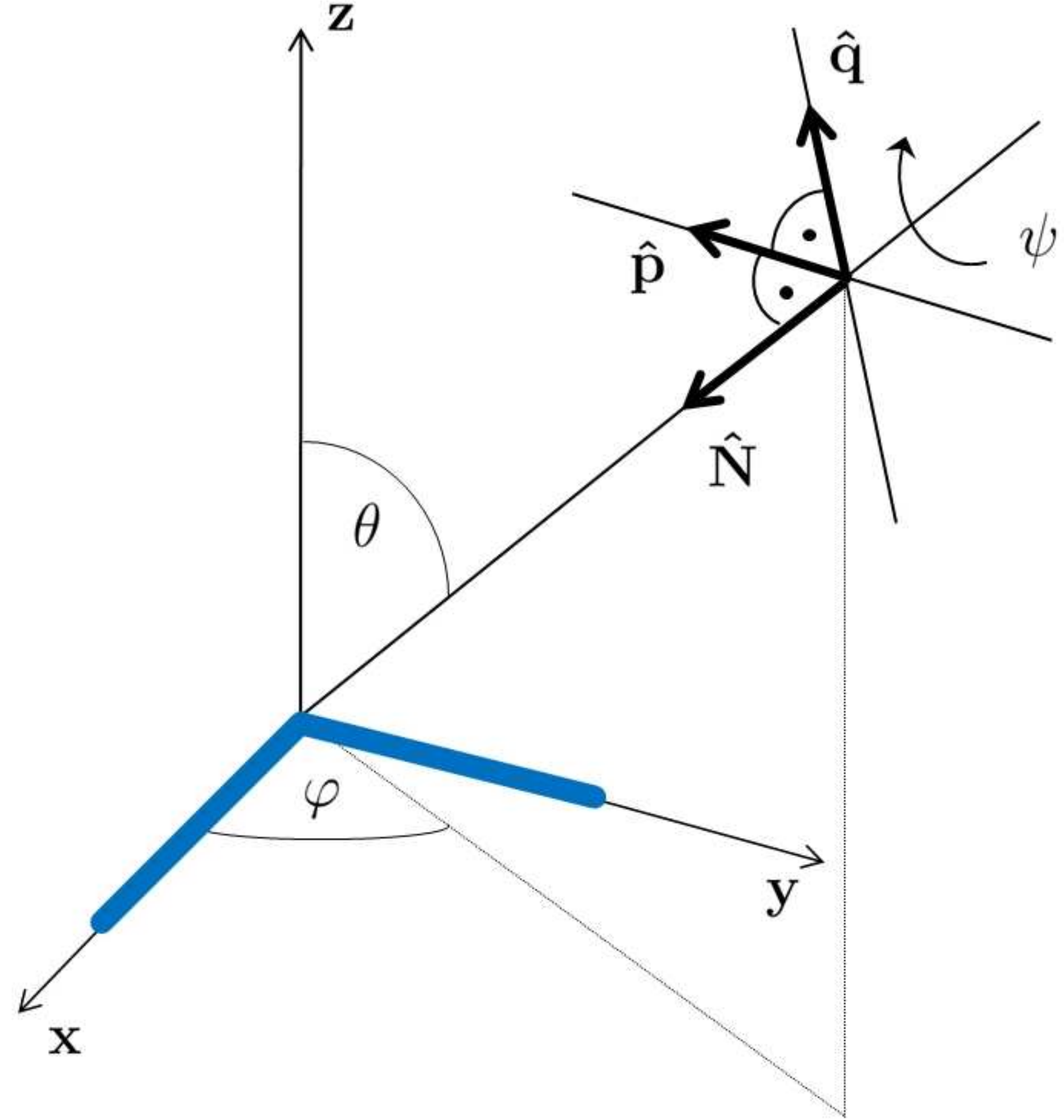}
\end{center}
\caption{\footnotesize The relative orientation of the detector
frame $(\mathbf{{x}},\mathbf{{y}},\mathbf{{z}})$---the
$\mathbf{{x}}$ and $\mathbf{{y}}$ axes are aligned with the detector
arms---and the radiation frame $(\mathbf{\hat{N},\hat{p},\hat{q})}$,
along with the polar angles $\theta $ and $\varphi$, are shown. }
\label{DetectorFrame}
\end{figure}

The polarization states can be given, with respect to the
orthonormal radiation frame $(\mathbf{\hat{N},\hat{p},\hat{q})}$, as
\cite{FC}
\begin{equation}
h_{+}=\frac{1}{2}\left(
\hat{p}_{i}\hat{p}_{j}-\hat{q}_{i}\hat{q}_{j}\right)
h_{ij}^{TT},\quad h_{{\times }}=\frac{1}{2}\left( \hat{p}_{i}\hat{q}_{j}+%
\hat{q}_{i}\hat{p}_{j}\right) h_{ij}^{TT}.  \label{pstatesgeneral}
\end{equation}
In the applied linear approximation the strain produced by the
binary system at the detector can be given as the combination
\begin{equation}
h(t)=F_{+}h_{+}(t)+F_{\times }h_{\times }(t),
\end{equation}%
where the antenna pattern functions $F_{+}$\ and $F_{\times }$ are given as
\begin{eqnarray}
F_{+} &=&-\frac{1}{2}\left( 1+\cos ^{2}\theta \right) \cos 2\varphi
\cos
2\psi -\cos \theta \sin 2\varphi \sin 2\psi , \\
F_{\times } &=&\frac{1}{2}\left( 1+\cos ^{2}\theta \right) \cos
2\varphi \sin 2\psi -\cos \theta \sin 2\varphi \cos 2\psi
\end{eqnarray}%
for ground-based interferometers, with Euler angles $\theta ,\varphi
$ and $\psi$ relating the radiation frame and the
detector frame $(\mathbf{x},\mathbf{y},\mathbf{z})$ as it is
indicated on Fig.\,\ref{DetectorFrame}. Note that in spite of the
fact that the polarization states do depend on the choice made for
the triad elements $\mathbf{\hat{p}}$ and $\mathbf{\hat{q}}$, due to
the compensating rotation in the polarization angle $\psi$, the
strain measured at a detector remains intact. The binary is called
to be optimally oriented, i.e.\,the detector sensitivity is maximal,
if $\theta =0$ or $\pi $, and the angle of inclination $\iota =0$.

In determining the radiation field far from the source, i.e.\,the
evaluation of all the general expressions in Eq.\,(\ref{Wform}), one
needs to know the precise motion of the bodies composing the binary
system. In the generic case the orbit of the binary system acquires
precession (due to spin effects) and shrinking (due to radiation
reaction) during the time evolution. In the adiabatic approach,
applied in the post-Newtonian setup, it is assumed that the time
scales of precession and shrinkage are both long compared to the
orbital period until the very late stage of evolution. The
acceleration of the reduced one-body system follows from the
conservation of the energy momentum (the geodesic equation in the
perturbed spacetime with harmonic coordinates) \cite{Kidder},
\begin{equation}
\mathbf{a}=\mathbf{a}_{N}+\mathbf{a}_{PN}+\mathbf{a}_{SO}+\mathbf{a}_{2PN}+%
\mathbf{a}_{SS}+\mathbf{a}_{RR}^{BT} \inserted{+ \mathbf{a}_{PNSO} +
\mathbf{a}_{3PN} + \mathbf{a}_{RR1PN} + \mathbf{a}_{RRSO} +
\mathbf{a}_{RRSS} }, \label{accel}
\end{equation}%
where $\mathbf{a}_{N}$, $\mathbf{a}_{PN}$, $\mathbf{a}_{SO}$,
$\mathbf{a}_{2PN}$, $\mathbf{a}_{SS}$ and $\mathbf{a}_{RR}^{BT}$ are
the Newtonian, first post-Newtonian, spin-orbit, second
post-Newtonian, spin-spin and radiation reaction parts of the
acceleration, respectively \inserted{ \cite{Kidder}. Moreover,
${\mathbf{a}}_{PNSO}$ is the 1PN correction of the spin-orbit term
}\cite{TOO,FBB}\inserted{, ${\mathbf{a}}_{3PN}$ is the 1PN
contribution \cite{MoraWill}, ${\mathbf{a}}_{RR1PN}$ is the 1PN
correction to the radiation reaction term \cite{IyerWill}, and the
$\mathbf{a}_{RRSO}$ \cite{WillPNSO} and $\mathbf{a}_{RRSS}$
\cite{WangWill} terms are the spin-orbit and spin-spin contributions
to radiation reaction.} The analytic form of these contributions are
given in terms of the kinematic variables in Appendix B [see
equations (B.1)-(B.\inserted{11})].

\section{Using the CBwaves software}\label{CBwaves}
The software package contains man pages, a readme file and it
consists of a single executable file. The source, the i686 and
x86\_64 binary packages can be downloaded from the homepage of the
RMKI Virgo Group \cite{cbwavesdownload}. The parameters necessary
for the determination of the initial conditions are passed by a
human readable and editable configuration file. In order to
automatize and make the mass production of this type of
configuration files possible a generator script is also included.
The comprehensive list of all the possible configuration file
parameter and their detailed explanation can be found in the man
page. The most important input parameters are the initial separation
$\mathbf{r}=\mathbf{x}_1-\mathbf{x}_2=r\hat{\mathbf{n}}$, the masses
$m_i$, the magnitude $s_i$ of the specific spin vector
$\mathbf{s}_i$ and the initial eccentricity $e$. Note that instead
of the individual spin vectors $\mathbf{S}_1$ and $\mathbf{S}_2$ the
specific spin vectors $\mathbf{s}_i$---defined by the relations
$\mathbf{S}_i=\mathbf{s}_i m_i^2$---are applied. The magnitude $s_i$
of the specific spin vectors $\mathbf{s}_i=(s_{ix},s_{iy},s_{iz})$
are $s_i=\sqrt{s_{ix}^2+s_{iy}^2+s_{iz}^2}$ and it is assumed that
$0<s_i<1$ for a black hole while $0<s_i<0.7$ for most neutron star
models \cite{Kidder}.

\bigskip

Since simple analytic expressions are available, the determination
of the initial values of various parameters is performed with high
precision in case of circular orbits. The situation is different for
eccentric orbits, where the initial speed of the bodies is
determined iteratively by successive approximation to ensure that
the orbit possesses the required eccentricity after the first half
orbit in its orbital evolution. The currently implemented
approximation is set to yield initial data for the eccentricity with
$0.01 \%$ precision but the accuracy can be increased to any
desirable value of precision.

It is also worth to be mentioning that in order to make the
submission of the software to research clusters straightforward we
provide a Condor \cite{condor} job description file generator
script, as well. All these above listed features make this code an
easy-to-use, fully fledged gravitational wave generator, which
already produced some very interesting and promising results
discussed in the following sections.

\section{Results}\label{results}

Since analytic formulas are available for the motion and radiation
of circular non-spinning binaries, it was straightforward to start
our investigation with these systems, and focus our attention to
more complicated configurations in the succeeding subsections.

\subsection{Non-spinning, circular waveforms}\label{non-spin}

\begin{figure}[ht!]
\begin{tabular}{cc}
\includegraphics[width=0.35\textwidth, angle=-90]{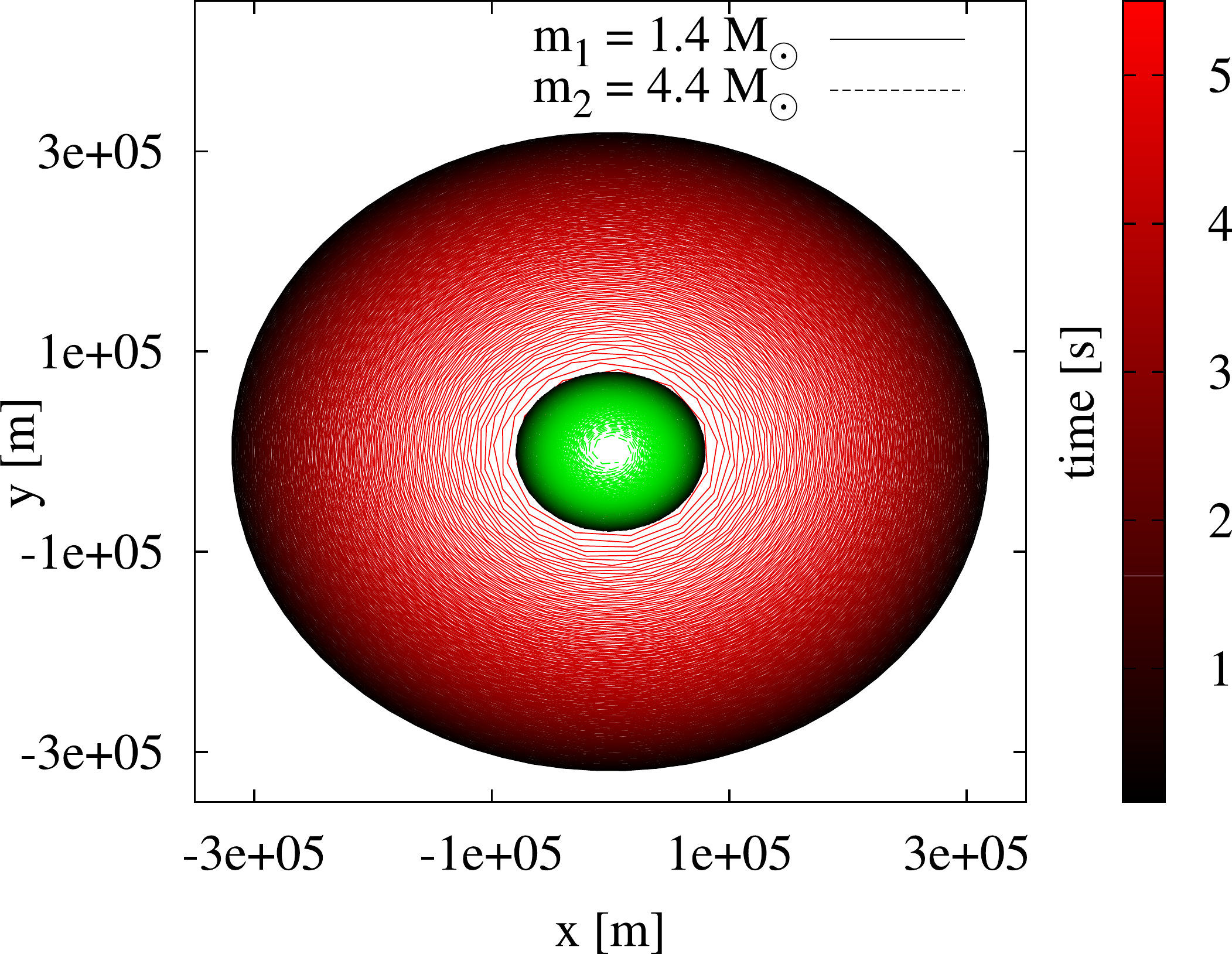} &
\includegraphics[width=0.35\textwidth, angle=-90]{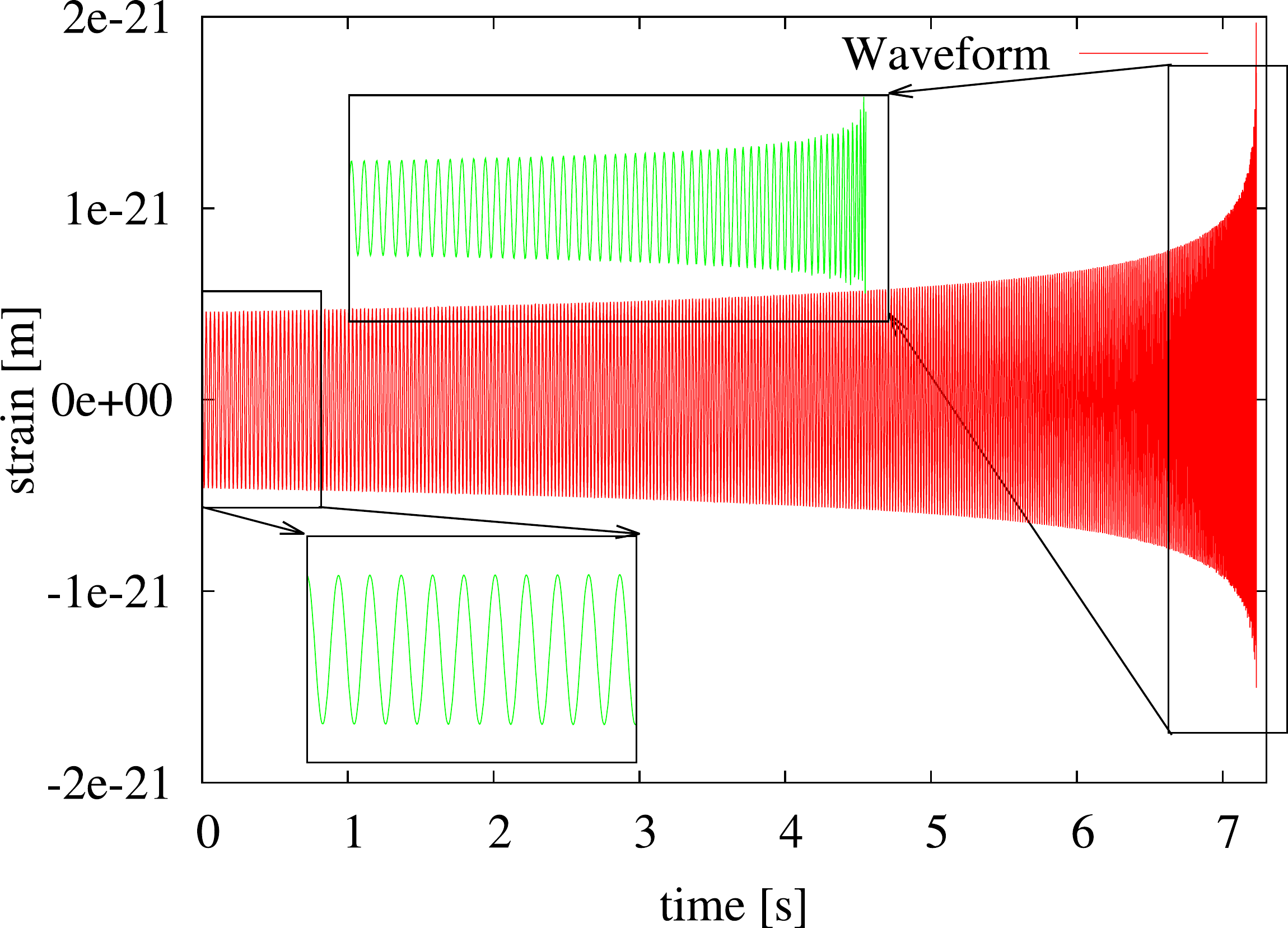}
\end{tabular}
\caption{\footnotesize The orbital evolution (left panel) and the yielded waveform (right panel) of
a non-spinning, circular binary system with masses $m_1 = 1.4\ M_{\odot}$, $m_2
= 4.4\ M_{\odot}$ and initial frequency $f_{low} \approx 18\ \mathrm{Hz}$ at $t=0$. The color shading for
the red curve is to indicate the passing of time while the green one---not
indicated---evolves respectively.} \label{nonspin-circular}
\end{figure}

We started our studies by constructing a non-spinning, circular
waveform template bank to serve as a reference for the forthcoming
investigations. For an immediate example of such a circular orbit
with non-spinning bodies and for the pertinent emitted waveform with
a slow rise in the amplitude and frequency see
Fig.\,\ref{nonspin-circular}.

\begin{figure}[ht!]
\hskip-0.3cm
\begin{center}
\begin{tabular}{c}
\includegraphics[width=0.5\textwidth, angle=-90]{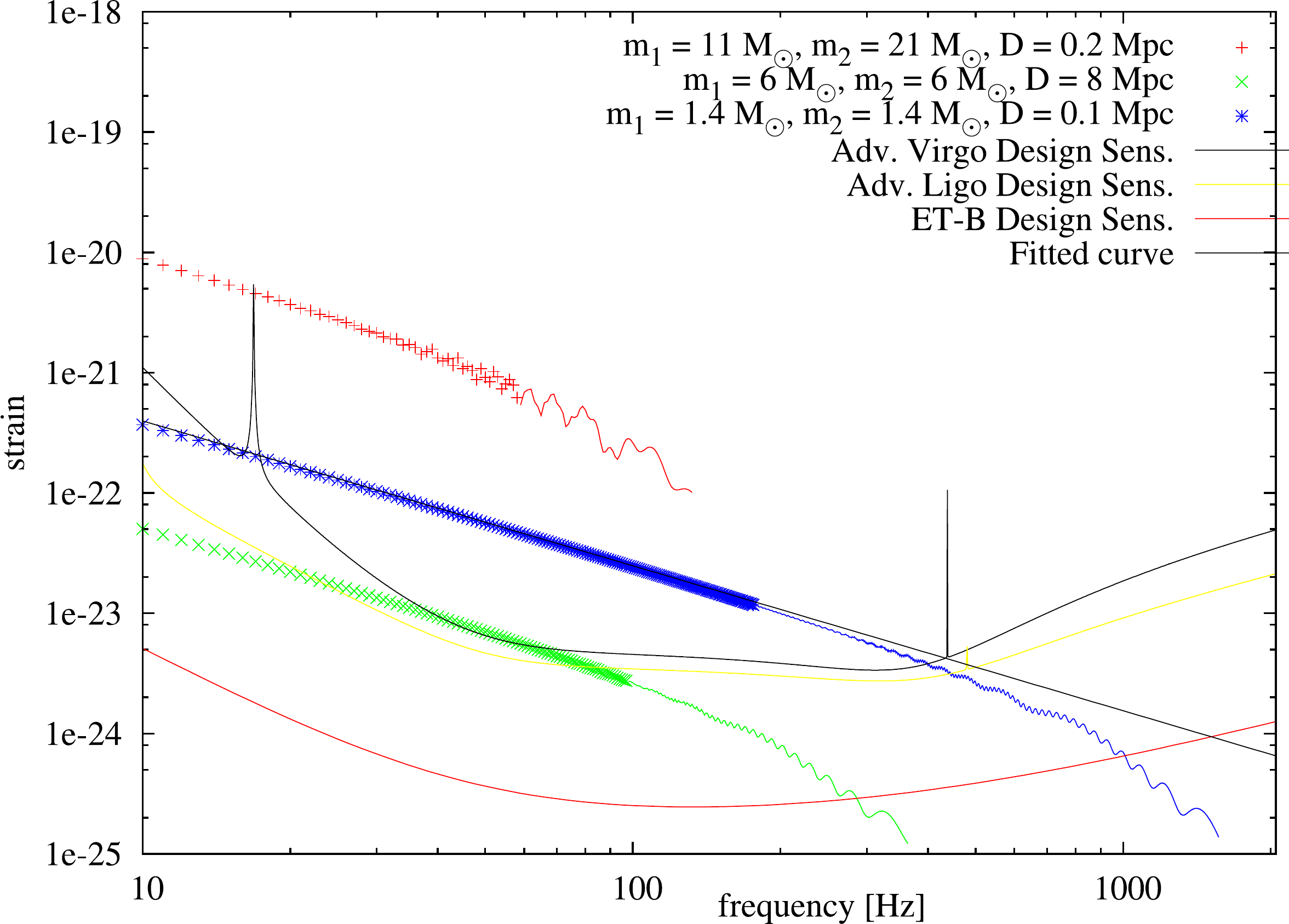} 
\end{tabular}
\end{center}
\caption{\footnotesize 
\color{black}{The spectral distribution of gravitational waves emitted by non-spinning circular binary
systems of various total masses and physical distances {are plotted with respect to} the design sensitivity curves of the  Advanced Virgo
\cite{advvirgo}, Advanced LIGO \cite{advligo} and the ET \cite{etsens} detectors. The spectra are shown from the initial
gravitational wave frequency ($f_{low} = 10\ \mathrm{Hz}$) up to the
frequency of the innermost stable circular orbit, $f_{isco}$.
The part {beyond} the limit of reliability of the
post-Newtonian approximation is indicated by thinner symbols.  {While the 40~Hz lower frequency cutoff
($f_{low}$) of current detectors limits the signal duration of a 1.4
- 1.4 $M_{\odot}$ binary to $\approx$ 26 seconds the value $f_{low}=10$~Hz for advanced detectors increases the signal length from the same binary up to $\approx$ 950 seconds!}}
} \label{circspectra}
\end{figure}

The detection pipelines based on the matched-filter methods are
primarily interested in the frequency domain representation and
phase/amplitude evolution of the waveforms both of which can easily
be obtained by making use of CBwaves. The spectral distribution of
such templates for various masses and distances are shown on
Fig.\,\ref{circspectra} with respect to the design sensitivity curve
of future interferometric gravitational wave detectors such as
Advanced LIGO \cite{advligo}, Advanced Virgo \cite{advvirgo} and the
Einstein Telescope \cite{etsens}. Note that the time interval while
the corresponding source will be visible by these detectors increase
significantly as the sensitivity is improved.

\begin{figure}[ht!]
\hskip-0.3cm
\begin{center}
\includegraphics[width=0.5\textwidth, angle=-90]{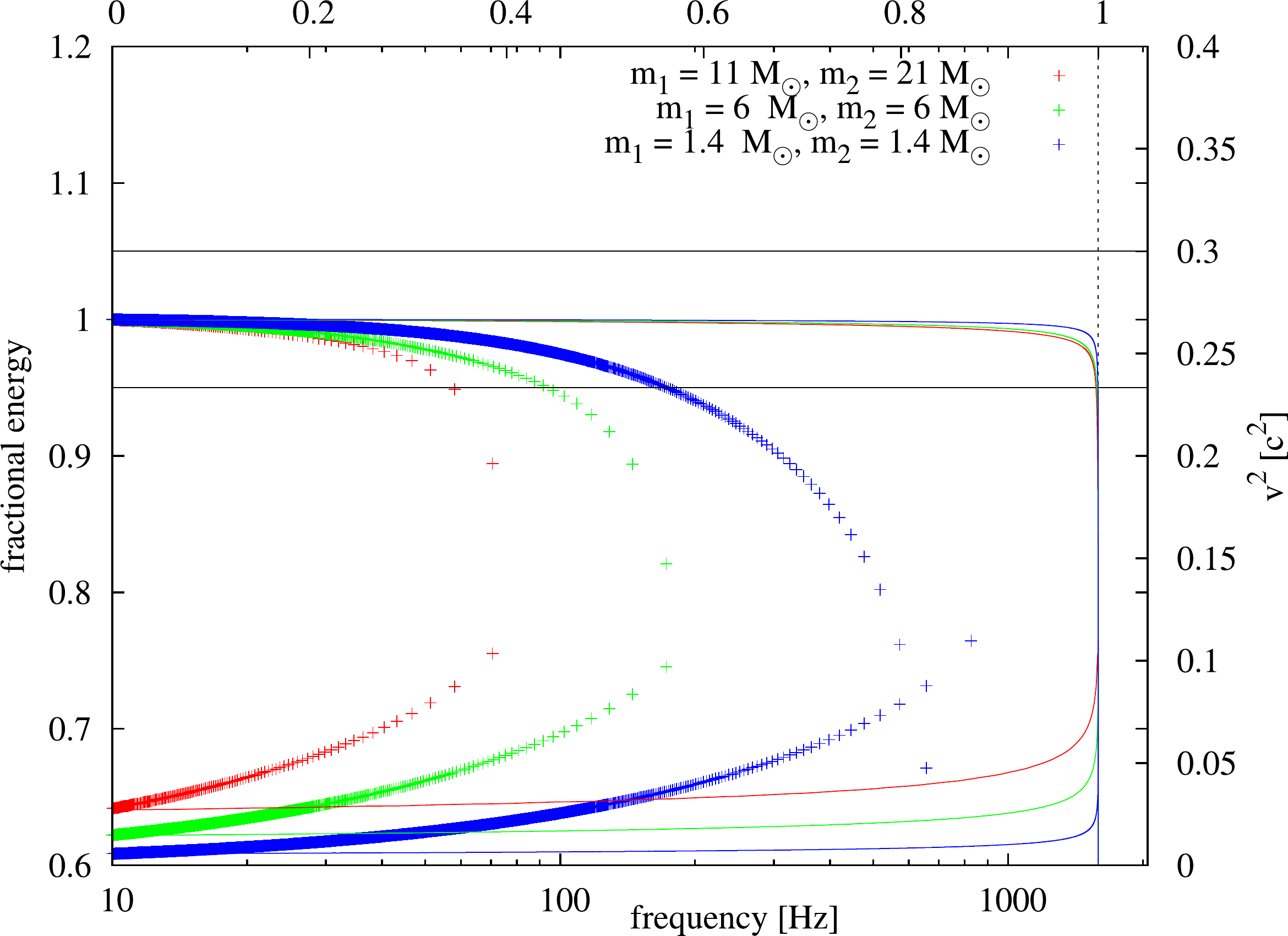}
\end{center}
\caption{\footnotesize
\inserted{On the upper part }\inserted{t}he frequency and ``fractional
time'' dependencies of the fractional energy\inserted{, defined by Eq.\,(\ref{FracEnergy0}),} are shown for the\inserted{ very same} three non-spinning circular binary systems \inserted{as on  Fig.\,\ref{circspectra}}. (The fractional time is \inserted{determined by dividing the} \inserted{temporal} time \inserted{by} the \inserted{total} evolution time\inserted{ of the considered system}.)
\inserted{The two thin horizontal lines, around the value $1$, indicate
the excess of $\pm5\%$ error in the fractional energy}.
\inserted{On the lower part} the \inserted{analogous} frequency and fractional time
dependencies of the PN parameter $\epsilon\inserted{\sim v^2}$, are indicated.
\inserted{On both the upper and lover parts the frequency and fractional time dependencies are indicated by crosses and solid lines, respectively.}
\inserted{Notice that the frequency and fractional time dependencies of both of the monitored quantities differ significantly. Simply, by inspecting the fractional time dependence it is visible that the energy balance relation holds
almost for the entire orbital evolution. However, from data analyzing point of view
it is the frequency domain behavior what counts and the frequency dependence is much less satisfactory especially if the frequency ranges of our current detectors are also taken into account.}
} \label{circspectra2}
\end{figure}

Recall that the simplest analytic models, like the stationary phase
approximation (see e.g. \cite{SPA}), provide estimates for the
amplitude evolution of the gravitational wave spectra as a function
of frequency \inserted{which, at leading order approximation, is}
proportional to $\sim f^{-7/6}$ \inserted{\cite{spawaves}}. Note that the
trustful part of the spectrum of the 1.4 $M_{\odot}$ - 1.4
$M_{\odot}$ NS binary system (shown on Fig. \ref{circspectra})
yields a fit of the form $|\tilde{h}(f)| = 1.019\cdot
10^{-19}*f^{-1.205}$, where the value $1.205$ in the exponent is a
bit larger but it is in 3$\%$ agreement with the analytically
derived value $7/6$.

\subsection{\inserted{The validity of the adiabatic approximation}}\label{non-spin-adiabb}

\inserted{Within the post-Newtonian community there are certain expectations concerning the suitability of the adiabatic approximation in determining the succeeding phases. More specifically, it is expected that whenever the time scales of both the precession and shrinkage of the orbits are long when they are compared to the orbital period the adiabatic approximation is appropriate. This, in particular, means that quasi-circular inspiral orbits are expected to be well approximated by nearly circular ones with a slowly shrinking radius (see for e.g. Section IV in \cite{Kidder}). In such an adiabatic approximation the rate of the inspiral is expected to be properly determined by the relation}
\begin{equation}\label{adiab}
\inserted{\frac{d{r}_{ad}}{dt}=\frac{{dE}/{dt}}{{dE}/{dr}}\,,}
\end{equation}
\inserted{where the terms $dE/dt$ and $dE/dr$ are assumed to be evaluated by applying Eq.~(\ref{dEdt}) - (\ref{dEdtPNSO}) and (\ref{Energy}), respectively, along with instantaneous substitution of the  temporal values of all variables involved in the corresponding expressions. Note that in the particular case of PN order applied in \cite{Kidder} the rate of inspiral ${d{r}_{ad}}/{dt}$ can be given as Eq.~(4.12) of \cite{Kidder}.}

\begin{figure}[ht!]
\begin{center}
\includegraphics[width=0.75\textwidth]{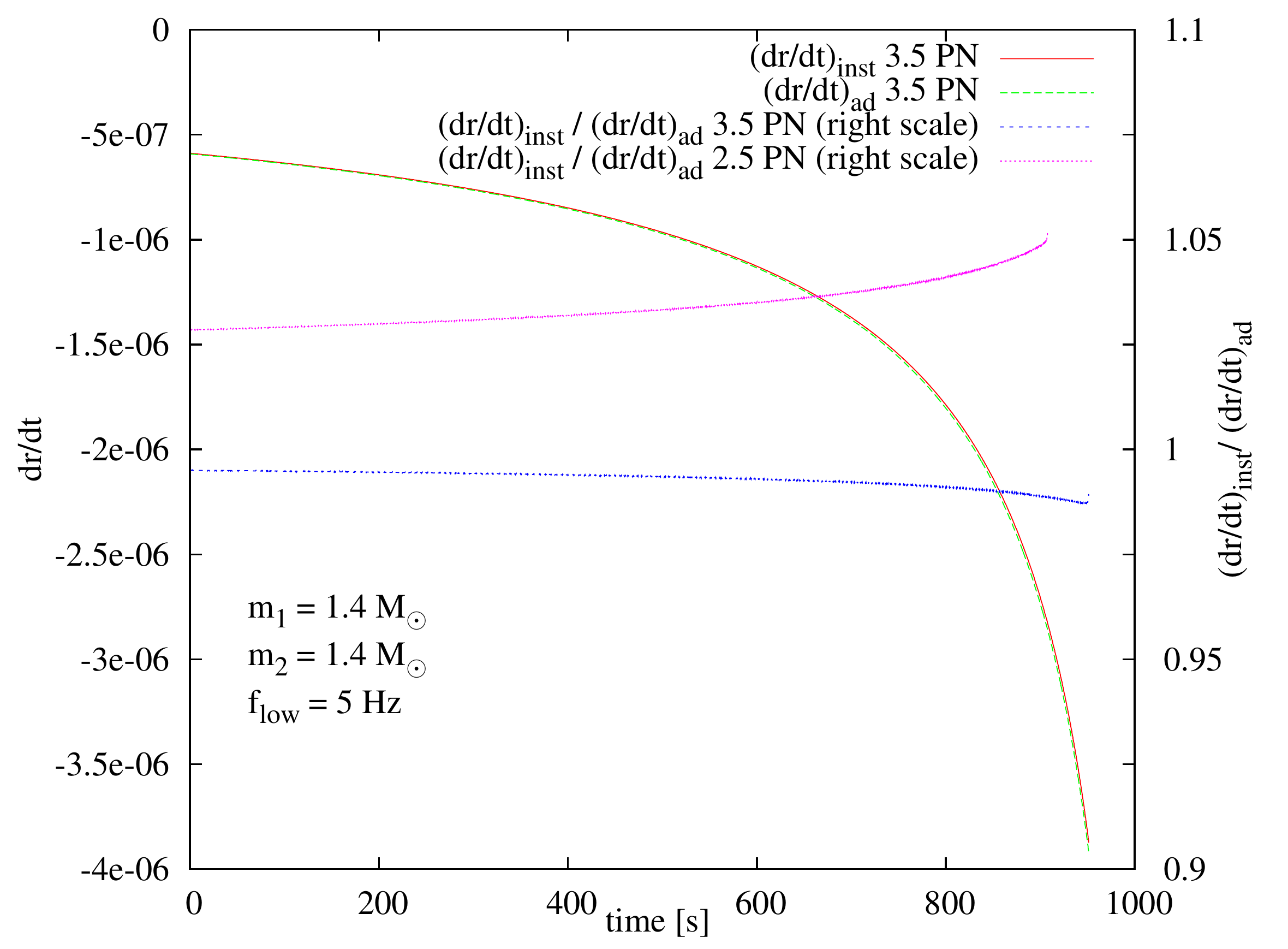}
\end{center}
\caption{\footnotesize
\inserted{
\color{black}The time dependence of the rates of inspiral ${d{r}_{inst}}/{dt}$ and ${d{r}_{ad}}/{dt}$, along with their ratios are plotted in two cases. First by applying the 2.5PN order setup as in \cite{Kidder} and second by using the highest possible PN orders implemented in CBwaves. \color{black}{The graphs of these ratios---beside making the improvements related to use of higher PN orders to be transparent---do indicate that the rate of inspiral is merely negligibly faster for adiabatic evolution.}
}} \label{adiabradius}
\end{figure}

\inserted{We have carried out the investigation of the validity of the adiabatic approximation by monitoring the ratio $({d{r}_{inst}}/{dt})/({d{r}_{ad}}/{dt})$ of the rate of the instantaneous inspiral ${d{r}_{inst}}/{dt}$, determined by using CBwaves in time evolution, and the rate of the adiabatic inspiral ${d{r}_{ad}}/{dt}$ evaluated as described above.} {\color{black}On Fig.~\ref{adiabradius} the corresponding ratios are plotted for both using the approximation considered in Section IV in \cite{Kidder} and the approximation involving all the PN terms implemented in CBwaves. According to the graphs on Fig.\,\ref{adiabradius} it is visible that in both of the monitored cases the adiabatic approximation yields almost the same (less then $5\%$) decrease of the rate of inspiral than the corresponding time evolution
if we used the highest possible PN orders implemented in CBwaves. The graphs of these ratios $({d{r}_{inst}}/{dt})/({d{r}_{ad}}/{dt})$ make also transparent the slight improvements related to use of higher PN orders.}

\subsection{The gauge freedom in the radiation reaction term}\label{rad-reac}

The radiation reaction is determined by assuming that the energy
radiated to infinity is balanced by the an equivalent loss of energy
of the binary system. It is know for long that the relative
acceleration term appearing in the radiation reaction expressions is
not unique. Indeed, it was already shown in \cite{IyerWill} that at
2.5PN order there is a two-parameter family of freedom in specifying
the radiation reaction terms such that for any choice of these two
parameters the loss of energy and angular momentum is in accordance
with the quadrupole approximation of energy and angular momentum
fluxes. This freedom corresponds to possible coordinate
transformations at 2.5PN and it represent\inserted{s} a residual
gauge freedom that is not fixed by the energy balance method. In
spite of the two-parameter freedom in the literature two specific
choices of the radiation reaction terms are applied. One of them was
derived from the Burke-Thorne radiation reaction potential
\cite{Kidder,BT1,BT2}
\begin{equation}
\mathbf{a}_{RR}^{BT} = {\frac{8}{5}}\eta
{\frac{G^{2}m^{2}}{c^{5}r^{3}}}\left\{
\dot{r}\mathbf{\hat{n}}\left[ 18v^{2}+{\frac{2}{3}}{\frac{Gm}{r}}-25\dot{r}%
^{2}\right] -\mathbf{v}\left[
6v^{2}-2{\frac{Gm}{r}}-15\dot{r}^{2}\right] \right\}\,, \label{aRRBT}
\end{equation}
while the other is the Damour-Deruelle radiation reaction formula
\cite{DDPLA,MoraWill}
\begin{equation}
\mathbf{a}_{RR}^{DD} ={\frac{8}{5}}\eta
{\frac{G^{2}m^{2}}{c^{5}r^{3}}}\left\{ \dot{r}\mathbf{\hat{n}}\left[
3v^{2}+{\frac{17}{3}}{\frac{Gm}{r}}\right] - \mathbf{v}\left[
v^{2}+3{\frac{Gm}{r}}\right] \right\}\,.
\end{equation}
Despite of the explicit functional differences in these two
radiation reaction terms---based on the above recalled argument
ending up with the gauge freedom in determining the motion of the
bodies---it is held (see, e.g.\,\cite{IyerWill}) that the energy
balance relation has to be insensitive to the specific form of the
applied radiation reaction term. To check the validity of this claim
we implemented both of these radiation reaction terms in CBwaves.

We have found that whenever a suitable coordinate transformation of the form
\begin{equation}\label{trafo}
\mathbf{x}'=\mathbf{x} + \delta_{\mathbf{x}_{2.5PN}}\,,
\end{equation}
(for the precise form of the correction term
$\delta_{\mathbf{x}_{2.5PN}}$ see equations (22) and (23) of
\cite{ZengWill}) and all the related implications are taken into
account in the total energy expression
\begin{equation}
E_{tot}=E_{N}+E_{PN}+E_{SO}+E_{2PN}+E_{SS}+E_{RR}\,, \label{Energy0}
\end{equation}
where $E_{RR}$ stands for the radiation reaction term\inserted{, see
Eq.~(\ref{dEdt}) of appendix B.}, i.e.\,the energy associated with the emitted
gravitational wave, in accordance with the claims in
\cite{IyerWill}, the energy balance relations are insensitive to the
specific choices of the parameters in the radiation reaction terms.

\begin{figure}[htb]
\begin{center}
\includegraphics[width=0.75\textwidth]{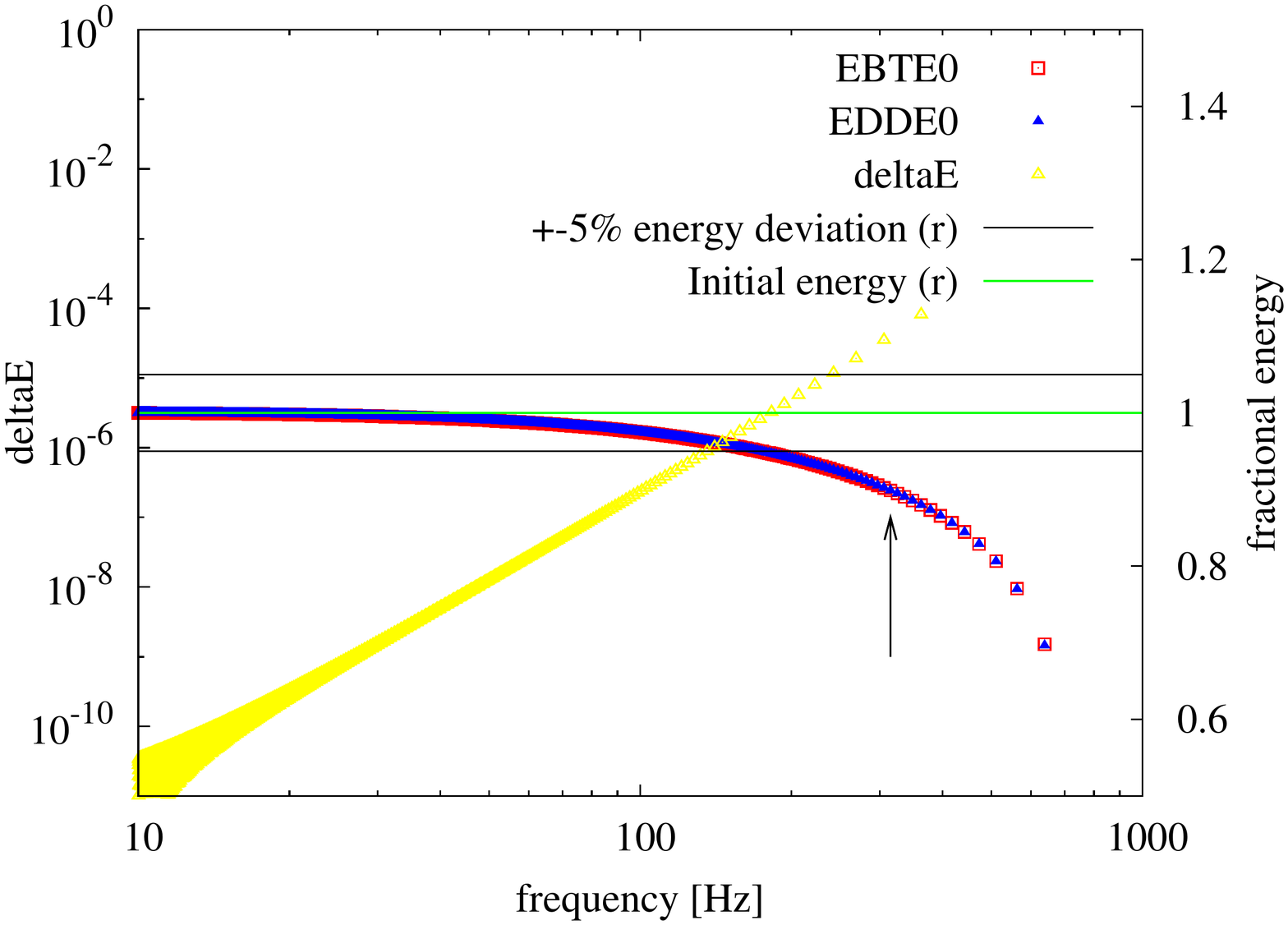}
\end{center}
\caption{\footnotesize The frequency dependence of the relative
difference $\delta E=(E^{BT}-E^{DD})/E^{BT}$ is shown (yellow). In
addition, the frequency dependence of the fractional energies
$E^{BT}_{tot}/E_0$ and $E^{DD}_{tot}/E_0$ for a non-spinning binary
on circular orbit with masses $m_1=m_2=1.4M_{\odot}$ are plotted.
The location where the deviation of the fractional energies exceeds
the value $10 \%$ is indicated by an upward pointing arrow. At that
point $\epsilon\sim 0.05$.} \label{RRformula}
\end{figure}

\inserted{In carrying out the pertinent investigations} instead of the energy balance relation it is more informative to consider the evolution of the fractional energy of the system
\begin{equation}
\frac{E_{tot}(f)}{E_0}\,,  \label{FracEnergy0}
\end{equation}
where $E_0$ denotes the initial value of the total energy
$E_0=E_{tot}(f_{low})$. Thus, the fractional energies
$E^{BT}_{tot}/E_0$ and $E^{DD}_{tot}/E_0$ correspond to the
alternative use of the radiation reaction terms in
Eq.\,(\ref{Energy0}) proposed by Burke-Thorne and by
Damour-Deruelle. On Fig.\,\ref{RRformula} the frequency dependence
of the fractional energies $E^{BT}_{tot}/E_0$ and $E^{DD}_{tot}/E_0$
and that of the relative difference $\delta
E=(E^{BT}-E^{DD})/E^{BT}$ are plotted. It was found that $\delta E$
is less than $0.0001\%$ even at the frequency, around 380~Hz, where
the energy balance relation becomes inaccurate as the error of both
of the fractional energies $E^{BT}_{tot}/E_0$ and $E^{DD}_{tot}/E_0$
exceeds $~12 \%$ there for the simplest possible case\inserted{ for circular orbits} where neither of the involved bodies have spin.

\subsection{Domain of validity of the PN approximation}\label{dom-val}

It is widely held within the post-Newtonian community \inserted{(see e.g.\,section 9.5 of \cite{Blanchet06})} that the
applied approximations are reliable up to the frequency of the
innermost stable circular orbit given as $f_{isco} =
c^3/(6\sqrt{6}\pi G\,m)$, where $c$ is the speed of light, $m = m_1
+ m_2$ is the total mass of the system while $G$ stands for the
gravitational constant. As it was mentioned above all these
expectations on the range of applicability of the post-Newtonian
expansion is based on the assumption of adiabaticity, i.e.\,it is
assumed that the time scales of precession and shrinkage are both
long compared to the orbital period until the very late stage of the
evolution.

On contrary to these expectations there are more and more
indications that the post-Newtonian approximations leaves its range
of applicability once the post-Newtonian parameter $\epsilon\sim
(v/c)^2\sim Gm/(rc^2)$ reaches the values $\epsilon\sim 0.08-0.1$.
In particular, the graphs on Figs.\,\ref{circspectra}, on Fig.\,\ref{RRformula}, as well as, on the left panel of
Figs.\,\ref{eccevolfig} clearly justify that as soon as the value of
the PN parameter gets to be close to $0.08-0.1$ a significant violation
of the energy balance relation---\inserted{monitored} via the fractional
energy ${E_{tot}(f)}/{E_0}$ on these plots---starts to show up
regardless whether the motion of the binary is as simple as being
circular or complicated with the inclusion of spin(s) and/or
eccentricity.

{As our conclusions are on contrary to the conventional
expectations it is important to emphasize that the observed
violation of the energy balance relation was found to be robust with
respect to the variation of the parameters of the investigated
binary systems. In addition, although the results are numerical
ones, on the one hand, the convergence rate of the code was
justified to be fourth order and, on the other hand, all the shown
results are insensitive to the size of the applied time steps in the
sense that the included figures yielded with the use of $dt=1/(256
kHz)$ are already identical to those with $dt=1/(16 kHz)$.}

Note {also that our} conclusions are in accordance with the claims
of \cite{janna} where\inserted{---although the energy balance relation were not applied or evaluated in either way---} simply the relative significance of \inserted{various} higher order contributions were monitored. It is of convincing that in \cite{janna} by inspecting merely the anomalous growth of \inserted{some of} the higher order PN contributions the same range of applicability with $\epsilon\lessapprox 0.08-0.1$ had been found.

It is worth to be mentioned here that the observed violation of the
energy balance relation gets to be transparent only on the plots
made in the frequency domain while by inspecting merely the
corresponding time domain plots (see Fig.\,\ref{circspectra2}) one
might be ready to conclude that the energy balance relation holds
almost \inserted{for} the entire orbital evolution. In this respect it
is important to be emphasized that from data analyzing point of view
it is the frequency domain behavior what \inserted{plays the central role}. In addition, it is
really unfavorable that the observed loss of accuracy is getting to be more and more
significant as we are approaching the sensitivity ranges of our
current\inserted{ly upgraded} ground based GW detectors. \inserted{It would be useful to know whether the fully general relativistic simulations may be free of the analogous type of violations of the energy balance relations, when they are monitored in the frequency domain.}

\subsection{Eccentric motions}\label{ecc-motion}

Due to radiation reaction the motion of eccentric binaries are
expected to be circularized. By making use of CBwaves we examined
the basic features of this circularization process and compared our
findings to the pertinent results of the literature.

Before reporting about our pertinent results it is worth to be
mentioned that there are several attempts (see, e.g.
\cite{gpv3,yunes,MGS}), aiming to provide analytic expressions for
the instantaneous value of the eccentricity. On contrary to our
expectations neither of these analytic expressions were found to be
satisfactory except for certain very narrow parameter intervals and
in most of the cases these analytic expressions yielded completely
inconsistent values everywhere else. We have found, however, that
the simplest possible geometric definition of the eccentricity,
referring to the main characteristic parameters of the orbit---i.e.,
the minimal and maximal separations of the bodies---, always yields
completely satisfactory result. This geometric, although not
instantaneous, eccentricity is defined as
\begin{equation}
e = \frac{r_{max}-r_{min}}{r_{max}+r_{min}},
\end{equation}
where $r_{max}$ and $r_{min}$ denote the maximum and minimum
distances between the two masses, i.e.\,the distances at the succeeding
`turning' points.

\subsubsection{Frequency modulation}

\begin{figure}[ht!]
\begin{tabular}{cc}
\includegraphics[width=0.45\textwidth]{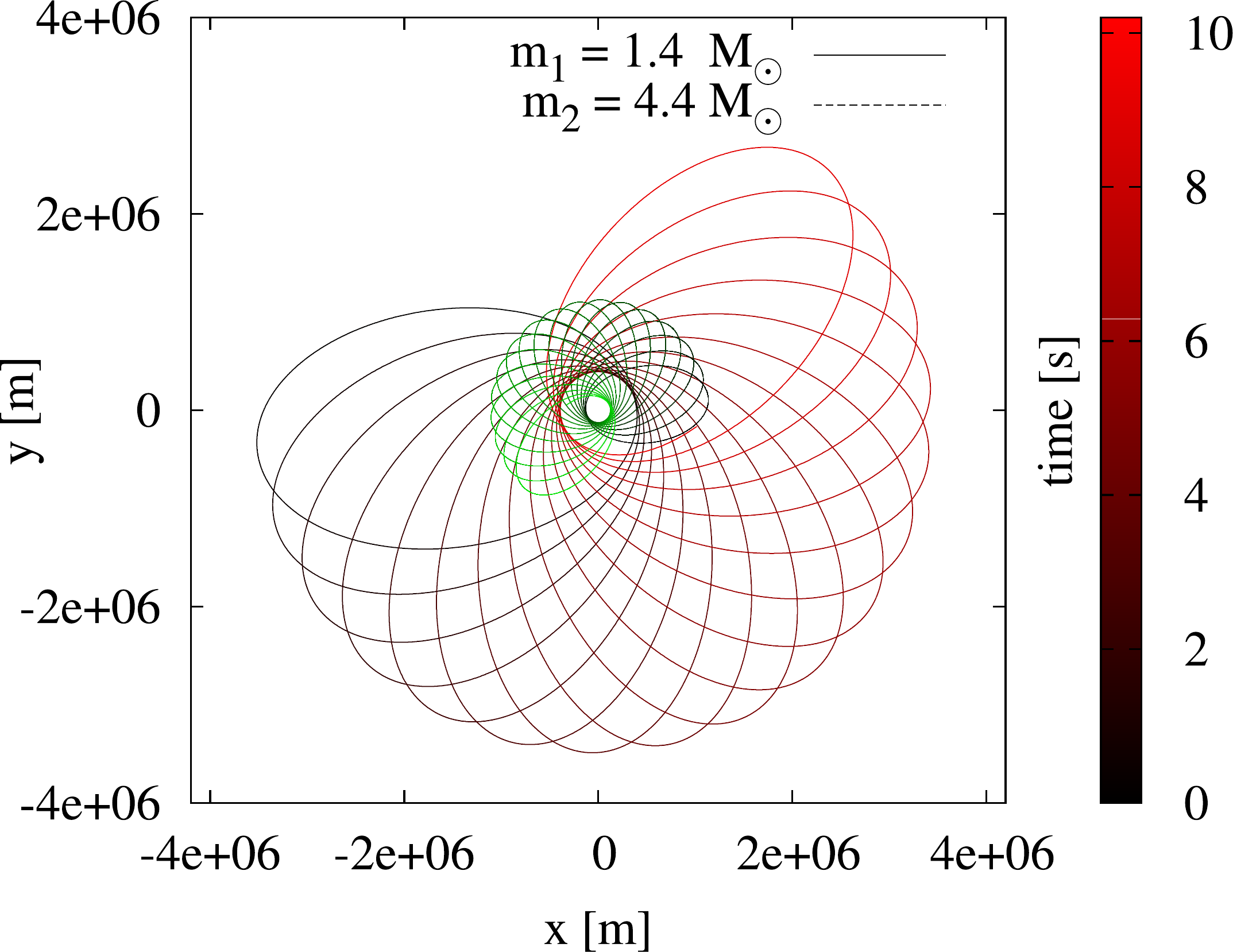} &
\includegraphics[width=0.50\textwidth]{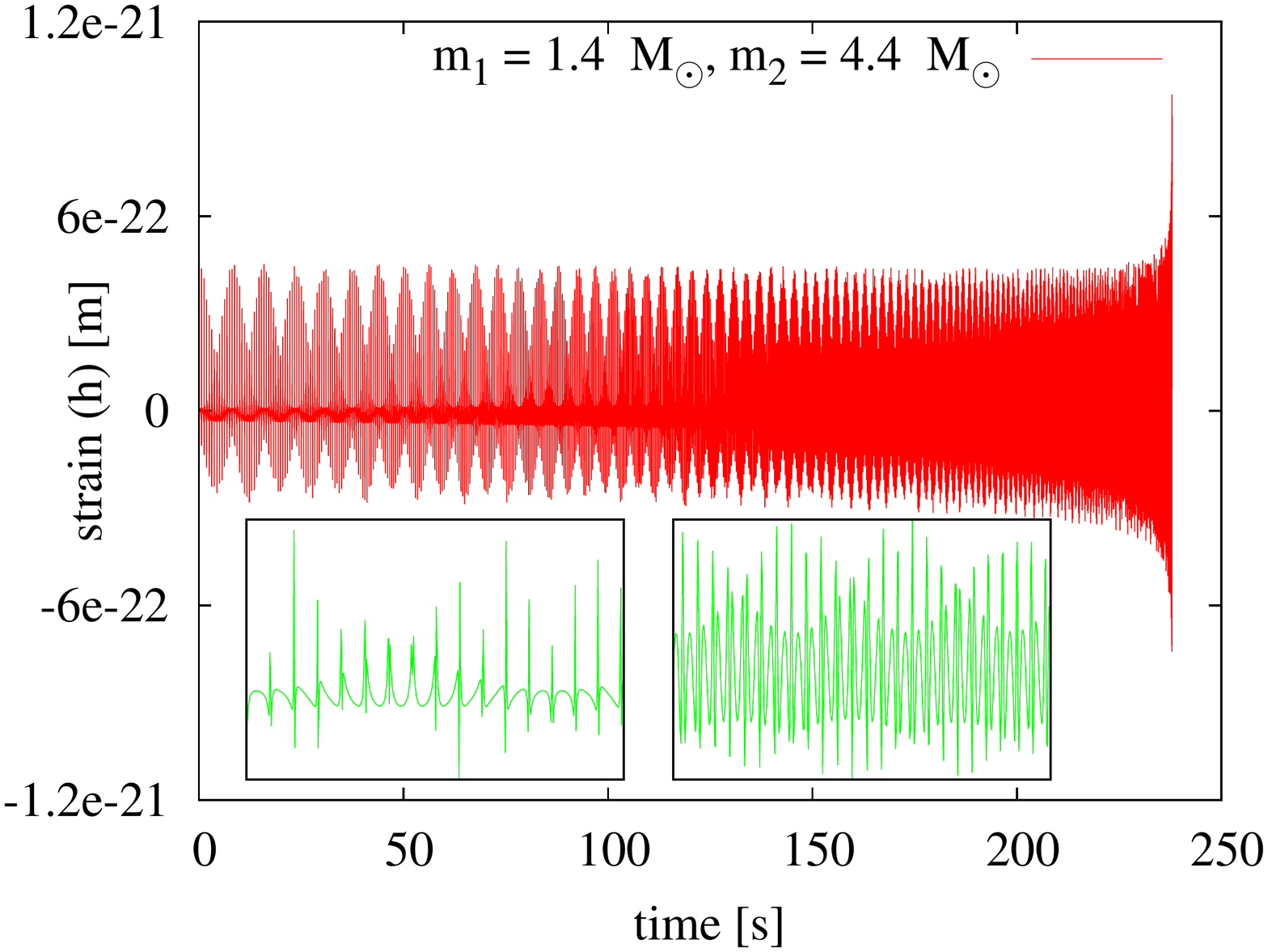}
\end{tabular}
\caption{\footnotesize The orbital evolution (left panel) and the emitted waveform (right panel)
of a non-spinning, eccentric binary system with masses $m_1 = 1.4\ M_{\odot}$ and $m_2 = 4.4\ M_{\odot}$, and with initial eccentricity $e_{flow} = 0.8$. The evolution starts at frequency
$f_{low} \approx 18\ \mathrm{Hz}$. The color shading for the red and green curves is to indicate the passing of time.}
\label{nospineccfig}
\end{figure}

Let us start by investigating the evolution of a highly eccentric
binary system with masses $m_1 = 1.4\ M_{\odot}$ and $m_2 = 4.4\
M_{\odot}$, and with initial eccentricity $e_{flow} = 0.8$ at
initial frequency $f_{low} \approx 18\ \mathrm{Hz}$. It is clearly
visible that due to non-negligible eccentricity the waveform suffers
simultaneous amplitude and frequency modulations. On
Fig.\,\ref{nospineccfig} a short interval of the orbital evolution
and pertinent waveform is shown for this eccentric binary system.

\subsubsection{Evolution of the eccentricity}

\begin{figure}[ht!]
\begin{tabular}{cc}
\includegraphics[width=0.48\textwidth]{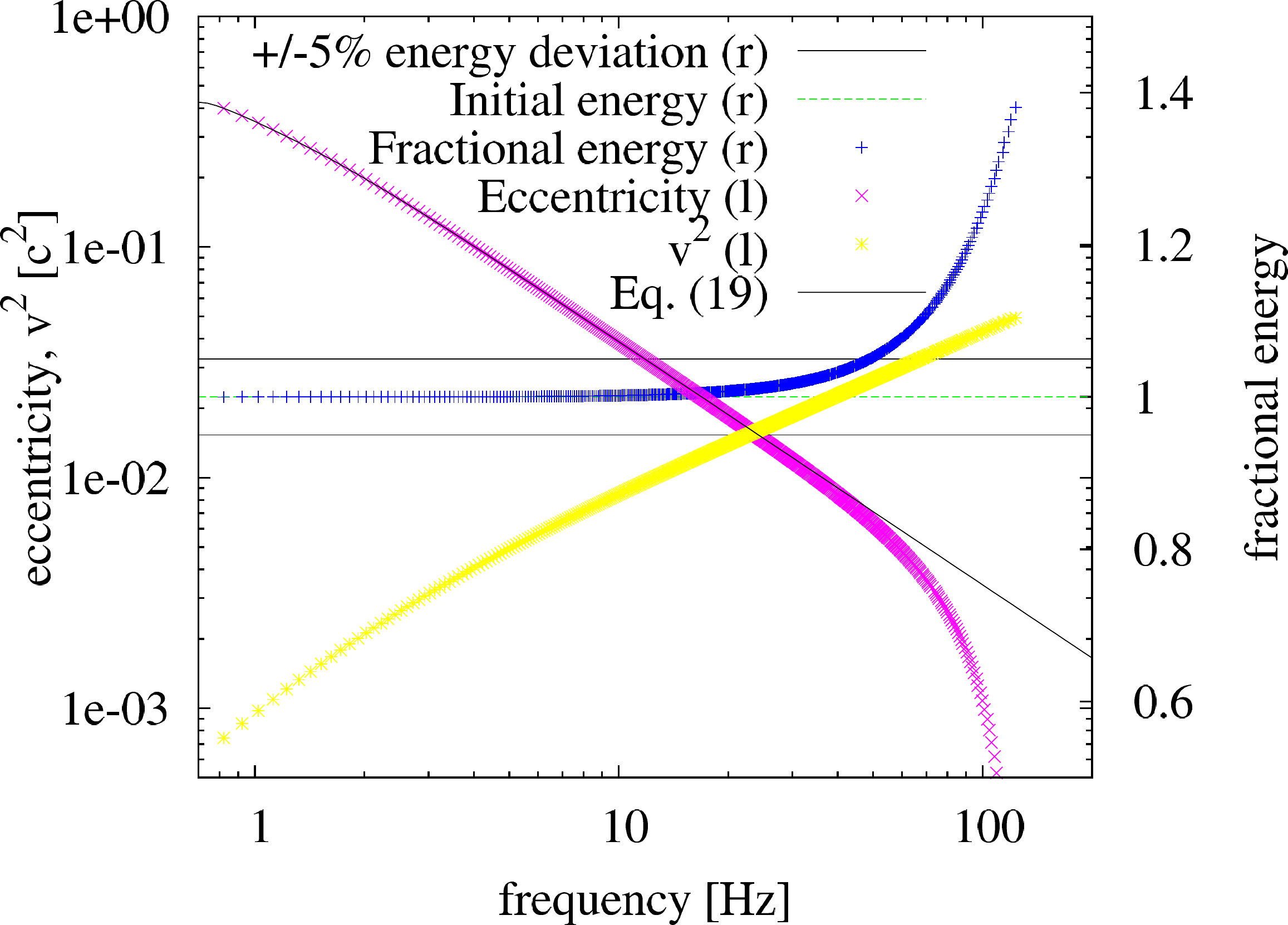} &
\includegraphics[width=0.48\textwidth]{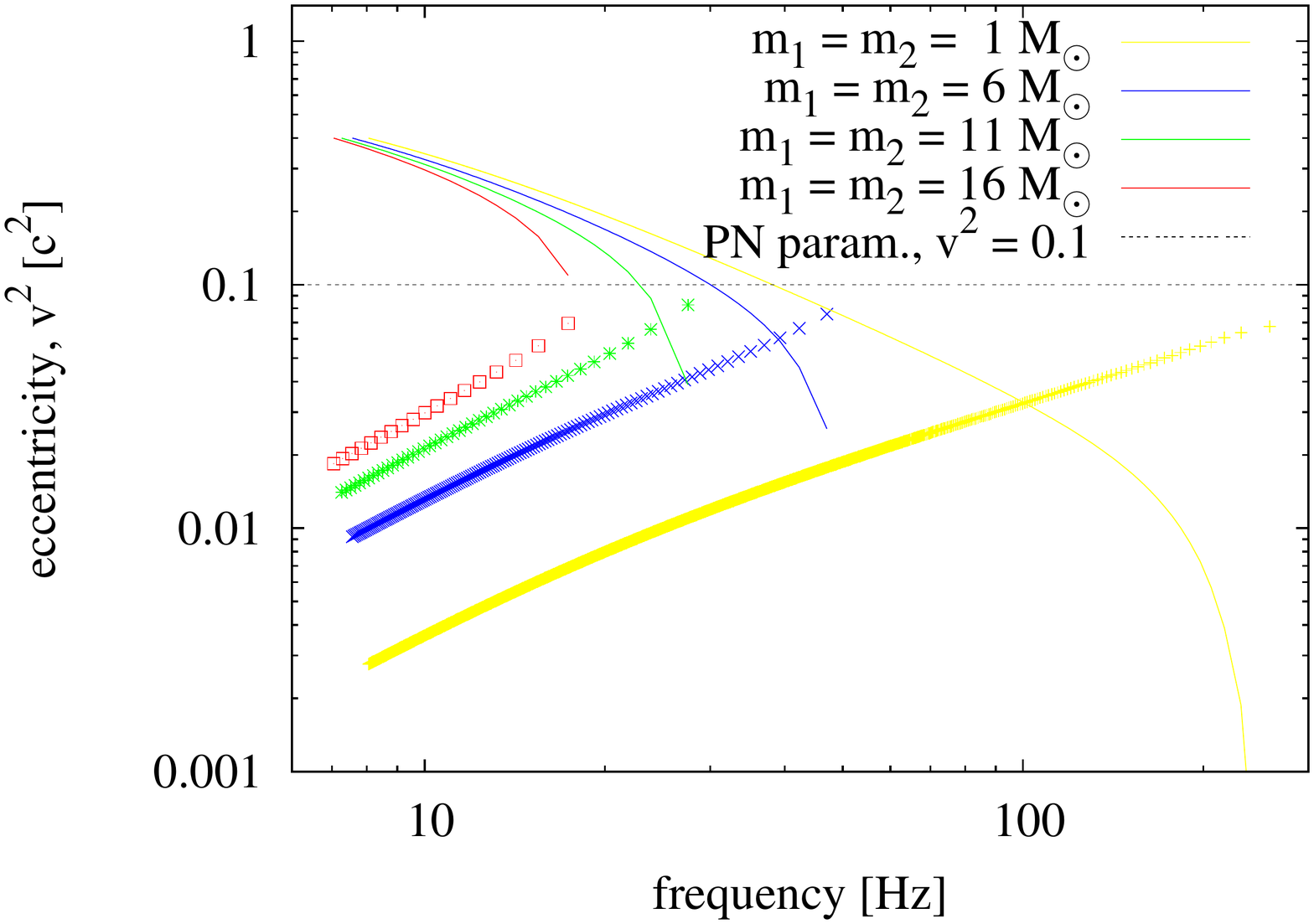}
\end{tabular}
\caption{\footnotesize {\it Left panel:} The evolution of the
eccentricity as a function of the frequency of a binary neutron star
system with masses $m_1 = m_2 = 1.4\ M_{\odot}$, and with initial
eccentricity $e_{flow} = 0.4$ at initial frequency $f_{low} \approx
0.5\ \mathrm{Hz}$. The fractional energy of the system (including
radiated energy) and the value of the post-Newtonian expansion
parameter $v^2 \approx m/r$ are shown. \color{black}{The latter is plotted} in order to indicate the
range of the validity of the applied approximation in the late
inspiral phase. {\it Right panel:} The evolution of the eccentricity
as a function of the frequency for binary systems with various total
masses. Here the post-Newtonian expansion parameter $v^2 \approx
m/r$ are also shown \color{black}{separately} for each total mass in order to indicate the
range of the validity of the PN approximation. } \label{eccevolfig}
\end{figure}

\begin{figure}[ht!]
\begin{tabular}{cc}
\includegraphics[width=0.35\textwidth, angle=-90]{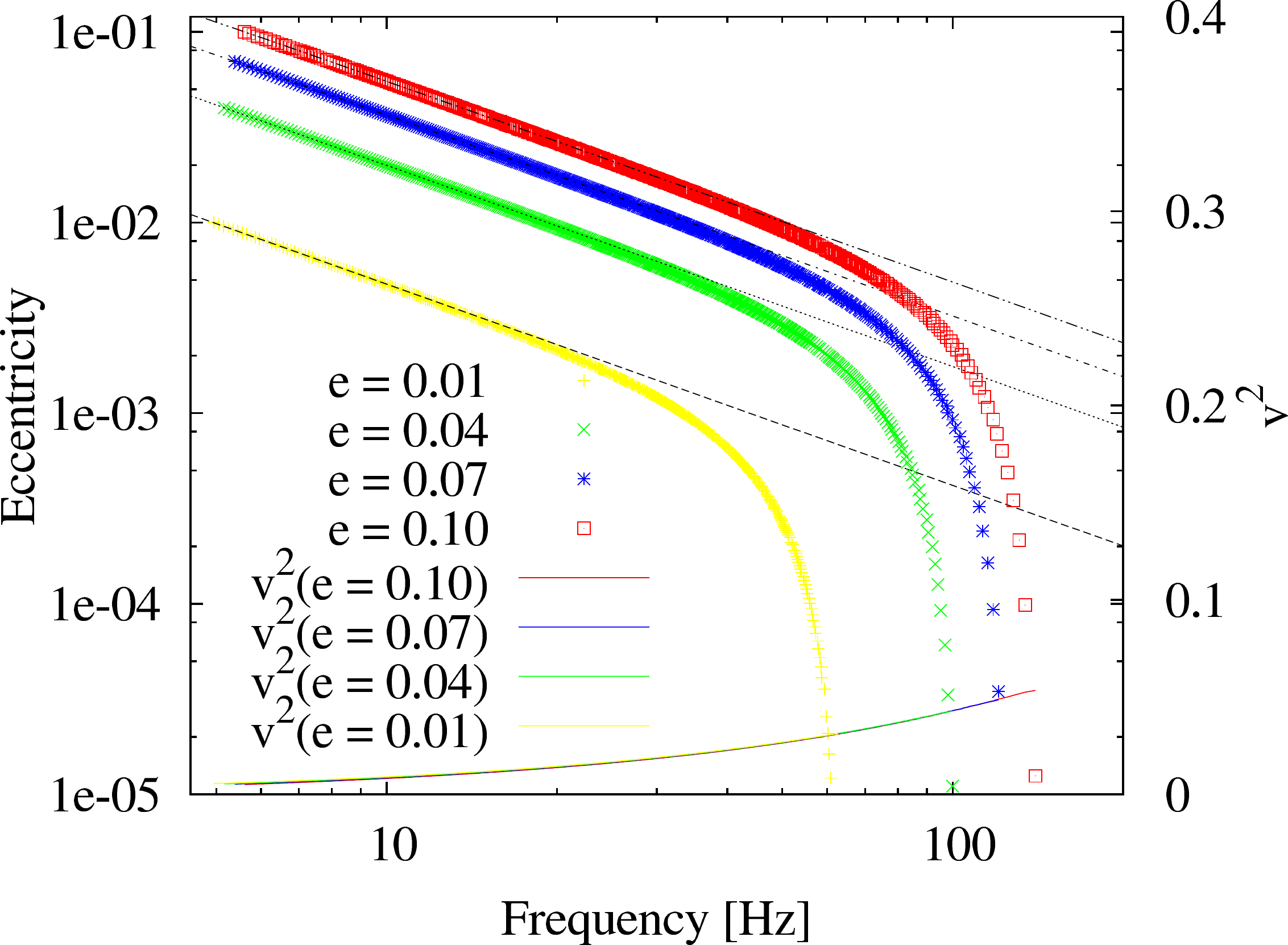} &
\includegraphics[width=0.35\textwidth, angle=-90]{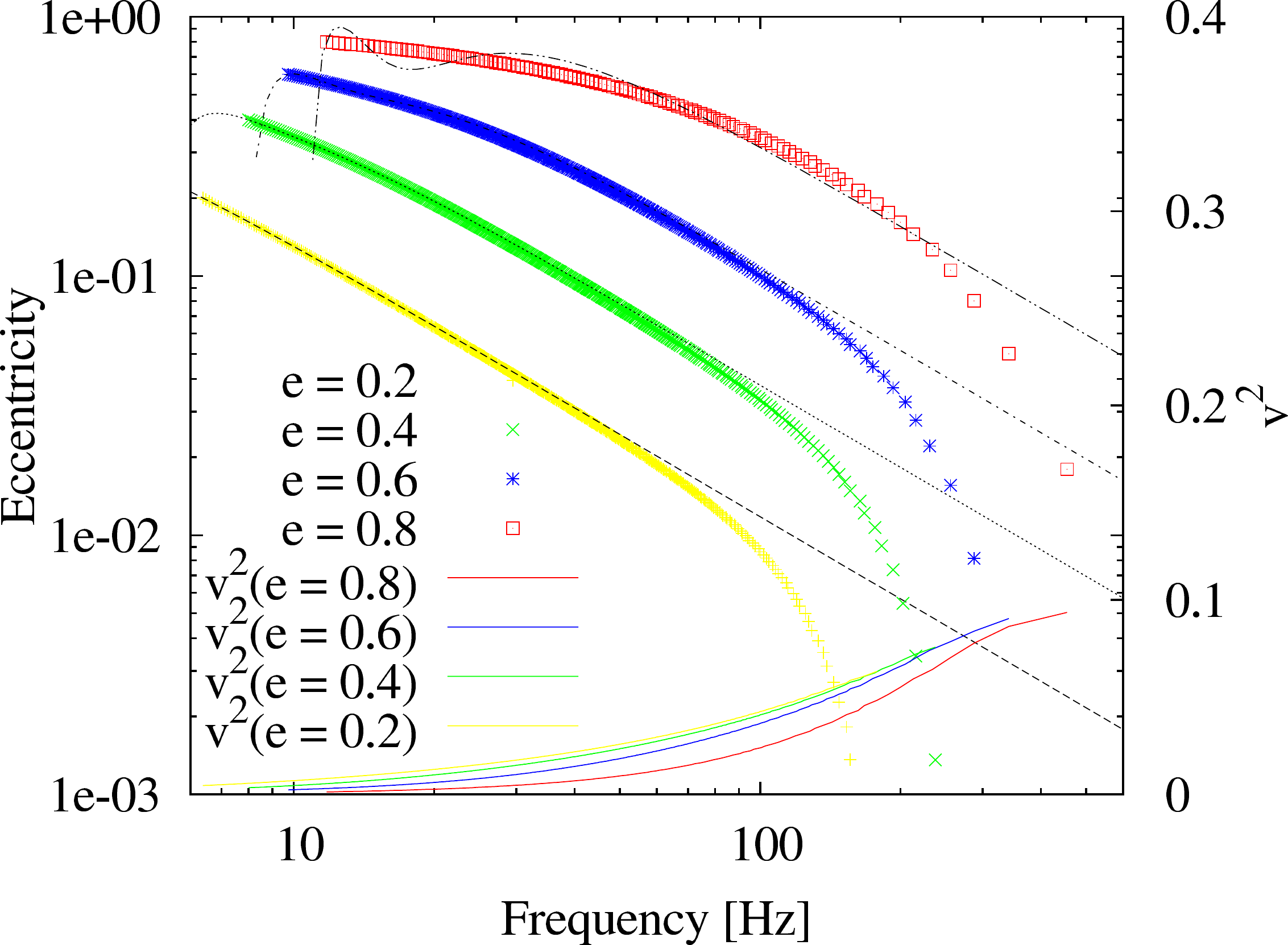}
\end{tabular}
\caption{\footnotesize The evolution of the eccentricity for a
binary system with masses $m_1 = m_2 = 1.4\ M_{\odot}$ and $f_{low}
\approx 5~Hz$ initial orbital frequency for small (left panel) and
high (right panel) values of initial orbital eccentricity. The \color{black}{thin black lines indicate}  the theoretical estimates \color{black}{given} by Eq.\,(\ref{eccequation}). The solid line shows the theoretical estimates as dictated by Eq.\,(\ref{eccequation}). The evolution of the post-Newtonian expansion parameter $v^2 \approx m/r$ is also shown with the same color for each eccentricity.} \label{eccevolscan}
\end{figure}

The gravitational wave detection pipelines are mainly using circular
waveform templates. This approach is based on the assumption that
the binaries are circularized quickly so that by the time the
emitted GWs enter the lower part of the frequency band of the
detectors the orbits are---with good approximation---circular.
Therefore it is crucial to check the validity of this assumption
within the post-Newtonian approximation implemented in CBwaves. The
relevant figures are presented by the two panels of
Fig.\,\ref{eccevolfig}. On the left panel the evolution of the
eccentricity as a function of frequency for such a binary NS system
is shown. For various total masses a comparison to the analytic
formula
\begin{eqnarray}
e=e_0\cdot\chi^{-\frac{19}{18}}\cdot\Big(
1+\frac{3323}{1824}\,e_0^2\,(1-\chi^{-\frac{19}{9}}) +
\frac{15994231}{6653952}\,e_0^4\,\Big(1-\frac{66253974}{15994231}\,\chi^{-\frac{19}{9}}
+ \nonumber \\
+ \frac{50259743}{15994231}\,\chi^{-\frac{38}{9}}\Big) +
\frac{105734339801}{36410425344}\,e_0^6\,\Big(1-\frac{1138825333323}{105734339801}\,
\chi^{-\frac{19}{9}} + \nonumber \\
+  \frac{2505196889835}{105734339801}\,\chi^{-\frac{38}{9}}
- \frac{1472105896313}{105734339801}\,\chi^{-\frac{19}{3}}\Big)\Big)
\label{eccequation},
\end{eqnarray}
derived in \cite{yunes}---see Eq.\,(3.11) therein---, where $\chi$ =
$f/f_0$, i.e.\,the ratio of the instantaneous and initial frequency,
while $e_0$ is the value of eccentricity at $f=f_0$, is also
indicated on both Figs.\,\ref{eccevolfig} and \ref{eccevolscan}.

Our numerical findings justify that the evolution of the
eccentricity as a function of frequency---at least in the early
phase---can be perfectly explained by the analytic estimates of
\cite{yunes}. However, it is important to note that towards the end
of the evolution in spite of the fact that the post-Newtonian
expansion parameter\inserted{, $\epsilon\sim(v/c)^2\sim Gm/(rc^2)$,} is still much below the critical upper bound $\sim 0.08-0.1$, where the PN approximation is supposed to
be valid, there is a non-negligible difference between the numerical
values and the analytical estimates. The explanation of this
\inserted{discrepancy} would deserve further investigations.

Notice, finally, that---as it is clearly transparent on the right panel of Fig.\,\ref{eccevolfig}---the
frequency dependence of the eccentricity is insensitive to the total mass or to the mass ratio.
This behavior is one of the universal properties of the investigated eccentric binary systems.

\subsubsection{Signal losses caused by eccentricity}

It is well know that the matched-filter method is the optimal one
when searching for known signals in noisy data \cite{matchedfilter}.
In practice this involves the set up of a so called template bank (a
collection of theoretical gravitational waveforms), which could
contain several hundreds of thousands of templates in the parameter
space of the physical model. Then each element of this template bank
is matched against the data. If the density of these templates is
high enough (see, e.g.\,\cite{spacing}) and the templates are based on the same physical model as the \inserted{anticipated}
signal, then it is straightforward to find the signal provided that
it has sufficient strength (amplitude). The situation is different
when the physical model behind the template bank differs from that
of the \inserted{arriving} signal\inserted{.} \inserted{It may happen that} the later is unknown or \inserted{or the signals come from} eccentric binary systems \inserted{while we are applying a} circular
template bank.

\inserted{According to the basic set up in matched filtering the overlap $\mathcal{O}_{s,t}$ between the time series of an expected signal, $s$, and an element, $t$, of the template bank is defined as}
\begin{equation}\label{cs}
\inserted{\mathcal{O}_{s,t} = \frac{ (s | t ) }{ \sqrt{(s | s )(t | t )} }\,,}
\end{equation}
\inserted{where the product ``$(.,.)$'' is given as}
\begin{equation}
\inserted{(a | b ) = 2 \int_{f_{min}}^{f_{max}}\frac{ \tilde{a}^*(f)\tilde{b}(f) + \tilde{a}(f)\tilde{b}^*(f)}{S_n(f)}df\,.}
\end{equation}
\inserted{In the latter equation $\tilde{a}, \tilde{b}$ stand for the frequency domain representation of the
quantities $a, b$ given in the time domain, while $S_n(f)$ is the spectral density of the detector noise.
{In addition, in determining the value of overlap one has to correct the above defined  quantity for the phase and duration differences of the expected signal, $s$, and an element, $t$, of the template bank. This, in practice, is done by marginalizing $\mathcal{O}_{s,t}$ over template phase and end time /coalescence time.}}

\inserted{As the signal-to-noise-ratio (SNR) is proportional to the numerator on the r.h.s of Eq.~(\ref{cs}) in case of significant drop of the overlap an analogous decrease of SNR may also be anticipated.}

\medskip

In ideal cases (when the \inserted{arriving signal and some members of the applied template bank are sufficiently close to each other}) the overlap is \inserted{also close to the value} 1. It often occurs that some \inserted{members of a} template bank give rise to higher value of overlaps with the
signal than the ones with \inserted{closer physical} parameters. Because of
this it is useful to \inserted{monitor} the so called {\it fitting factor}
\cite{ffactor} which is the maximum of the overlaps of all the
element of the template bank with the expected signal. While for
parameter estimation the overlap is the important quantity, for
detection purposes the fitting factor has more relevance.

Since a considerably large fraction of binaries \inserted{may} retain at least a tiny residual eccentricity during evolution it is important to ask in what extent this eccentricity may affect the detection performance of matched-filtering. \inserted{The investigation of the overlap of the eccentric waveforms with circular templates was done first in \cite{exc} based on the use of the initial LIGO sensitivity. It was reported in \cite{exc} that a tiny residual eccentricity may significantly decrease of the value of overlap.}

\inserted{By making use of the highly accurate wave generating setup of CBwaves we carried out an analogous investigation of the drop of overlap by applying the spectral density of the detector noise relevant for the advanced Virgo. In this respect our results are complementary to that of \cite{exc}. We have found that in consequence of the increase of the sensitivity of the advanced Virgo the drop of overlap is even more significant than it was for initial LIGO basically because the signals and templates spend longer period within the sensitivity ranges of \color{black} advanced Virgo.}

\begin{figure}[ht!]
\begin{center}
\includegraphics[width=0.5\textwidth, angle=-90]{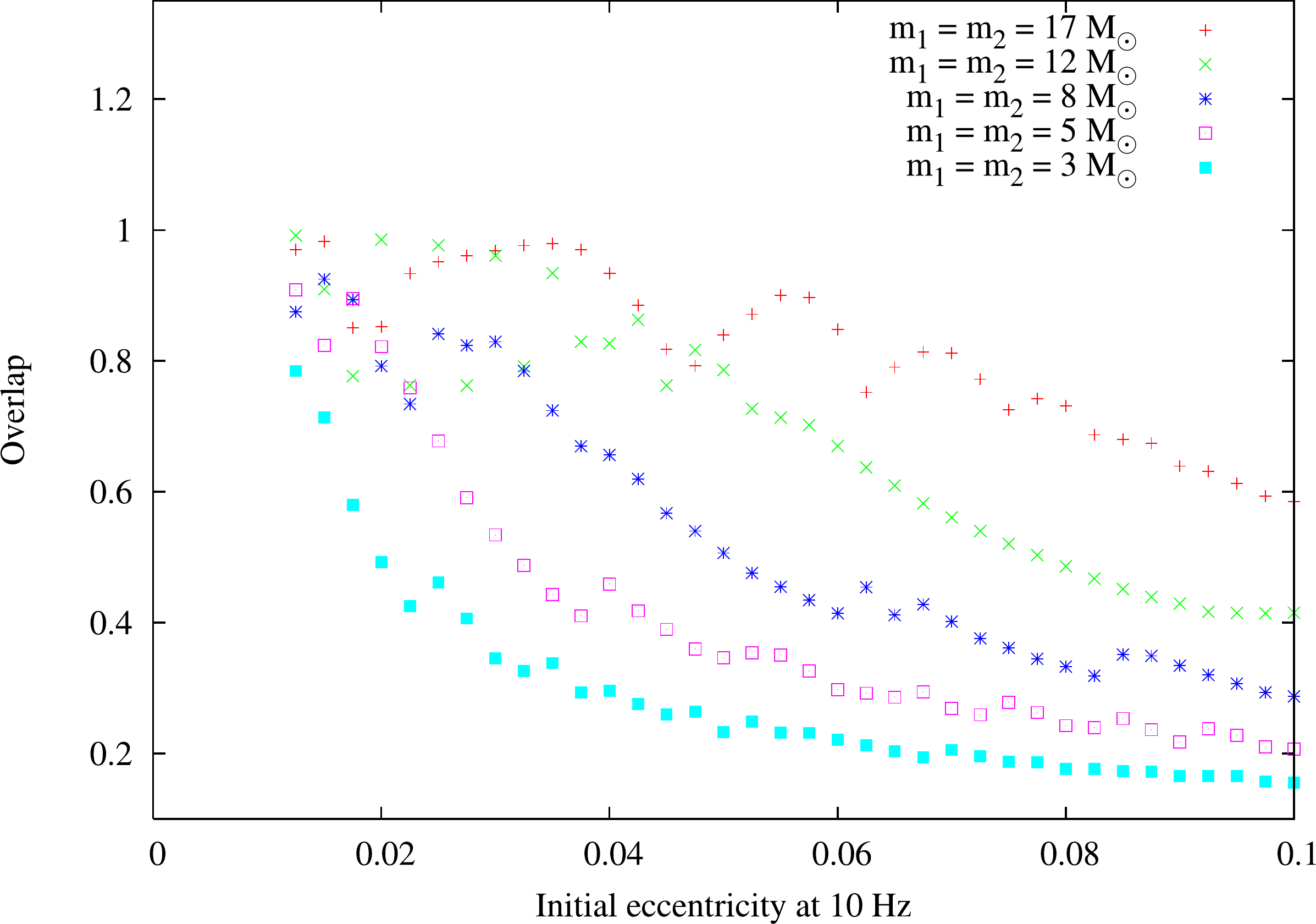}
\end{center}
\caption{\footnotesize The overlap
between circular templates and eccentric \inserted{signals emitted by sources possessing the same total masses as the circular templates} is shown as the
function of the initial eccentricity. \inserted{The applied} initial
frequency \inserted{was} \inserted{1}0~Hz.} \label{snrdropp}
\end{figure}

As a result of our pertinent investigations the overlap
\inserted{between circular templates and eccentric} \inserted{signals, emitted by sources possessing the same total masses as the circular templates{\color{black}, both of which are with initial frequency 10~Hz}, is shown as the function of the initial eccentricity on Fig.\,\ref{snrdropp}.}

It is clearly visible that tiny residual eccentricities lead to
considerably large SNR loss. This loss of SNR is also found to be
larger for smaller mass binaries. Accordingly the circular template
banks are found to be very ineffective in identifying binaries even
with negligible residual orbital eccentricity.

\subsection{Spinning binary systems}\label{spinning-binaries}

\begin{figure}[ht!]
\begin{tabular}{cc}
\includegraphics[width=0.48\textwidth]{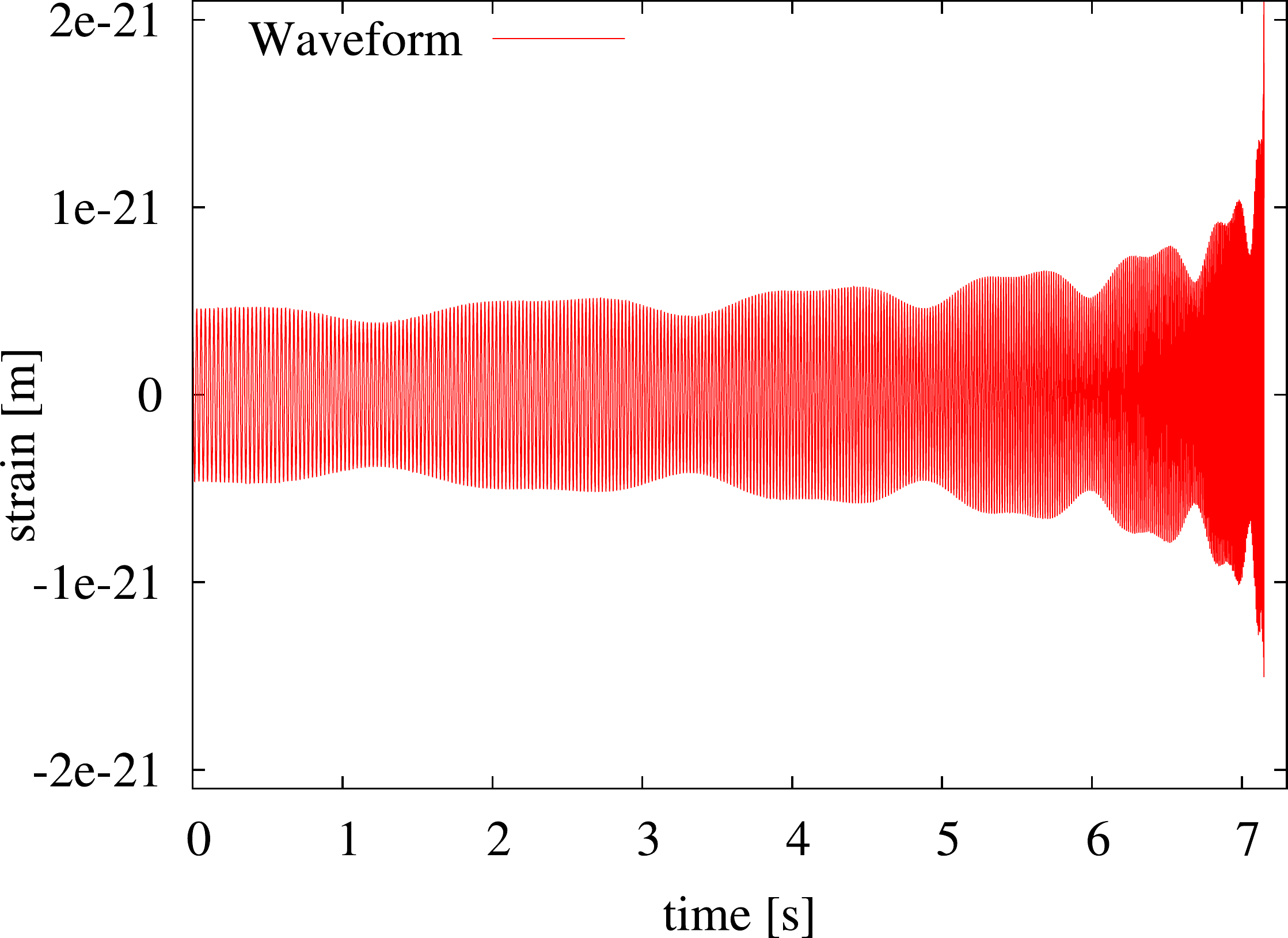} &
\includegraphics[width=0.48\textwidth]{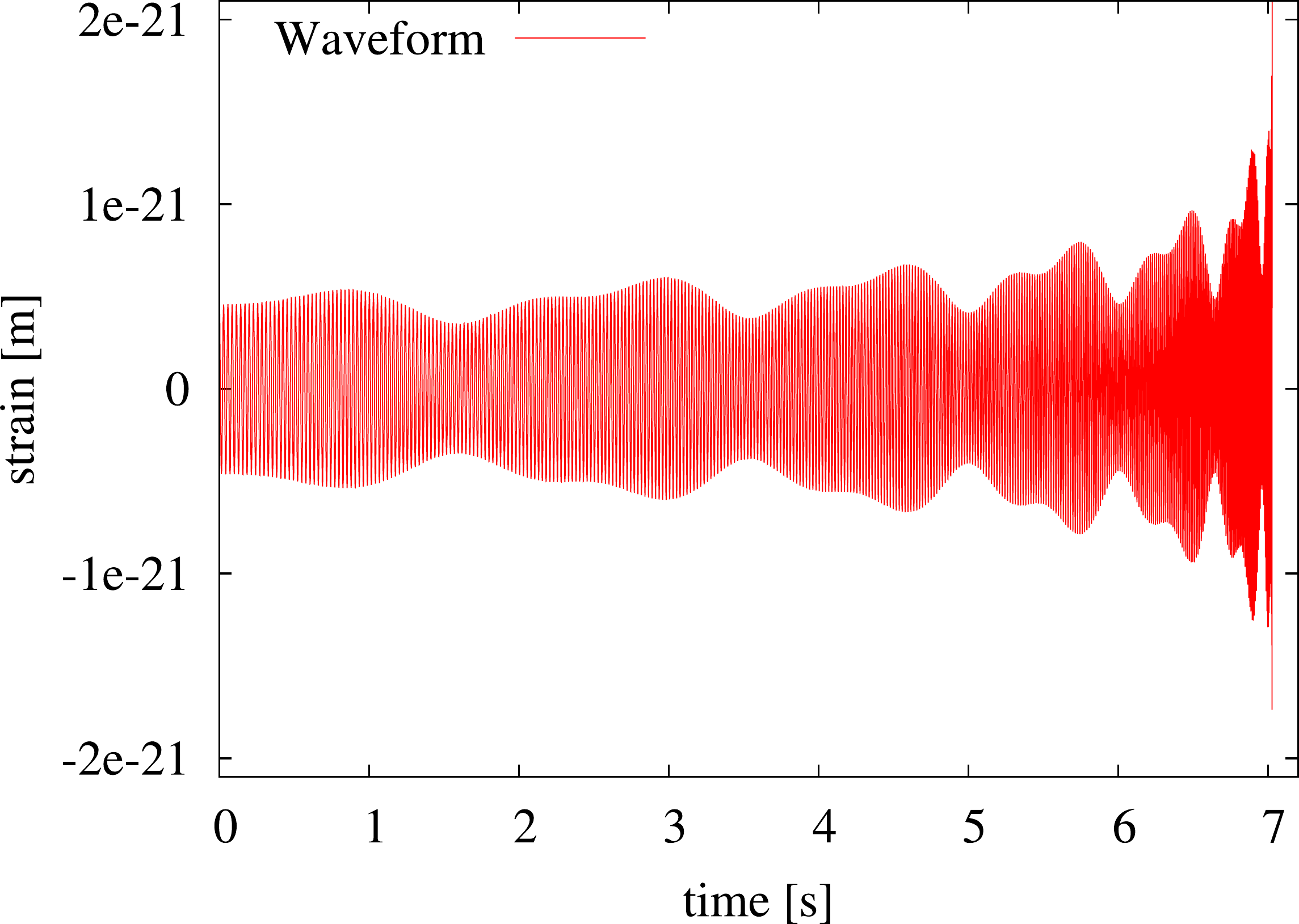}
\end{tabular}
\caption{\footnotesize  The waveforms emitted by binary systems with
masses $m_1 = m_2 = 3.4\ M_{\odot}$ and with specific spin vector(s)
on the left panel for a single spin with $s_{1x} = 1.0$, while on
the right panel for a double spinning binary with $s_{1x} = 1.0$ and
$s_{2y} = 1.0$ are shown. In both cases the initial frequency is
$f_{low} \sim 18\,\mathrm{Hz}$ and it is straightforward to
recognize the amplitude modulation caused by the rotation of the
orbital plane.} \label{spinningfig}
\end{figure}

Whenever at least one of the bodies possesses spin---due to the
precession of the orbital plane---the emitted gravitational wave
acquires a considerable large amplitude modulation {even in the
simplest possible case with zero initial eccentricity}. Waveforms of
single spinning and double spinning binary systems of this type are
shown on Fig.\,\ref{spinningfig}. On the left panel only one of the
bodies possesses spin with specific spin vector $s_{1x} = 1.0 $, and
the spin vector is perpendicular to the orbital angular momentum,
while on the right panel both of the bodies possess spin with
specific spin vectors $s_{1x} = 1.0$ and $s_{2y}=1.0$. It is
straightforward to recognize the yielded amplitude modulations on
both panels.

\subsection{Generic waveforms for spinning and eccentric binaries}

As indicated above the CBwaves software is capable to determine the
evolution and the waveforms of completely general spinning and
eccentric binaries. The simultaneous effect of amplitude and
frequency modulation gets immediately transparent. Such generic
orbits and waveforms are shown on Figs.\,\ref{doublespineccearly}
and \ref{doublespineccfinal} for the early phase and late inspiral
evolution of the system, respectively.

The early phase of the gravitational wave emitted by strongly
eccentric binary systems possesses burst type character. Fitting
analytic formula to all the possible waveforms of these type seems
not to be feasible. Nevertheless, they can be investigated with the
help of the CBwaves software, making thereby possible the
construction of detection pipelines, algorithms and their efficiency
studies for these types of events.

\subsection{The evolution of eccentricity for spinning binaries}

\begin{figure}[ht!]
\begin{center}
\includegraphics[width=0.7\textwidth]{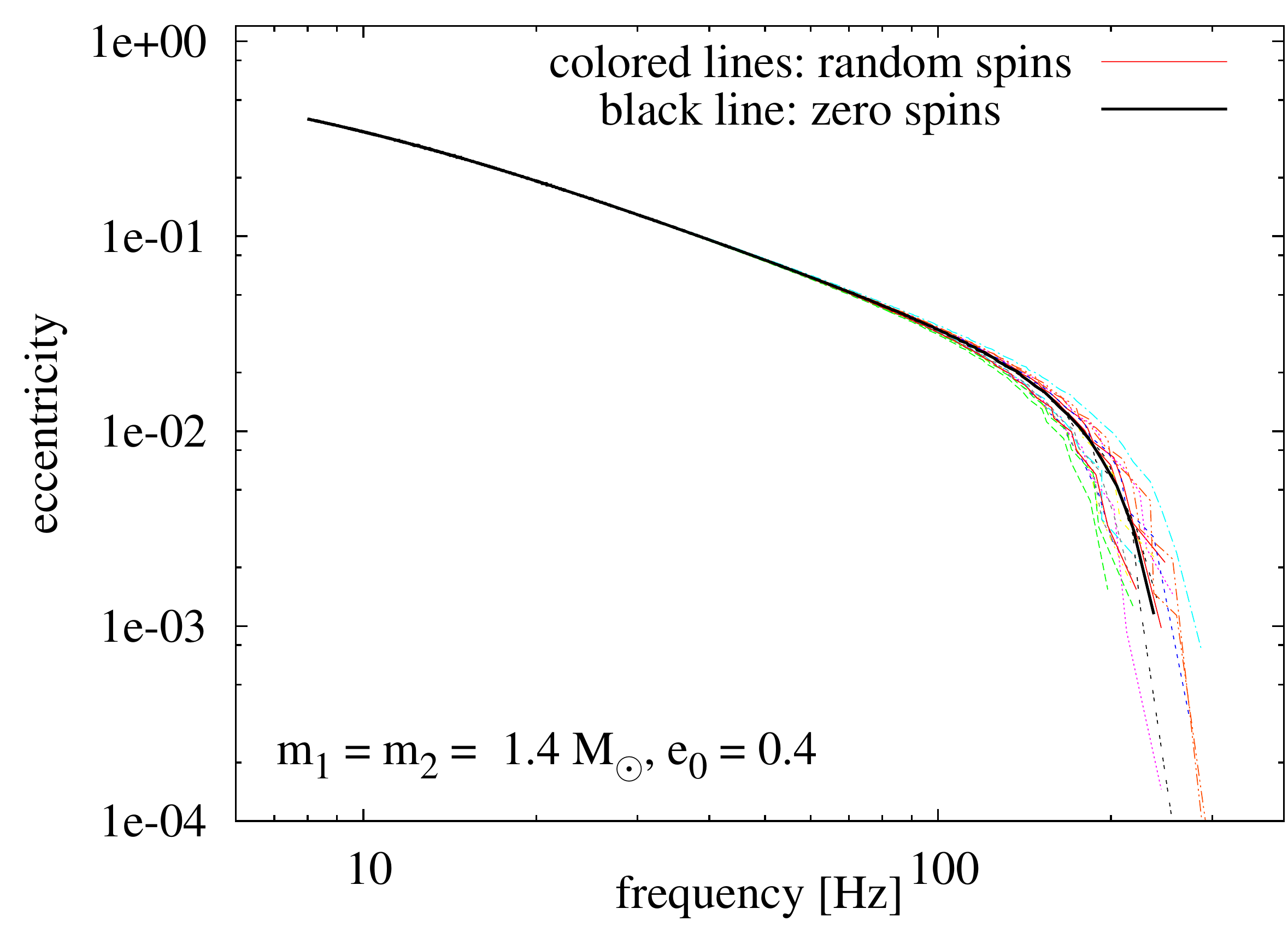}
\end{center}
\caption{\footnotesize  {\color{black}The time evolutions of the eccentricity relevant for NS-NS binary systems is shown assuming no or randomly oriented spins for the constituents. If spin is involved the magnitude of the specific spin vector, as usual, was chosen to be $0.7$. It is visible that the values of eccentricity relevant for the evolution with randomly oriented spins (indicated by thin color lines) remain always very close to the values (indicated by a black solid line) relevant for the same binary with no spin at all.}} \label{spinning-eccentric-fig}
\end{figure}

{\color{black}As it was emphasized already CBwaves was developed to accurately evolve and determine the waveforms emitted by generic spinning binary configurations moving on possibly eccentric orbits. Based on these capabilities it seems to be important to investigate how the evolution of eccentricity may be affected by the presence of spin or spins of the constituents. This short subsection is to present our pertinent results which can be summarized by claiming that the evolution of eccentricity is highly insensitive to the presence of spin. On Fig.\,\ref{spinning-eccentric-fig} the time dependence of the eccentricity is plotted for a NS-NS binary. It is clearly visible that the evolutions of the eccentricity relevant for spinning NS-NS binaries with randomly oriented spins (they are indicated by thin color lines) remain always very close to the evolution relevant for the same type of binary (it is indicated by a black solid line) with no spin at all. Although the evolution shown on Fig.\,\ref{spinning-eccentric-fig} is relevant for systems with specific masses $m_1=m_2=1.4\,M_\odot$ and initial eccentricity $e=0.4$ the qualitative behavior is not significantly different, i.e., the evolution of eccentricity is found to be insensitive to the presence of spin(s), for other binaries with various masses and for other values of the initial eccentricity.}

\subsubsection{Strongly eccentric binary systems}

\begin{figure}[ht!]
\begin{tabular}{cc}
\includegraphics[width=0.48\textwidth]{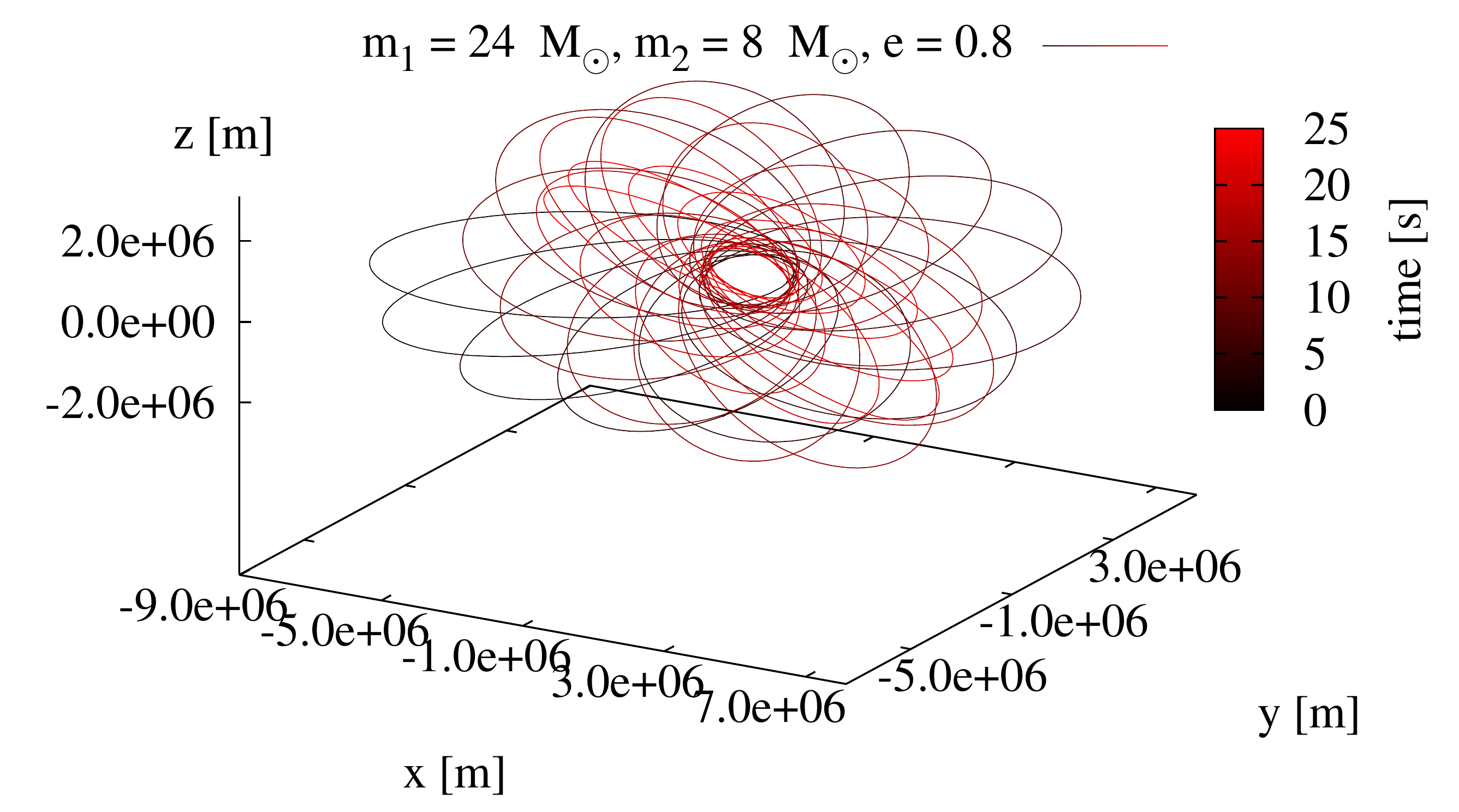} &
\includegraphics[width=0.48\textwidth]{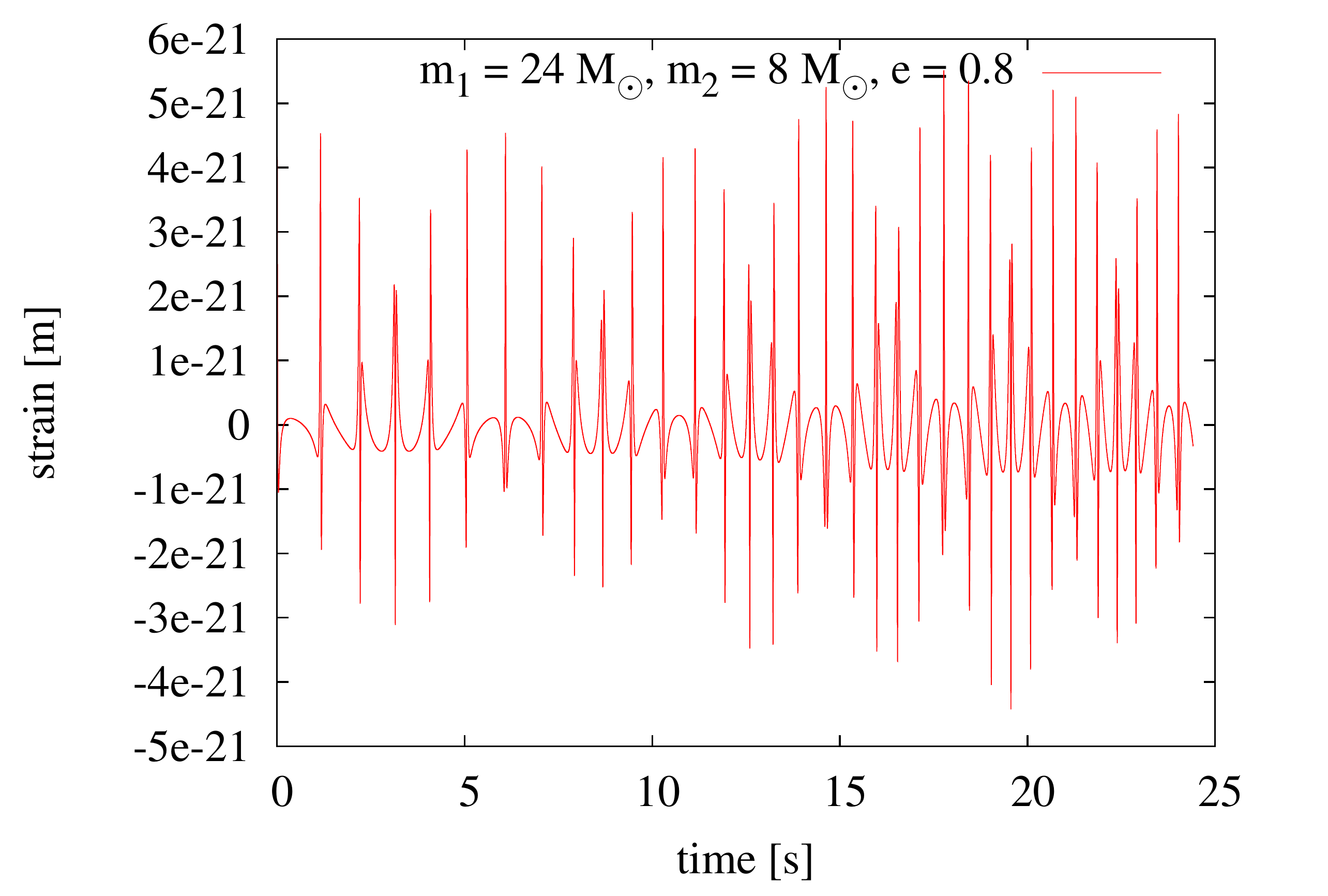}
\end{tabular}
\caption{\footnotesize Orbital motion (left panel) and emitted
waveform (right panel) of a strongly eccentric, double spinning,
precessing binary system in the early phase of the orbital evolution
with masses $m_1 = 24\,M_{\odot}, m_2 = 8\,M_{\odot}$\inserted{, with eccentricity $e=0.8$} and with
specific spin vectors $s_{1x} = 0.7, s_{1z}=0.7$ and $s_{2x} = 0.7,
s_{2z}=-0.7$.} \label{doublespineccearly}
\end{figure}

\begin{figure}[ht!]
\begin{tabular}{cc}
\includegraphics[width=0.48\textwidth]{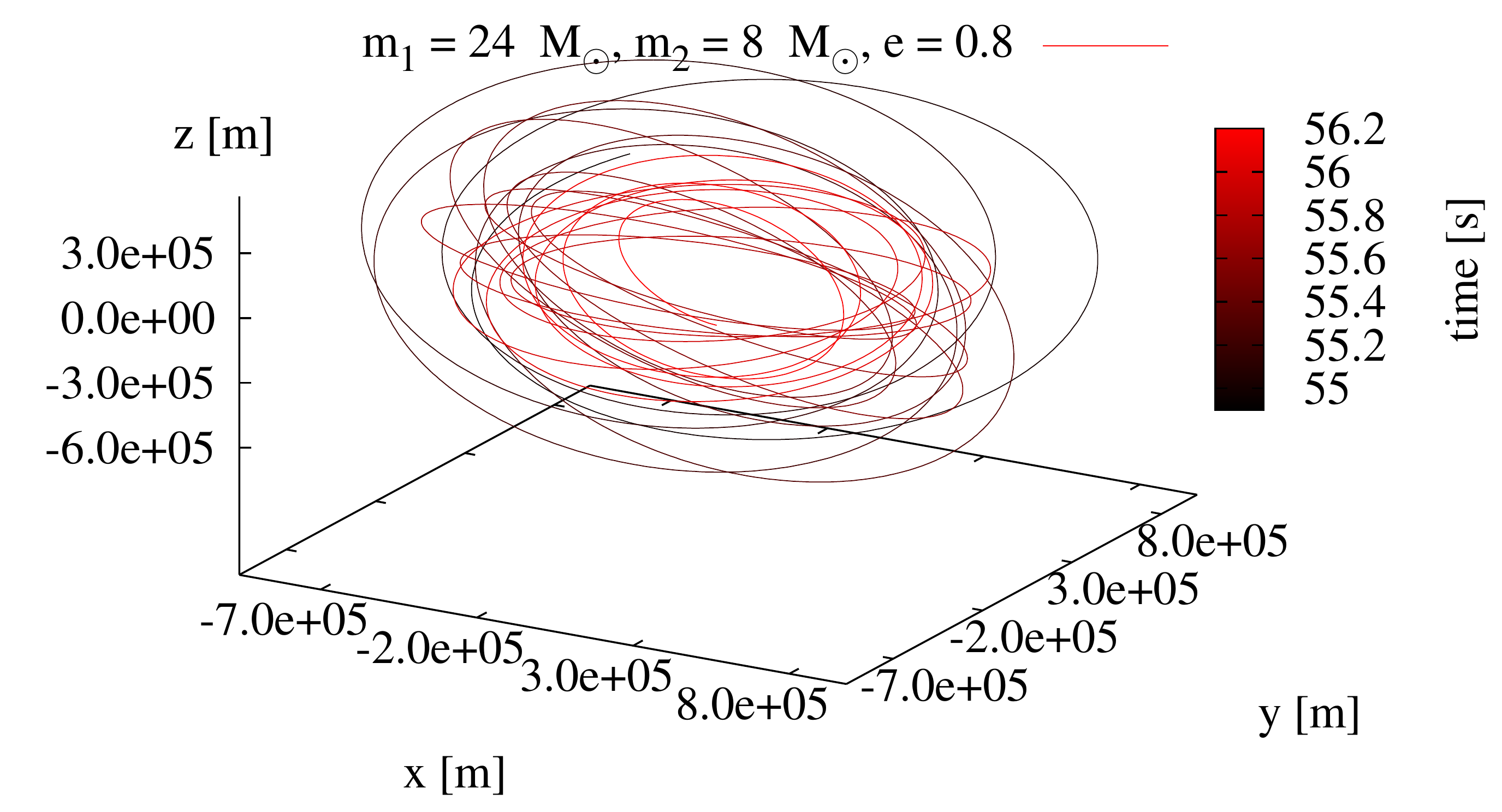} &
\includegraphics[width=0.48\textwidth]{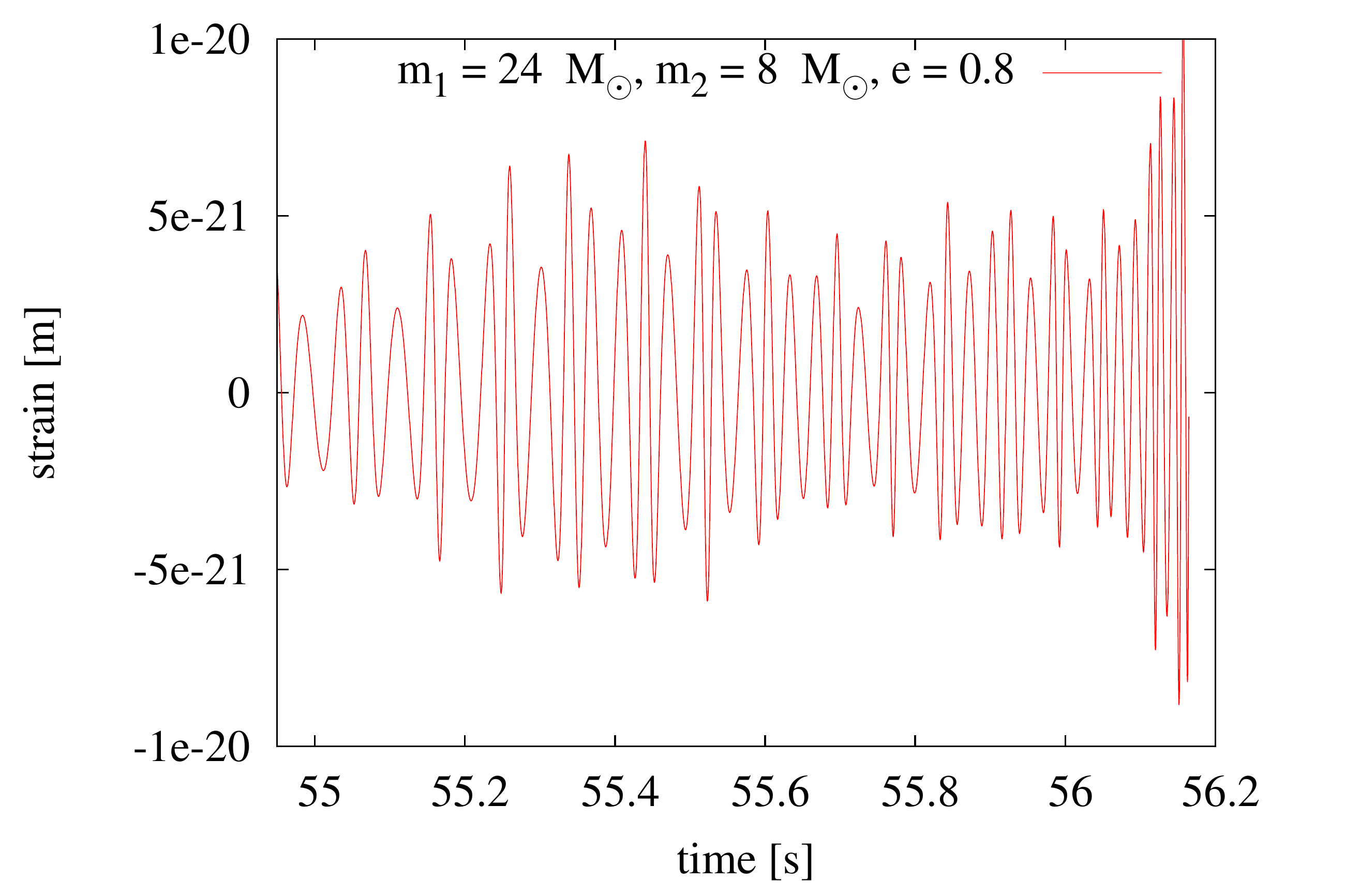}
\end{tabular}
\caption{\footnotesize In the final phase of the inspiral the
orbital evolution (left panel) and emitted waveform (right panel)
are shown for the highly eccentric, double spinning, precessing
binary system as depicted on Fig.\,\ref{doublespineccearly}.}
\label{doublespineccfinal}
\end{figure}

It is expected that strongly eccentric black hole binaries in the
halo of the galactic supermassive black hole are formed in
consequence of many-body interactions \cite{exc}. The amplitude and
the frequency of the gravitational waves emitted by such systems
changes significantly during the inspiral due to the circularization
effect. In Fig.\,\ref{doublespineccearly} the orbital evolution and
the time dependence of the amplitude of the waveform---in a
relatively early phase of its evolution---are shown for a binary
system with masses $m_1 = 24\,M_{\odot}, m_2 = 8\,M_{\odot}$\inserted{, with eccentricity $e=0.8$} and
with specific spin vectors $s_{1x} = 0.7, s_{1z}=0.7$ and $s_{2x} =
0.7, s_{2z}=-0.7$.
By the circularization of the orbit the features of the waveform
changes significantly and it becomes more and more similar to that
of a simple circular, spinning binary system. The result of this
process is shown---in the very final phase of the inspiral
process---on Fig.\,\ref{doublespineccfinal} for the system as on
Fig.\,\ref{doublespineccearly}.

\bigskip

\begin{figure}[ht!]
\begin{center}
\includegraphics[width=0.7\textwidth]{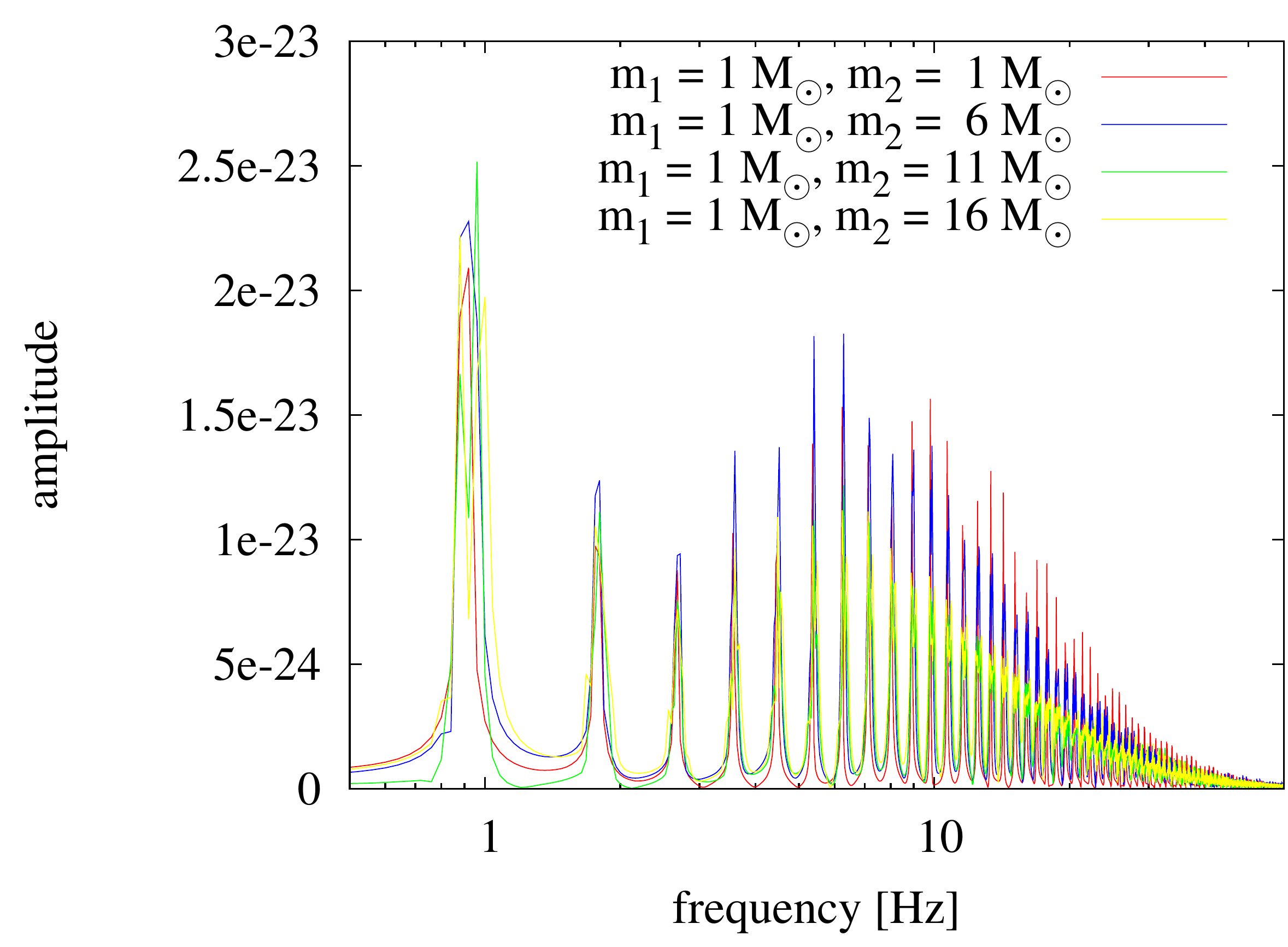}
\end{center}
\caption{\footnotesize The specific characteristics of the power
normalized frequency-domain waveforms of binaries with initial
eccentricity $e_{flow} = 0.8$, initial
frequency $f_{low} = 0.5\,\mathrm{Hz}$ and {with masses $m_1=1 M_{\odot}$ and $m_2 = 1 M_{\odot}, 6 M_{\odot}, 11 M_{\odot}, 16 M_{\odot}$, respectively,} are shown. 
} \label{burstfftmass}
\end{figure}

\begin{figure}[ht!]
\begin{tabular}{c}
\includegraphics[width=0.35\textwidth,angle=-90]{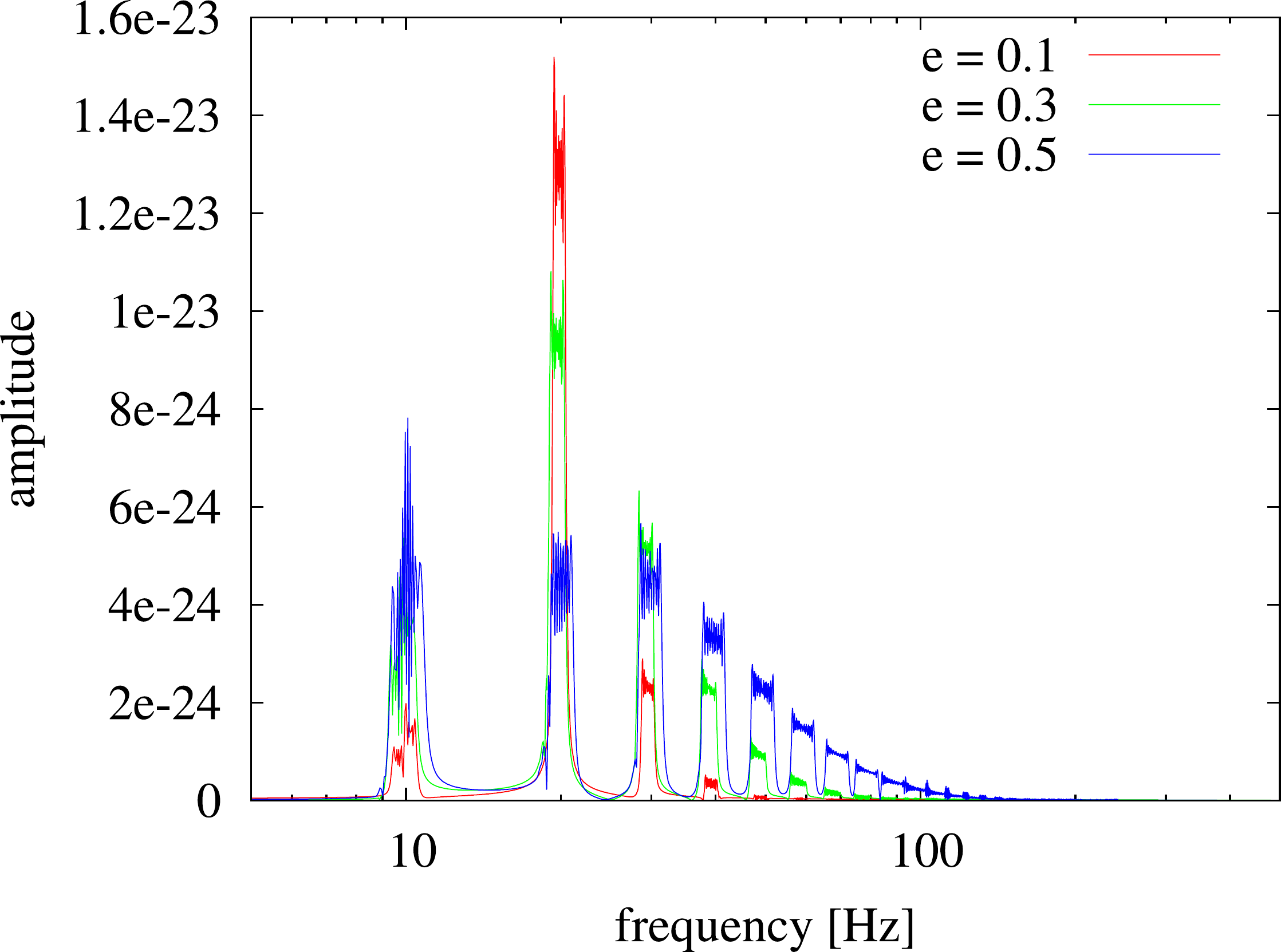}
\includegraphics[width=0.35\textwidth,angle=-90]{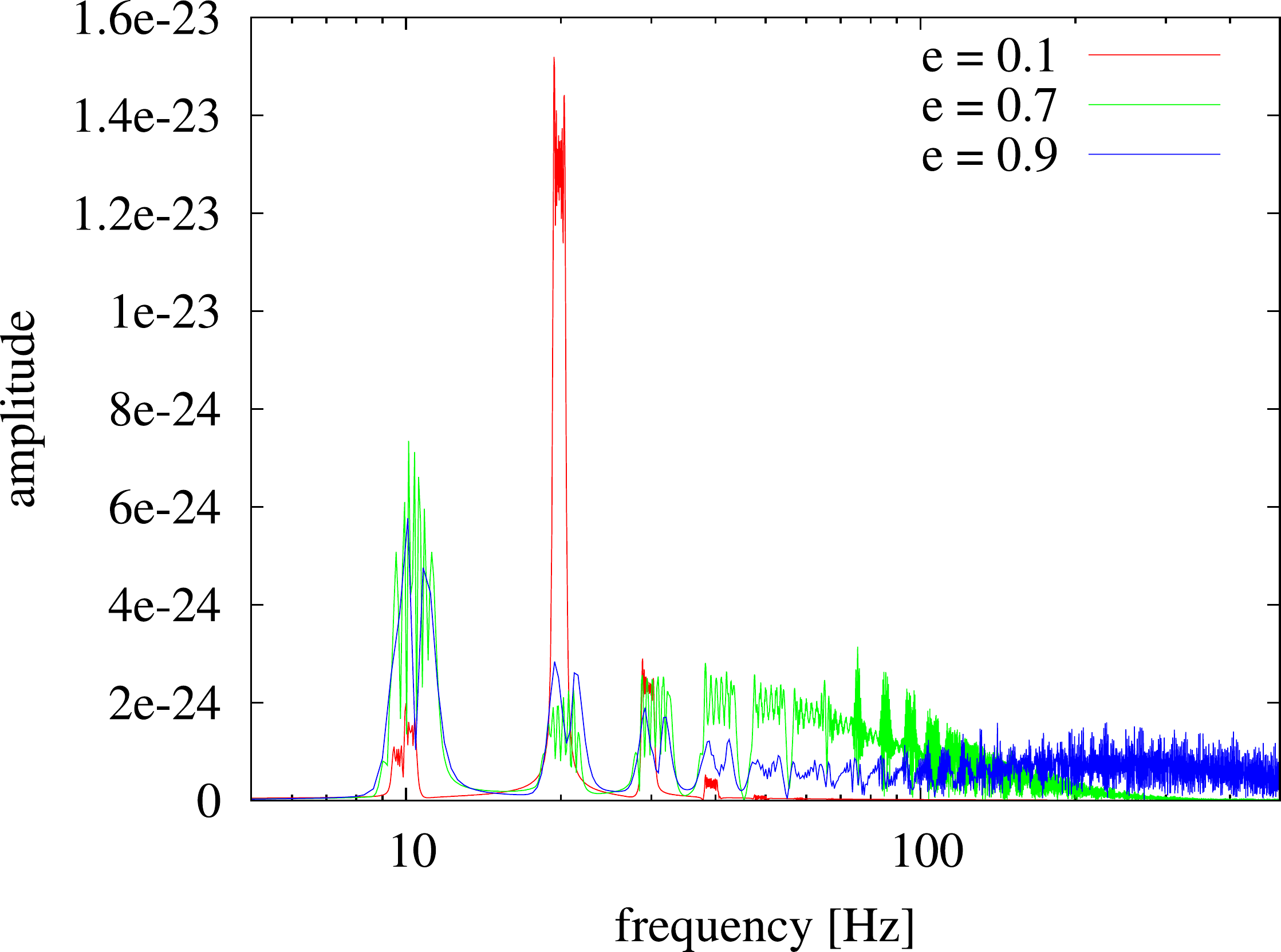}
\end{tabular}
\caption{\footnotesize The frequency-domain waveforms emitted by
binary systems with masses $m_{1} = m_{2} = 1.4 M_{\odot}$, and
initial orbital frequency $f_{low} = 10 \mathrm{Hz}$ are shown for
various lower (left panel) and higher (right panel) values of the
initial eccentricity. (On the right panel the frequency-domain
waveform with initial eccentricity $e=0.1$ is repeated to assist
comparison.)} \label{burstfftecchigh}
\end{figure}

From data analyzing respects it is of crucial importance that the
frequency-domain waveforms of highly eccentric binary systems
possess very specific characteristics which may be used to determine
the physical parameters of the system. For an immediate examples see
Fig.\,\ref{burstfftmass} on which various frequency-domain waveforms
are depicted for binary systems each with fixed initial
eccentricity, $e_{flow} = 0.8$, and initial frequency, $f_{low} =
0.5$~Hz. {The masses are chosen such that $m_1=1 M_{\odot}$ while $m_2$ takes either of the values $1 M_{\odot}, 6 M_{\odot}, 11 M_{\odot}, 16 M_{\odot}$, respectively.} Note that the waveforms shown on
Fig.\,\ref{burstfftmass} (and also on Fig.\,\ref{burstfftecchigh})
are normalized such that each waveform possesses equal power. The
various waveforms differ only in amplitude---which translates to
effective distance---and in the ratio of power present in the head
(twice of the orbital frequency) and the tail of the frequency
distribution.

\inserted{Interestingly,} as the initial eccentricity of the
system is increased the time domain waveform---as
expected---approaches an idealized Dirac-delta-like function while
the frequency-domain waveform becomes more and more broadband. An
\inserted{important} consequence of this \inserted{behavior} is that the waveforms spread over several---although not adjacent---frequency bands which provides a
unique imprint to them. This means that despite being present on
several frequencies it remains frequency limited, which may help in
constructing sensible detection pipelines robust against transient
noises. From the fractional power present at various frequency bands
the eccentricity of a system may be deduced, which is of great
importance from parameter estimation point of view. Waveforms for
moderately and highly eccentric binary systems are shown on the left
and right panels of Fig.\,\ref{burstfftecchigh}, respectively.

\section{Summary}\label{Summary}

Our main aim in writing up this paper was to introduce our general
purpose gravitational waveform generator software, called CBwaves,
which is capable to simulate waveforms emitted by binary systems on
closed or open orbits with possessing spin(s) and/or orbital
eccentricity. It was done by direct integration of the equation of
motion of the post-Newtonian expansion thereby the software can
easily be extended up to any desired order of (known) precision. The \inserted{open}
source of current version, which is accurate up to 3.5PN order, may be downloaded
from \cite{cbwavesdownload}.

\medskip

\inserted{There may be several other similar waveform generating software based on the use of post-Newtonian framework. Nevertheless, with the exception of our own code and the one that had been applied in recent scientific investigations as reported in \cite{janna} the authors are not aware of any other software  as, if they exist, they are not available for public use. It is also important to be mentioned here that even the approach applied in \cite{janna} is not fully complete in the sense that the acceleration term refereed as "${\bf a}_{PNSO}$" in our framework is missing from the description of the generic motion of the involved bodies. In addition, in \cite{janna} the waveforms were implemented and evaluated only up to the lowest possible level of quadrupole approximation. }

\inserted{In this respect, it is important to be emphasized that CBwaves was developed with the intention to provide a reliable tool that is fully complete involving all the possible terms that are compatible with the consistent description of the considered generic GW sources within the applied PN approximation. In particular, all the terms concerning the accelerations, the energy and the radiation were implemented which are needed to guarantee the capability of CBwaves to accurately evolve and determine the waveforms emitted by generic spinning binary configurations regardless whether they are moving on eccentric closed orbits or on open ones.}

\medskip

With the help of CBwaves we have investigated some of the characteristics of the orbital evolution of various binary configurations and the emitted waveforms. The relevance of the results obtained for the next generation of interferometric gravitational wave detectors was also underlined. While currently allowable parametric density of a general configuration template bank practically limits the direct use of the yielded waveforms in the detection pipelines, \inserted{we hope that} these generic waveforms will be useful in various parameter estimation studies.

\medskip

As already emphasized with the help of CBwaves various physically interesting scenarios have been investigated. The main results covered by the present paper are as follows:
\begin{itemize}

\item \inserted{\color{black}Investigating the validity of the adiabatic approximation it was found that the adiabatic approximation yields almost the same, at 3.5PN level, less then $2\%$ faster decrease than the time evolution. }

\item We have justified that the energy balance relation is indeed insensitive to the specific form of the applied radiation reaction term.

\item \inserted{\color{black}Using the designed sensitivity of the advanced Virgo detector} it is demonstrated that circular template banks are \inserted{even less} effective \inserted{than the initial detectors were} in identifying binaries possess\inserted{ing} only a tiny residual orbital eccentricity.

\item \inserted{The ranges of correspondences, along with} some of the discrepancies, between the analytic and numerical description of the evolution of eccentric binaries, along with some universal properties characterizing their evolution{\color{black}, and its insensitivity for the inclusion of spins}, were pointed out.

\item By inspecting the energy balance relation it was shown that, on contrary to the general expectations,  the post-Newtonian approximations loose its accuracy once the post-Newtonian parameter gets beyond its critical value $\epsilon\sim 0.08-0.1$.

\item By studying gravitational waves emitted during the early phase of the evolution of strongly eccentric binary systems it was found that they possess very specific characteristics which may make the detection of these type of binary systems to be feasible.
\end{itemize}

\medskip

\begin{figure}[ht!]
\begin{tabular}{c}
\includegraphics[width=0.95\textwidth]{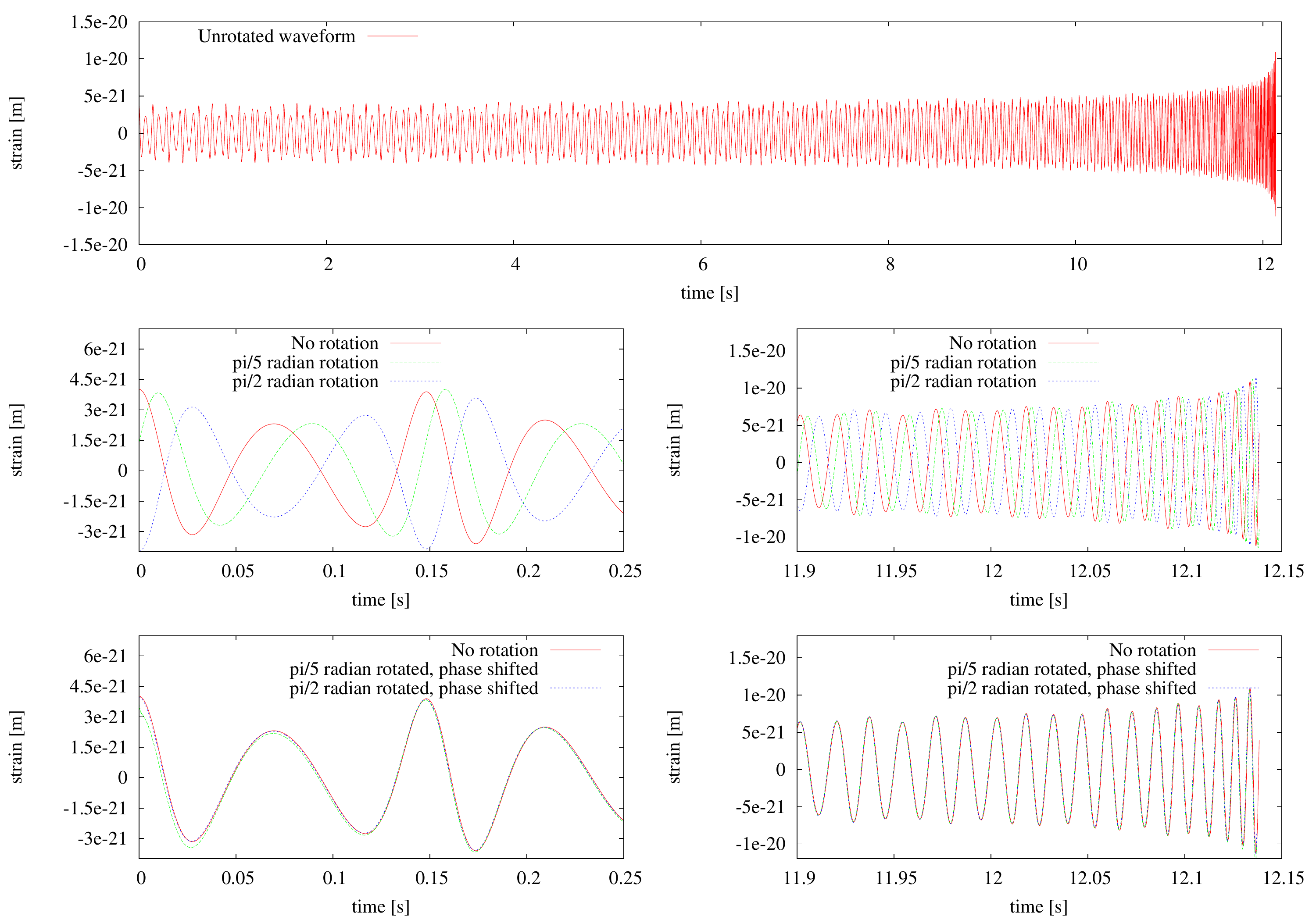}
\end{tabular}
\caption{\footnotesize The $\hat \varphi$ (rotation angle) dependence
of the waveform emitted by the binary system with $m_1$ = $m_2$ = 10
$M_{\odot}$, $s_1$ = 0.7 , $s_2$ = 0 and with eccentricity $e$ = 0.2
is shown for different amount of initial rotation around the total
angular momentum vector. The full waveform (top), the early stage
(middle left) the late phase (middle right), the phase shifted early
stage (bottom left) and the phase shifted late stage (bottom right)
of the inspiral process. (Note that the small initial discrepancies
on the last but one plot has no physical relevance as it is an
artifact of inherent inaccuracy of the method used in determining
the phase shift.)} \label{doublespinrotecc}
\end{figure}

From data analyzing point of view it is crucial to know which of the
involved parameters are essential. In accordance with the discussion
on pages  6-7 in \cite{BCV} some of the relative angles between the
vectors determining the initial configurations are of this type and
they are relevant in determining the motion of the involved bodies
and in computing the emitted waveform. Nevertheless, whatever are
the initial conditions the lengths and the relative angles of the vectors comprising it remain intact under a rigid rotation of the binary as a whole in space around any spatial vector of free choice. Consequently, the emitted waveform should not
change more than dictated by the simultaneous replacement of the direction
of observation. In the particular case when the base vector of such a rigid rotation is also distinguished
by the dynamics---such as the orbital angular momentum $\mathbf{L}$
in \cite{BCV}, or the total angular momentum
$\mathbf{J}=\mathbf{L}+\mathbf{S}_1+\mathbf{S}_2$ in our case as
$\dot{\mathbf{J}}\approx 0$ holds up to 2PN order---and the binary is optimally oriented, i.e.\,the direction of observation $\hat{\bf N}$ is chosen to be parallel to $\mathbf{J}$ and it is orthogonal to the plane spanned by the  detector arms (see Figs.\,\ref{SourceFrame} and \ref{DetectorFrame}), only a simple phase shift, with angle $\hat \varphi$, is expected to show up in the waveform as a response to the rigid rotation by angle $\hat \varphi$ around $\mathbf{J}$. It is straightforward to see that in this particular case the considered rigid rotation of the source could be replaced by a rigid rotation of the detector arms by angle $-\hat \varphi$. Fig.\,\ref{doublespinrotecc} is to provide justification of these
expectations on which the rotation angle dependence of the emitted
waveforms and the compensation by appropriate phase
shifts are shown in case of a binary system with $m_1$ = $m_2$ = 10
$M_{\odot}$, $s_1$ = 0.7 , $s_2$ = 0 and with initial eccentricity
$e$ = 0.2.
\medskip

Note finally that there are various physically interesting problems
which may be investigated with the help of CBwaves. The most
immediate ones include a systematic study of the effect of the {\it
spin supplementary conditions} \cite{SSC1,SSC2,SSC3} and the
investigation of the time evolution of open binary systems, in
particular, that of the spin flip phenomenon. The results of the
corresponding studies will be published elsewhere.

\section{Acknowledgments}
This research was supported in parts by a VESF postdoctoral
fellowship to the RMKI Virgo Group for the period 2009-2011, and by
the Hungarian Scientific Research Fund (OTKA) Grant No. 67942.

\vfill\eject

\appendix
\section*{Appendix A}

\renewcommand{\theequation}{A.\arabic{equation}}
\setcounter{equation}{0}
\renewcommand{\thesubsection}{A.\arabic{subsection}}
\setcounter{subsection}{0}

\subsection{The radiation field}

The gravitational waveform generated by a compact binary system is expressed by a sum of contributions originating from different PN
orders. The particular form of the contributions listed in Eq.\,(\ref{Wform}) can be found in \cite{Kidder}, but for conveniences
they are also summarized below. Accordingly, the quadrupole term and higher order relativistic corrections read as \small
\begin{eqnarray}
Q^{ij} &=&2\left[ v^{i}v^{j}-{\frac{Gm}{r}}n^{i}n^{j}\right] , \\
P^{0.5}Q^{ij} &=&{\frac{\delta m}{cm}}\left\{ 3{\frac{Gm}{r}}\left[
2n^{(i}v^{j)}-\dot{r}n^{i}n^{j}\right] (\mathbf{\hat{N}\cdot
\hat{n}})+\left[
{\frac{Gm}{r}}n^{i}n^{j}-2v^{i}v^{j}\right] (\mathbf{\hat{N}\cdot v}%
)\right\} , \\
PQ^{ij} &=&{\frac{1}{3c^{2}}}(1-3\eta)\left\{ 4{\frac{Gm}{r}}\left[ 3\dot{r}%
n^{i}n^{j}-8n^{(i}v^{j)}\right] (\mathbf{\hat{N}\cdot \hat{n}})(\mathbf{\hat{%
N}\cdot v})+2\left[ 3v^{i}v^{j}-{\frac{Gm}{r}}n^{i}n^{j}\right] (\mathbf{%
\hat{N}\cdot v})^{2}\right. \\
&&\left. +{\frac{Gm}{r}}\left[ (3v^{2}-15\dot{r}^{2}+7{\frac{Gm}{r}}%
)n^{i}n^{j}+30\dot{r}n^{(i}v^{j)}-14v^{i}v^{j}\right]
(\mathbf{\hat{N}\cdot \hat{n}})^{2}\right\}
+{\frac{4}{3}}{\frac{Gm}{r}}\dot{r}(5+3\eta
)n^{(i}v^{j)} \nonumber \\
&&+\left[ (1-3\eta )v^{2}-{\frac{2}{3}}(2-3\eta
){\frac{Gm}{r}}\right] v^{i}v^{j}+{\frac{Gm}{r}}\left[ (1-3\eta
)\dot{r}^{2}-{\frac{1}{3}}(10+3\eta
)v^{2}+{\frac{29}{3}}{\frac{Gm}{r}}\right] n^{i}n^{j}, \nonumber \\
P^{1.5}Q^{ij} &=&{\frac{\delta m}{mc^{3}}}(1-2\eta )\Biggl\{{\frac{1}{4}}{%
\frac{Gm}{r}}\Biggl[(45\dot{r}^{2}-9v^{2}-28{\frac{Gm}{r}}%
)n^{i}n^{j}+58v^{i}v^{j}-108\dot{r}n^{(i}v^{j)}\Biggr](\mathbf{\hat{N}\cdot
\hat{n}})^{2}(\mathbf{\hat{N}\cdot v}) \\
&&+{\frac{1}{2}}\left[ {\frac{Gm}{r}}n^{i}n^{j}-4v^{i}v^{j}\right] (\mathbf{%
\hat{N}\cdot v})^{3}+{\frac{Gm}{r}}\Biggl[{\frac{5}{4}}(3v^{2}-7\dot{r}^{2}+6%
{\frac{Gm}{r}})\dot{r}n^{i}n^{j}-{\frac{1}{6}}(21v^{2}-105\dot{r}^{2}
 \nonumber \\
&&+44{\frac{Gm}{r}})n^{(i}v^{j)}-{\frac{17}{2}}\dot{r}v^{i}v^{j}\Biggr](%
\mathbf{\hat{N}\cdot \hat{n}})^{3}+{\frac{3}{2}}{\frac{Gm}{r}}\left[
10n^{(i}v^{j)}-3\dot{r}n^{i}n^{j}\right] (\mathbf{\hat{N}\cdot \hat{n}})(%
\mathbf{\hat{N}\cdot v})^{2}\Biggr\} \nonumber \\
&&+{\frac{\delta m}{mc^{3}}}{\frac{1}{12}}{\frac{Gm}{r}}(\mathbf{\hat{N}%
\cdot \hat{n}})\Biggl\{n^{i}n^{j}\dot{r}\left[ \dot{r}^{2}(15-90\eta
)-v^{2}(63-54\eta )+{\frac{Gm}{r}}(242-24\eta )\right]  \nonumber \\
&&-\dot{r}v^{i}v^{j}(186+24\eta )+2n^{(i}v^{j)}\left[
\dot{r}^{2}(63+54\eta
)-{\frac{Gm}{r}}(128-36\eta )+v^{2}(33-18\eta )\right] \Biggr\}  \nonumber  \\
&&+{\frac{\delta m}{mc^{3}}}(\mathbf{\hat{N}\cdot v})\Biggl\{{\frac{1}{2}}%
v^{i}v^{j}\left[ {\frac{Gm}{r}}(3-8\eta )-2v^{2}(1-5\eta )\right]
-n^{(i}v^{j)}{\frac{Gm}{r}}\dot{r}(7+4\eta )  \notag \\
&&-n^{i}n^{j}{\frac{Gm}{r}}\left[ {\frac{3}{4}}(1-2\eta )\dot{r}^{2}+{\frac{1%
}{3}}(26-3\eta ){\frac{Gm}{r}}-{\frac{1}{4}}(7-2\eta )v^{2}\right]
\Biggr\}\,,  \nonumber
\end{eqnarray}%
\normalsize where $\mathbf{r}=\mathbf{x_{1}}-\mathbf{x_{2}}$,
$\mathbf{v}={d\mathbf{r}/dt} $, $\mathbf{\hat{n}}={\mathbf{r}/r}$,
$m=m_{1}+m_{2}$, $\delta m=m_{1}-m_{2}$, $\eta =\mu /m$ and the
derivative with respect to time is indicated by an overdot. \inserted{The
$P^{2}Q^{ij}$ contribution to the waveform is \cite{WW}} \small
\begin{eqnarray}
\inserted{ P^{2}Q^{ij} }&=& \frac{1}{c^4}\biggr[{1 \over 60 }
(1-5\eta+5\eta^2) \biggl\{ 24({\bf \hat N \cdot v})^4 \biggl[ 5 v^i
v^j - {m \over r} {\hat n}^i {\hat n}^j \biggr]
\nonumber \\
&& +{m \over r} ({\bf \hat N \cdot \hat n})^4 \biggl[ 2 \left( 175
{m \over r} - 465 \dot r^2 + 93 v^2 \right) v^i v^j
 + 30 \dot r \left( 63 \dot r^2 - 50{m \over r} - 27 v^2 \right)
{\hat n}^{(i}v^{j)}
\nonumber \\
&& + \left(1155 {m \over r} \dot r^2 - 172 \left({m \over
r}\right)^2 - 945 \dot r^4 - 159 {m \over r} v^2 + 630 \dot r^2 v^2
- 45 v^4 \right) {\hat n}^i {\hat n}^j \biggr]
\nonumber \\
&& +24 {m \over r} ({\bf \hat N \cdot \hat n})^3 ({\bf \hat N \cdot
v}) \biggl[ 87 \dot r v^i v^j + 5 \dot r \left( 14 \dot r^2 - 15 {m
\over r} - 6v^2 \right) {\hat n}^i {\hat n}^j
\nonumber \\
&& + 16 \left( 5 {m \over r} - 10 \dot r^2 + 2v^2 \right) {\hat
n}^{(i} v^{j)} \biggr] +288 {m \over r} ({\bf \hat N \cdot \hat n})
({\bf \hat N \cdot v})^3 \biggl[ \dot r {\hat n}^i {\hat n}^j - 4
{\hat n}^{(i} v^{j)} \biggr]
\nonumber \\
&& +24 {m \over r} ({\bf \hat N \cdot \hat n})^2 ({\bf \hat N \cdot
v})^2 \biggl[ \left( 35 {m \over r} - 45 \dot r^2 + 9 v^2 \right)
{\hat n}^i {\hat n}^j
               - 76 v^i v^j + 126 \dot r {\hat n}^{(i} v^{j)}
\biggr] \biggr\}
\nonumber \\
&& + {1 \over 15} ({\bf \hat N \cdot v})^2 \biggl\{
            \biggl[ 5 ( 25-78\eta+12\eta^2 ) {m \over r}
                    - (18 - 65 \eta + 45 \eta^2 ) v^2
\nonumber \\
&&
                    + 9 ( 1 - 5 \eta + 5 \eta^2 ) \dot r^2
            \biggr] {m \over r} {\hat n}^i {\hat n}^j
          +3\biggl[ 5 ( 1 - 9\eta + 21\eta^2 ) v^2
                   -2 ( 4 - 25 \eta + 45 \eta^2 ) {m \over r}
            \biggr] v^i v^j
\nonumber \\
&&
          + 18 ( 6 - 15 \eta - 10 \eta^2 ) {m \over r} \dot r {\hat
n}^{(i} v^{j)} \biggr\} +{1 \over 15}({\bf \hat N \cdot \hat
n})({\bf \hat N \cdot v}){m \over r} \biggl\{
           \biggl[ 3 ( 36-145\eta+150\eta^2 ) v^2 \nonumber \\
&&
                  -5 ( 127 - 392 \eta + 36 \eta^2 ) {m \over r}
                -15( 2 - 15 \eta + 30 \eta^2 ) \dot r^2
           \biggr] \dot r {\hat n}^i {\hat n}^j \nonumber \\
&&
             + 6 (98 - 295 \eta - 30 \eta^2 ) \dot r v^i v^j
         +2\biggl[ 5 ( 66 - 221\eta + 96 \eta^2 ) {m \over r}\nonumber \\
&&
                  -9 ( 18 -  45\eta - 40 \eta^2 ) \dot r^2
                  -  ( 66 - 265\eta +360 \eta^2 ) v^2
           \biggr] {\hat n}^{(i} v^{j)}
\biggr\}
\nonumber \\
&& +{1 \over 60}({\bf \hat N \cdot \hat n})^2 {m \over r} \biggl\{
\biggl[ 3   (33- 130\eta + 150\eta^2) v^4
                + 105( 1 - 10 \eta + 30 \eta^2 ) \dot r^4
\nonumber \\
&&
                + 15 (181-572 \eta + 84 \eta^2) {m \over r} \dot r^2
                -    (131-770 \eta + 930\eta^2) {m \over r} v^2
\nonumber \\
&&
                - 60 (  9- 40 \eta +  60\eta^2) v^2 \dot r^2
                -  8 (131-390 \eta +  30\eta^2) \left( {m \over r} \right)^2
           \biggr] {\hat n}^i {\hat n}^j
\nonumber \\
&&
        + 4 \biggl[     (12+   5\eta - 315\eta^2) v^2
                   -9   (39- 115\eta -  35\eta^2) \dot r^2
                   +5   (29- 104\eta +  84\eta^2) {m \over r}
            \biggr] v^i v^j
\nonumber \\
&&
        + 4 \biggl[15   ( 18-  40\eta -  75\eta^2) \dot r^2
                   -5   (197- 640\eta + 180\eta^2) {m \over r }
\nonumber \\
&&
                   +3   (21- 130\eta + 375\eta^2) v^2
            \biggr] \dot r {\hat n}^{(i} v^{j)}
\biggr\}
\nonumber \\
&& + {1 \over 60} \biggl\{ \biggl[    (467+780\eta-120\eta^2) {m
\over r} v^2
                - 15( 61- 96\eta+ 48\eta^2) {m \over r} \dot r^2
\nonumber \\
&&
                -   (144-265\eta-135\eta^2) v^4
                +  6( 24- 95\eta+ 75\eta^2) v^2 \dot r^2
\nonumber \\
&&
                -  2(642+545\eta          ) \left( {m \over r} \right)^2
                - 45(  1-  5\eta+  5\eta^2) \dot r^4
         \biggr] {m \over r} {\hat n}^i {\hat n}^j
\nonumber \\
&&
       + \biggl[  4 ( 69+ 10\eta-135\eta^2) {m \over r} v^2
                - 12(  3+ 60\eta+ 25\eta^2) {m \over r} \dot r^2
\nonumber \\
&&
                + 45(  1-  7\eta+ 13\eta^2) v^4
                - 10( 56+165\eta- 12\eta^2) \left( {m \over r} \right)^2
         \biggr] v^i v^j
\nonumber \\
&&
       +4\biggl[  2 ( 36+  5\eta- 75\eta^2) v^2
                - 6 (  7- 15\eta- 15\eta^2) \dot r^2
                + 5 ( 35+ 45\eta+ 36\eta^2) {m \over r}
         \biggr] {m \over r} \dot r {\hat n}^{(i} v^{j)}
\biggr\}\biggr] \, . \nonumber
\end{eqnarray}
\normalsize

The analogous expressions for spin contributions are
\inserted{\cite{Kidder}}
 \small
\begin{eqnarray}
PQ_{SO}^{ij}\!\!\!\!&=&\!\!\!\! {\frac{2G}{cr^{2}}}(\mathbf{\Delta
\times \hat{N}}
)^{(i}n^{j)}, \label{PQSO} \\
P^{1.5}Q_{SO}^{ij}
\!\!\!\!&=&\!\!\!\!{\frac{2G}{c^{2}r^{2}}}\Biggl\{n^{i}n^{j}\left[ (
\mathbf{\hat{n}\times v})\mathbf{\cdot }(12\mathbf{S}+6{\frac{\delta
m}{m}} \mathbf{\Delta })\right] -n^{(i}\left[ \mathbf{v\times
}(9\mathbf{S}+5{\frac{
\delta m}{m}}\mathbf{\Delta })\right] ^{j)} \label{P15QSO} \\
&&+\left[ 3\dot{r}(\mathbf{\hat{N}\cdot
\hat{n}})-2(\mathbf{\hat{N}\cdot v}) \right] \left[
(\mathbf{S}+{\frac{\delta m}{m}}\mathbf{\Delta })\mathbf{ \times
\hat{N}}\right] ^{(i}n^{j)}-v^{(i}\left[ \mathbf{\hat{n}\times }(2
\mathbf{S}+2{\frac{\delta m}{m}}\mathbf{\Delta })\right] ^{j)}  \notag \\
&&+\dot{r}n^{(i}\left[ \mathbf{\hat{n}\times
}(12\mathbf{S}+6{\frac{\delta m }{m}}\mathbf{\Delta })\right]
^{j)}-2(\mathbf{\hat{N}\cdot \hat{n}})\left[ (
\mathbf{S}+{\frac{\delta m}{m}}\mathbf{\Delta })\mathbf{\times
\hat{N}} \right] ^{(i}v^{j)}\Biggr\}, \nonumber \\
\inserted{ P^{2}Q^{ij}_{SS} } \!\!\!\!&=&\!\!\!\! \inserted{ -
\frac{6G}{c^2\mu r^3} \left\{ n^i n^j \left[ ({\bf S_1 \cdot S_2}) -
5 ({\bf \hat n \cdot S_1})({\bf \hat n \cdot S_2}) \right] +
2n^{(i}S_1^{j)} ({\bf \hat n \cdot S_2}) + 2n^{(i}S_2^{j)} ({\bf
\hat n \cdot S_1}) \right\}. } \label{P2QSS}
\end{eqnarray}%
where $\mathbf{S}=\mathbf{S}_{1}+\mathbf{S}_{2}$ and
$\mathbf{\Delta}=m(\mathbf{S}_{2}/m_{2}-\mathbf{S}_{1}/m_{1})$.

In black hole perturbation theory and in numerical simulations the
radiation field is frequently given in terms of spin weighted
spherical harmonics. As the injection of numerical templates also
requires this type of expansion \cite{LSCCommon} CBwaves does
contain a module evaluating some of the spin weighted spherical
harmonics. The relations we have applied in generating the
components read as
\begin{equation*}
MH_{lm}=\oint {}^{-2}Y_{lm}^{\ast }(\iota ,\phi )\left( rh_{+}-irh_{{\times }%
}\right) d\Omega ,
\end{equation*}%
where, for example,%
\begin{eqnarray*}
^{-2}Y_{2\pm 2} &=&\sqrt{\frac{5}{64\pi }}\left( 1\pm \cos \iota
\right)
^{2}e^{\pm 2i\phi }, \\
^{-2}Y_{2\pm 1} &=&\sqrt{\frac{5}{16\pi }}\sin \iota \left( 1\pm
\cos \iota
\right) e^{\pm i\phi }, \\
^{-2}Y_{20} &=&\sqrt{\frac{5}{32\pi }}\sin ^{2}\iota \ .
\end{eqnarray*}%
$h_{+}^{(lm)}$ and $h_{{\times }}^{(lm)}$ are defined as%
\begin{equation*}
rh_{+}^{(lm)}(t)-irh_{{\times }}^{(lm)}(t)=MH_{lm}(t)\,.
\end{equation*}%
Note that these modes of $rh_{+}$ and $rh_{{\times }}$ are used for
injections \cite{LSCCommon}.

\appendix
\section*{Appendix B}

\renewcommand{\theequation}{B.\arabic{equation}}
\setcounter{equation}{0}
\renewcommand{\thesubsection}{B.\arabic{subsection}}
\setcounter{subsection}{0}

\subsection{Equations of motion}

The various order of relative accelerations, as listed in
Eq.\,(\ref{accel}), can be given as \small
\begin{eqnarray}
\mathbf{a}_{N} &=&-{\frac{Gm}{r^{2}}}\mathbf{\hat{n}}, \\
\mathbf{a}_{PN} &=&-{\frac{Gm}{c^{2}r^{2}}}\left\{
\mathbf{\hat{n}}\left[ (1+3\eta )v^{2}-2(2+\eta
){\frac{Gm}{r}}-{\frac{3}{2}}\eta \dot{r}^{2}\right]
-2(2-\eta )\dot{r}\mathbf{v}\right\} , \\
\mathbf{a}_{SO} &=&{\frac{G}{c^{2}r^{3}}}\left\{ 6\mathbf{\hat{n}}[(\mathbf{%
\hat{n}}\times \mathbf{v})\mathbf{\cdot }(\mathbf{S}
+{\mbox{\boldmath$\sigma$}})]
-[\mathbf{v}\times (4\mathbf{S}+3{\mbox{\boldmath $\sigma$}})]+3%
\dot{r}[\mathbf{\hat{n}}\times (2\mathbf{S}+{\mbox{\boldmath $\sigma$}}%
)]\right\} , \label{aSO}\\
\mathbf{a}_{2PN} &=&-{\frac{Gm}{c^{4}r^{2}}}\biggl\{\mathbf{\hat{n}}\biggl[{%
\frac{3}{4}}(12+29\eta )({\frac{Gm}{r}})^{2}+\eta (3-4\eta )v^{4}+{\frac{15}{%
8}}\eta (1-3\eta )\dot{r}^{4} \\
&&-{\frac{3}{2}}\eta (3-4\eta )v^{2}\dot{r}^{2}-{\frac{1}{2}}\eta (13-4\eta )%
{\frac{Gm}{r}}v^{2}-(2+25\eta +2\eta ^{2}){\frac{Gm}{r}}\dot{r}^{2}\biggr] \nonumber \\
&&-{\frac{1}{2}}\dot{r}\mathbf{v}\left[ \eta (15+4\eta
)v^{2}-(4+41\eta
+8\eta ^{2}){\frac{Gm}{r}}-3\eta (3+2\eta )\dot{r}^{2}\right] \biggr\}, \nonumber \\
\mathbf{a}_{SS} &=&-{\frac{3G}{c^{2}\mu r^{4}}}\biggl\{\mathbf{\hat{n}}(%
\mathbf{S}_{1}\cdot \mathbf{S}_{2})+\mathbf{S}_{1}(\mathbf{\hat{n}}\cdot \mathbf{S}_{2})+%
\mathbf{S}_{2}(\mathbf{\hat{n}}\cdot \mathbf{S}_{1})-5\mathbf{\hat{n}}(\mathbf{\hat{n}}%
\cdot \mathbf{S}_{1})(\mathbf{\hat{n}}\cdot \mathbf{S}_{2})\biggr\}\,, \label{aSS} \\
\mathbf{a}_{RR}^{BT} &=& {\frac{8}{5}}\eta
{\frac{G^{2}m^{2}}{c^{5}r^{3}}}\left\{
\dot{r}\mathbf{\hat{n}}\left[ 18v^{2}+{\frac{2}{3}}{\frac{Gm}{r}}-25\dot{r}%
^{2}\right] -\mathbf{v}\left[
6v^{2}-2{\frac{Gm}{r}}-15\dot{r}^{2}\right] \right\}\,, \label{aRRBT2}
\end{eqnarray}
\normalsize where ${\mbox{\boldmath
$\sigma$}}=(m_{2}/m_{1})\mathbf{S}_{1}+(m_{1}/m_{2})\mathbf{S}_{2}$.
Note also that the above form of $\mathbf{a}_{SO}$ tacitly presumes
the use of the covariant spin supplementary condition, $S_{A}^{\mu
\nu }{u_{A}}_{\nu }=0$, where $u_{A}^{\mu }$ is the four-velocity of
the center-of-mass world line of body $A$, with $A=1,2$. Finally, as
discussed above the term $\mathbf{a}_{RR}^{BT}$ refers to the
radiation reaction expression derived from a Burke-Thorne type
radiation reaction potential \cite{IyerWill, ZengWill}.

\inserted{Higher order corrections to the acceleration are given as}
\begin{eqnarray}
\inserted{ \mathbf{a}_{PNSO} }&=& \frac{G}{c^4r^3} \bigg\{
\mathbf{\hat{n}} \bigg[(\mathbf{\hat{n}\times v})\mathbf{\cdot
}\mathbf{S} \bigg(-30 \eta \dot{r}^2  + 24 \eta
v^2 - \frac{G m}{r} (38 + 25 \eta) \bigg) \nonumber \\
&& + \frac{\delta m}{m} (\mathbf{\hat{n}\times v})\mathbf{\cdot
}\mathbf{\Delta} \bigg(-15 \eta \dot{r}^2 + 12 \eta v^2 - \frac{G
m}{r} (18 + \frac{29}{2} \eta) \bigg) \bigg]
\nonumber\\
&& + \dot{r} \mathbf{v} \bigg[ (\mathbf{\hat{n}\times
v})\mathbf{\cdot }\mathbf{S} (-9 + 9 \eta) + \frac{\delta m}{m}
(\mathbf{\hat{n}\times v})\mathbf{\cdot
}\mathbf{\Delta} (-3 + 6 \eta) \bigg] \nonumber \\
&& + \mathbf{\hat{n}} \times \mathbf{v} \bigg[\dot{r} ({\bf v\cdot
S}) (-3 + 3 \eta) - 8 \frac{G m}{r} \eta ({\bf \hat n \cdot S}) -
\frac{\delta m}{m} \bigg(4 \frac{G
m}{r} \eta ({\bf \hat n \cdot \mathbf{\Delta}}) + 3 \dot{r} ({\bf v \cdot \mathbf{\Delta}})  \bigg) \bigg] \nonumber\\
&& + \dot{r} \mathbf{\hat{n}} \times \mathbf{S} \bigg[-\frac{45}{2}
\eta \dot{r}^2 + 21 \eta v^2 - \frac{G m}{r} (25 + 15 \eta) \bigg]
\nonumber \\
&&+ \frac{\delta m}{m} \dot{r} \mathbf{\hat{n}} \times
\mathbf{\mathbf{\Delta}} \bigg[- 15 \eta \dot{r}^2  + 12 \eta v^2 -
\frac{Gm}{r} (9 + \frac{17}{2} \eta)\bigg] \nonumber \\
&& + \mathbf{v} \times \mathbf{S} \bigg[\frac{33}{2} \eta \dot{r}^2
+ \frac{G m}{r}(21 + 9 \eta) - 14 \eta v^2 \bigg] \nonumber \\
&& + \frac{\delta m}{m} \mathbf{v} \times \mathbf{\mathbf{\Delta}}
\bigg[9 \eta \dot{r}^2 - 7 \eta v^2 + \frac{G m}{r} (9 + \frac{9}{2}
\eta) \bigg] \bigg\} \, ,
\end{eqnarray}
\begin{eqnarray}
\inserted{ \mathbf{a}_{3PN} }&=& \frac{Gm}{c^3r^2} \bigg\{
\mathbf{\hat{n}} \bigg[ \left [16+\left(\frac{1399}{12}
-\frac{41}{16}\pi^{2} \right )\eta +\frac{71}{2}\eta^{2}\right
]{\left (\frac{Gm}{r}\right )}^{3} +\eta\left
[\frac{20827}{840}+\frac{123}{64}\pi^{2}-\eta^{2}\right ] {\left
(\frac{Gm}{r}\right )}^{2}v^{2}
\nonumber\\
& & -\left [1+\left (\frac{22717}{168}+\frac{615}{64}\pi^{2}\right
)\eta +\frac{11}{8}\eta^{2}-7\eta^{3}\right ] {\left
(\frac{Gm}{r}\right )}^{2}{\dot{r}}^{2}
\nonumber\\
& & -\frac{1}{4}\eta (11-49\eta+52\eta^{2})v^{6} +\frac{35}{16}\eta
(1-5\eta+5\eta^2 ){\dot{r}}^{6} - \frac{1}{4}\eta\left
(75+32\eta-40\eta^{2}\right )\frac{Gm}{r}v^{4}
\nonumber\\
& & - \frac{1}{2}\eta\left (158-69\eta-60\eta^{2}\right
)\frac{Gm}{r}{\dot{r}}^{4} +\eta\left (121-16\eta-20\eta^{2}\right
)\frac{Gm}{r}v^{2}{\dot{r}}^{2}
\nonumber\\
& & + \frac{3}{8}\eta\left (20-79\eta+60\eta^{2}\right
)v^{4}{\dot{r}}^{2} -\frac{15}{8}\eta\left
(4-18\eta+17\eta^{2}\right )v^{2}{\dot{r}}^{4}  \bigg]
\nonumber\\
&& +  \dot{r} \mathbf{v} \bigg[ \left [4+\left
(\frac{5849}{840}+\frac{123}{32}\pi^{2}\right )\eta
-25\eta^{2}-8\eta^{3}\right ]{\left (\frac{Gm}{r}\right )}^{2}
+\frac{1}{8} \eta\left(65-152\eta-48\eta^{2}\right )v^{4}
\nonumber\\
& &  +\frac{15}{8}\eta\left (3-8\eta-2\eta^{2}\right ){\dot{r}}^{4}
+\eta\left (15+27\eta+10\eta^{2}\right )\frac{Gm}{r}v^{2}
\nonumber\\
& &  -\frac{1}{6}\eta\left (329+177\eta+108\eta^{2}\right
)\frac{Gm}{r}\dot{r}^{2} -\frac{3}{4}\eta\left
(16-37\eta-16\eta^{2}\right )v^{2}\dot{r}^{2} \bigg] \bigg\} \, ,
\end{eqnarray}
\begin{eqnarray}
\inserted{ \mathbf{a}_{RR1PN} }&=& {\frac{8}{5}}\eta
{\frac{G^{2}m^{2}}{c^{7}r^{3}}}\bigg\{\dot{r}\mathbf{\hat{n}}\bigg[
\left(\frac{87}{14}-48\eta\right)v^{4} -
\left(\frac{5379}{28}-\frac{136}{3}\eta\right)v^{2}\frac{Gm}{r}
+\frac{25}{2}(1+5\eta)v^2\dot{r}^2 \nonumber\\
&&+ \left(\frac{1353}{4}-133\eta\right)\dot{r}^2\frac{Gm}{r}
-\frac{35}{2}(1-\eta)\dot{r}^4 +
\left(\frac{160}{7}+\frac{55}{3}\eta\right)\left(\frac{Gm}{r}\right)^2\bigg]
\nonumber\\
&& -\mathbf{v}\bigg[ -\frac{27}{14}v^{4} -
\left(\frac{4861}{84}+\frac{58}{3}\eta\right)v^{2}\frac{Gm}{r}
+\frac{3}{2}(13-37\eta)v^2\dot{r}^2 \nonumber\\
&&+ \left(\frac{2591}{12}+97\eta\right)\dot{r}^2\frac{Gm}{r}
-\frac{25}{2}(1-7\eta)\dot{r}^4 +
\frac{1}{3}\left(\frac{776}{7}+55\eta\right)\left(\frac{Gm}{r}\right)^2\bigg]
\bigg\}\,,
\end{eqnarray}
\begin{eqnarray}
\inserted{ \mathbf{a}_{RRSO} }&=& -\frac{G^2\eta m}{5c^7r^4}
\biggl \{ \frac{{\dot r}{\bf \hat{n}}}{\mu r} \left [ \left (
120v^2+280{\dot r}^2+453\frac{Gm}{r} \right ) {\bf {L}}_{\rm N}
\cdot {\bf \mathbf{S}} \right .
\nonumber \\
&& \left . + \left ( 120v^2+280{\dot r}^2+458\frac{Gm}{r} \right )
{\bf {L}}_{\rm N} \cdot {\mbox{\boldmath $\sigma$}} \right ]
\nonumber \\
&& + \frac{\bf v}{\mu r} \left [ \left ( 87v^2-675{\dot
r}^2-\frac{901}{3}\frac{Gm}{r} \right ) {\bf {L}}_{\rm N} \cdot {\bf
\mathbf{S}} + 4\left ( 18v^2-150{\dot r}^2 - 66\frac{Gm}{r} \right )
{\bf {L}}_{\rm N} \cdot {\mbox{\boldmath $\sigma$}} \right ]
\nonumber \\
&& - \frac{2}{3}{\dot r}{\bf v} \times {\bf \mathbf{S}} \left (
48v^2 + 15{\dot r}^2+364\frac{Gm}{r} \right ) + \frac{1}{3}{\dot
r}{\bf v} \times {\mbox{\boldmath $\sigma$}} \left ( 291v^2
-705{\dot r}^2-772\frac{Gm}{r} \right )
\nonumber \\
&& +\frac{1}{2}{\bf \hat{n}}\times {\bf \mathbf{S}} \left (
31v^4-260v^2{\dot r}^2+245{\dot r}^4 -\frac{689}{3}v^2\frac{Gm}{r} +
537{\dot r}^2\frac{Gm}{r} +\frac{4}{3}\frac{G^2m^2}{r^2} \right )
\nonumber \\
&& +\frac{1}{2}{\bf \hat{n}}\times {\mbox{\boldmath $\sigma$}} \left
( 115v^4-1130v^2{\dot r}^2+1295{\dot r}^4
-\frac{869}{3}v^2\frac{Gm}{r} + 849{\dot r}^2\frac{Gm}{r} +
\frac{44}{3}\frac{G^2m^2}{r^2} \right ) \biggr \} \,,
\end{eqnarray}
\begin{eqnarray}
\inserted{ \mathbf{a}_{RRSS} }&=& \frac{G^2}{c^7r^5} \biggl \{
{\bf \hat{n}} \biggl [ \biggl ( 287{\dot r}^2 - 99v^2 +
\frac{541}{5}\frac{Gm}{r} \biggr ){\dot r} (\mathbf{S}_{1}\cdot
\mathbf{S}_{2}) \nonumber \\
&& - \biggl (  2646{\dot r}^2 - 714 v^2 +\frac{1961}{5}\frac{Gm}{r}
\biggr ){\dot r} (\mathbf{\hat{n}}\cdot \mathbf{S}_{1})
(\mathbf{\hat{n}}\cdot \mathbf{S}_{2})
\nonumber \\
&& + \biggl ( 1029{\dot r}^2 - 123 v^2 + \frac{629}{10}\frac{Gm}{r}
\biggr ) \biggl((\mathbf{\hat{n}}\cdot \mathbf{S}_{1})
(\mathbf{\hat{v}}\cdot \mathbf{S}_{2}) + (\mathbf{\hat{n}}\cdot
\mathbf{S}_{2}) (\mathbf{\hat{v}}\cdot \mathbf{S}_{1}) \biggr)\nonumber \\
&& - 336 {\dot r} (\mathbf{\hat{v}}\cdot \mathbf{S}_{1})
(\mathbf{\hat{v}}\cdot \mathbf{S}_{2}) \biggr ] + {\bf v} \biggl [
\biggl ( \frac{171}{5} v^2 - 195 {\dot r}^2 - 67\frac{Gm}{r} \biggr
) (\mathbf{S}_{1}\cdot \mathbf{S}_{2})\nonumber \\
&& - \biggl ( 174 v^2 -1386 {\dot r}^2 - \frac{1038}{5}\frac{Gm}{r}
\biggr ) (\mathbf{\hat{n}}\cdot \mathbf{S}_{1})
(\mathbf{\hat{n}}\cdot \mathbf{S}_{2})
\nonumber \\
&& - 438 {\dot r} \biggl( (\mathbf{\hat{n}}\cdot \mathbf{S}_{1})
(\mathbf{\hat{v}}\cdot \mathbf{S}_{2}) + (\mathbf{\hat{n}}\cdot
\mathbf{S}_{2}) (\mathbf{\hat{v}}\cdot \mathbf{S}_{1}) \biggr) + 96
(\mathbf{\hat{v}}\cdot \mathbf{S}_{1}) (\mathbf{\hat{v}}\cdot
\mathbf{S}_{2}) \biggr ]
\nonumber \\
&& +  \biggl ( \frac{27}{10} v^2 - \frac{75}{2} {\dot r}^2 -
\frac{509}{30}\frac{Gm}{r} \biggr ) \biggl((\mathbf{\hat{v}}\cdot
\mathbf{S}_{2}) {\bf S}_1 + (\mathbf{\hat{v}}\cdot \mathbf{S}_{1})
{\bf S}_2 \biggr)
\nonumber \\
&& + \biggl (  \frac{15}{2} v^2 + \frac{77}{2} {\dot r}^2 +
\frac{199}{10}\frac{Gm}{r} \biggr ) {\dot r}
\biggl((\mathbf{\hat{n}}\cdot \mathbf{S}_{2}) {\bf S}_1 +
(\mathbf{\hat{n}}\cdot \mathbf{S}_{1}) {\bf S}_2 \biggr) \biggr \}
\, .
\end{eqnarray}

In general\inserted{, the accelerations} $\mathbf{a}_{SO}$ and $\mathbf{a}_{SS}$ are not confined
to the orbital plane thereby they yield a precession of th\inserted{is}
plane and, in turn, a\inserted{n amplitude and frequency} modulation of the observed signal. In
addition, spin vectors themselves precess according to their
evolution equations
\begin{eqnarray}
\mathbf{\dot{S}}_{i} &=&\frac{G}{c^{2}r^{3}}\left\{ \frac{4+3\zeta
_{i}}{2}\mathbf{L}_{N}-\mathbf{S}_{j}+3\left( \mathbf{\hat{n}\cdot
S}_{j}\right) \mathbf{\hat{n}}\right. \nonumber\\
&&\left. \inserted{+\frac{G^2\mu m}{c^5r^2}\left[\frac23\left(
\mathbf{v\cdot S}_{j}\right)+30\dot{r}\left( \mathbf{\hat{n}\cdot
S}_{j}\right)\right]\mathbf{\hat{n}}} \right\} \times
\mathbf{S}_{i}\,, \label{Sprec}
\end{eqnarray}
where $\mathbf{L}_{N}=\mu \mathbf{r}\times \mathbf{v}$ is the
Newtonian angular momentum and $\zeta _{i}=m_{j}/m_{i}$, with
$i,j=1,2$, $i\neq j$. \inserted{In Eq. (\ref{Sprec}), in addition to the standard spin-orbit and spin-spin terms \cite{Kidder}, the last expression stands for the 3.5PN spin-spin contribution \cite{WangWill}.}

\bigskip

The terms in the equations of motion, Eq.\,(\ref{accel}), up to 2PN
order can be deduced from a generalized Lagrangian which depends
only on the relative acceleration. From this Lagrangian the energy
$E$ and total angular momentum $\mathbf{J}$ of the system can be
computed which are known to be conserved up to 2PN order
\cite{Kidder}, i.e.\,in the absence of radiation reaction. The
conserved energy is given as
\inserted{\cite{Kidder},\cite{MoraWill},\cite{FBB}}\small
\begin{equation}
E=E_{N}+E_{PN}+E_{SO}+E_{2PN}+E_{SS} \inserted{+E_{3PN}+E_{PNSO}
}, \label{Energy}
\end{equation}%
\begin{eqnarray*}
E_{N} &=&\mu \left\{ {\frac{1}{2}}v^{2}-{\frac{Gm}{r}}\right\} , \\
E_{PN} &=&\frac{\mu }{c^{2}}\left\{ {\frac{3}{8}}(1-3\eta )v^{4}+{\frac{1}{2}%
}(3+\eta )v^{2}{\frac{Gm}{r}}+{\frac{1}{2}}\eta {\frac{Gm}{r}}\dot{r}^{2}+{%
\frac{1}{2}}\left({\frac{Gm}{r}}\right)^{2}\right\} , \\
E_{SO} &=&{\frac{G}{c^{2}r^{3}}}\mathbf{L}_{N}\mathbf{\cdot }
{\mbox{\boldmath$\sigma$}}, \\
E_{2PN} &=&\frac{\mu }{c^{4}}\biggl\{{\frac{5}{16}}(1-7\eta +13\eta
^{2})v^{6}-{\frac{3}{8}}\eta (1-3\eta ){\frac{Gm}{r}}\dot{r}^{4}+{\frac{1}{8}%
}(21-23\eta -27\eta ^{2}){\frac{Gm}{r}}v^{4} \\
&&+{\frac{1}{8}}(14-55\eta +4\eta ^{2})\left( {\frac{Gm}{r}}\right)
^{2}v^{2}+{\frac{1}{4}}\eta (1-15\eta ){\frac{Gm}{r}}v^{2}\dot{r}^{2}-{\frac{%
1}{4}}(2+15\eta )\left( {\frac{Gm}{r}}\right) ^{3} \\
&&+{\frac{1}{8}}(4+69\eta +12\eta ^{2})\left( {\frac{Gm}{r}}\right) ^{2}\dot{%
r}^{2}\biggr\}, \\
E_{SS} &=&{\frac{G}{c^{2}r^{3}}}\left\{ 3\left( \mathbf{\hat{n}\cdot S_{1}}%
\right) \left( \mathbf{\hat{n}\cdot S_{2}}\right) - \left(
\mathbf{S_{1}\cdot S_{2}}\right) \right\}, \\
\inserted{ E_{3PN} }&=&\frac{\mu}{c^6}\bigg\{ \left [\frac{3}{8}+
\frac{18469}{840}\eta\right ]{\left (\frac{Gm}{r}\right )}^{4}
 +\left [\frac{5}{4}-\left (\frac{6747}{280}-\frac{41}{64}\pi^{2}\right )\eta
 -\frac{21}{4}\eta^{2}+\frac{1}{2}\eta^{3}\right ]{\left (\frac{Gm}{r}\right )}^{3}v^{2}
\nonumber\\
 & &
 +\left [\frac{3}{2}+\left (\frac{2321}{280}-\frac{123}{64}\pi^{2}\right )\eta
 +\frac{51}{4}\eta^{2} +\frac{7}{2}\eta^{3}\right ]{\left (\frac{Gm}{r}\right )}^{3}\dot{r}^{2}
 \nonumber\\
 & &
+\frac{1}{128}\left (35-413\eta+1666\eta^{2}-2261\eta^{3}\right
)v^{8}
 +\frac{1}{16}(135-194\eta+406\eta^{2}-108\eta^{3}){\left (\frac{Gm}{r}\right )}^{2}v^{4}
 \nonumber\\
 & &
 +\frac{1}{16}(12+248\eta-815\eta^{2}-324\eta^{3}){\left (\frac{Gm}{r}\right )}^{2}v^{2}\dot{r}^{2}
 -\frac{1}{48}\eta(731-492\eta-288\eta^{2}){\left (\frac{Gm}{r}\right )}^{2}\dot{r}^{4}
 \nonumber\\
 & &
 +\frac{1}{16}(55-215\eta+116\eta^{2}+325\eta^{3})\frac{Gm}{r}v^{6}
 +\frac{1}{16}\eta(5-25\eta+25\eta^{2})\frac{Gm}{r}\dot{r}^{6}
 \nonumber\\
 & &
 -\frac{1}{16}\eta(21+75\eta-375\eta^{2})\frac{Gm}{r}v^{4}\dot{r}^{2}
 -\frac{1}{16}\eta(9-84\eta+165\eta^{2})\frac{Gm}{r}v^{2}\dot{r}^{4}
 \bigg\}, \\
\inserted{ E_{PNSO} }&=&\frac{G\mu}{2c^3r^2}(\mathbf{\hat{n}\times
v})\mathbf{\cdot }\bigg\{ \mathbf{\Delta}\frac{\delta
m}{m}\left[(1-5\eta)v^2 + (2+\eta)\frac{Gm}{r}\right] -
3\mathbf{S}\left[(1+\eta)v^2 +\eta {\dot r}^2-\frac{2Gm}{3r}\right]
\bigg\}.\label{EnergyPNSO}
\end{eqnarray*}%
\normalsize while the conserved total angular momentum is
\begin{equation}
\mathbf{J}=\mathbf{L}+\mathbf{S}\,,  \label{TotAngMom}
\end{equation}%
where \inserted{\cite{Kidder},\cite{MoraWill}}
\begin{equation}
\mathbf{L}=\mathbf{L}_{N}+\mathbf{L}_{PN}+\mathbf{L}_{SO}+\mathbf{L}_{2PN}
\inserted{+\mathbf{L}_{3PN} }, \label{AngMom}
\end{equation}%
and
\small
\begin{eqnarray*}
\mathbf{L}_{PN} &=&{\frac{\mathbf{L}_{N}}{c^{2}}}\left\{ {\frac{1}{2}}%
v^{2}(1-3\eta )+(3+\eta ){\frac{Gm}{r}}\right\} , \\
\mathbf{L}_{SO} &=&{\frac{\mu }{c^{2}m}}\Biggl\{{\frac{Gm}{r}}\mathbf{\hat{n}%
\times }\left[ \mathbf{\hat{n}\times }\left( 2\mathbf{S}
+{\mbox{\boldmath$\sigma$}}\right) \right]
-{\frac{1}{2}}\mathbf{v\times }\left(
\mathbf{v\times } {\mbox{\boldmath$\sigma$}} \right) \Biggr\}, \\
\mathbf{L}_{2PN} &=&{\frac{\mathbf{L}_{N}}{c^{4}}}\biggl\{{\frac{3}{8}}%
(1-7\eta +13\eta ^{2})v^{4}-{\frac{1}{2}}\eta (2+5\eta ){\frac{Gm}{r}}\dot{r}%
^{2} \\
&&+{\frac{1}{2}}(7-10\eta -9\eta ^{2}){\frac{Gm}{r}}v^{2}+{\frac{1}{4}}%
(14-41\eta +4\eta ^{2})\left( {\frac{Gm}{r}}\right) ^{2}\biggr\} ,\\
\inserted{ \mathbf{L}_{3PN} }
&=&{\frac{\mathbf{L}_{N}}{c^{6}}}\biggl\{ \left [ \frac{5}{2}-\left
(\frac{5199}{280}-\frac{41}{32}\pi^{2}\right
)\eta-7\eta^{2}+\eta^{3}\right ]{\left (\frac{Gm}{r}\right )}^{3}
\nonumber\\
 & &
 +\frac{1}{16}(5-59\eta+238\eta^{2}-323\eta^{3} )v^{6}
+\frac{1}{12}(135-322\eta+315\eta^{2}-108\eta^{3}){\left
(\frac{Gm}{r}\right )}^{2}v^{2}
\nonumber\\
 & &
+\frac{1}{24} (12-287\eta-951\eta^{2}-324\eta^{3}){\left
(\frac{Gm}{r}\right )}^{2}\dot{r}^{2}
+\frac{1}{8}(33-142\eta+106\eta^{2}+195\eta^{3})\frac{Gm}{r}v^{4}
\nonumber\\
 & &
 -\frac{1}{4} \eta (12-7\eta-75\eta^2)\frac{Gm}{r}v^{2}\dot{r}^{2}
+\frac{3}{8}\eta(2-2\eta-11\eta^{2})\frac{Gm}{r}\dot{r}^{4}
\biggr\}.
\end{eqnarray*}%
\normalsize Notice that at the applied level of PN approximation
there is no spin-spin contribution to $\mathbf{J}$.

\bigskip

The leading order radiative change of the conserved quantities $E$
and $\mathbf{J}$ is governed by the quadrupole formula
\cite{IyerWill}. To lowest 2.5PN order the instantaneous loss of
energy $E$ is given as \cite{Kidder}
\begin{eqnarray}
\frac{dE_{\inserted{N}}}{dt}=-\frac{8}{15}\frac{G^3m^2\mu^2}{c^5r^4}\left(12v^2
- 11\dot{r}^2 \right)\,,
\end{eqnarray}
while the radiative angular momentum loss is
\begin{eqnarray}
\frac{d\mathbf{J}_{\inserted{N}}}{dt}=-\frac{8}{5}\frac{G^2m\mu}{c^5r^3}\mathbf{L}_{N}\left(2v^2
- 3\dot{r}^2 + 2\frac{Gm}{r}\right)\,.
\end{eqnarray}
\inserted{The radiative change of the energy is expressed as}
\begin{eqnarray}\label{dEdt}
\inserted{ \frac{dE_{RR}}{dt} } &=&
\frac{dE_N}{dt}+\frac{dE_{PN}}{dt}+\frac{dE_{SO}}{dt}+\frac{dE_{2PN}}{dt}
+\frac{dE_{SS}}{dt}+\frac{dE_{2.5PN}}{dt}+\frac{dE_{PNSO}}{dt}\,.
\end{eqnarray}
\inserted{Note that, in practice, the total amount of radiated energy $E_{RR}$ has to be determined during the evolution by evaluating the integral in Eq. (\ref{dEdt}).
The higher order contributions to the energy loss are given as
\cite{Kidder},\cite{GopaIyer},\cite{Arun08},\cite{FBB2} }
\begin{eqnarray} \inserted{
\frac{dE_{PN}}{dt} } &=& - {2 \over 105} {G^3m^2 \mu^2 \over
c^7r^4}\biggl\{ (785-852\eta)v^4 -160(17-\eta) {Gm \over r}v^2  +
8(367-15\eta){Gm \over r} \dot r^2 \nonumber \\
&&  -2(1487-1392\eta)v^2 \dot r^2 + 3(687-620\eta) \dot r^4 +
16(1-4\eta)\left({Gm \over r}\right)^2 \biggr\},
\end{eqnarray}
\begin{eqnarray}
\inserted{ \frac{dE_{SO}}{dt} } = - {8 \over 15} {G^3m \mu \over
c^7r^6} \Bigg\{ {\bf L_N \cdot} \Bigg[ {\bf S} (78 \dot r^2 -80v^2
-8 {Gm \over r}) \mbox{} + {\delta m \over m}{\bf \Delta} (51 \dot
r^2 - 43v^2 + 4{Gm \over r} \Biggr] \Biggr\},
\end{eqnarray}
\begin{eqnarray}
\inserted{ \frac{dE_{2PN}}{dt} } &=&-{8 \over
15}\frac{G^3\,m^2\,\mu^2}{c^9\,r^4} \left\{
                  \frac{1}{42} (1692 - 5497\eta
                 + 4430\eta^2)v^6
              -\frac{1}{14} (1719 - 10278\eta
+ 6292\eta^2) v^4\dot{r}^2\right.\nonumber \\
              &&- \left.\frac{1}{21} (4446 - 5237\eta
+ 1393\eta^2)\frac{Gm}{r}\,v^4
              +\frac{1}{14} (2018 - 15207\eta
+ 7572\eta^2)v^2\dot{r}^4\right.\nonumber \\
              &&+ \left.\frac{1}{7} (4987 - 8513\eta
+ 2165\eta^2)\frac{Gm}{r}\,v^2\dot{r}^2\right.\nonumber \\
              &&+\left. \frac{1}{756} (281473 + 81828\eta
+ 4368\eta^2)\left(\frac{Gm}{r}\right)^2\,v^2\right.\nonumber \\
              &&- \left.\frac{1}{42} (2501 - 20234\eta
+ 8404\eta^2)\dot{r}^6
              -\frac{1}{63} (33510 - 60971\eta
+ 14290\eta^2)\frac{Gm}{r}\,\dot{r}^4\right.\nonumber \\
              &&- \left.\frac{1}{252} (106319 + 9798\eta
+ 5376\eta^2)\left( \frac{Gm}{r}\right)^2\,\dot{r}^2\right.\nonumber \\
              &&+\left. \frac{2}{63} (-253 + 1026\eta
- 56\eta^2) \left( \frac{Gm}{r}\right)^3\right\},
\end{eqnarray}
\begin{eqnarray}
\inserted{ \frac{dE_{SS}}{dt} } &=& - {4 \over 15} {G^3m \mu \over
c^7r^6} \biggl\{ - 3 ({\bf \hat n \cdot S_1})({\bf \hat n \cdot S_2}
) \left( 168 v^2 - 269 \dot r ^2 \right)  + 3 ({\bf S_1 \cdot S_2})
\left( 47 v^2- 55 \dot r ^2 \right) \nonumber \\ && + 71 ( {\bf v
\cdot S_1})({\bf v \cdot S_2} ) - 171 \dot r \left[ ({\bf v \cdot
S_1})({\bf \hat n \cdot S_2}) + ({\bf \hat n \cdot S_1})( {\bf v
\cdot S_2}) \right] \biggr\} ,
\end{eqnarray}
\begin{eqnarray}
\inserted{ \frac{dE_{2.5PN}}{dt} } &=& -\frac{32}{5} \frac{G^3 \mu^2
m^2}{c^{10} r^4} \dot{r}\eta\bigg( -\frac{12349}{210} \frac{Gm}{r}
v^4 +\frac{4524}{35} \frac{Gm}{r}
v^2\dot{r}^2 -\frac{2753}{126} \frac{G^2\, m^2}{r^2} v^2 \nonumber\\
&&-\frac{985}{14} \frac{Gm}{r} \dot{r}^4 + \frac{13981}{630}
\frac{G^2 m^2}{r^2} \dot {r}^2 -\frac{1}{315} \frac{G^3 m^3}{r^3}
\bigg)\,,
\end{eqnarray}
\begin{eqnarray}\label{dEdtPNSO}
\inserted{ \frac{dE_{PNSO}}{dt} } &=& -\frac{8}{105} \frac{G^3 \mu^2
m}{c^{10} r^5} \bigg\{ {(\mathbf{\hat{n}\times v})\mathbf{\cdot
}\mathbf{S}}\left[{\dot r}^4\left(
     {3144}\eta- {2244}\right)+\frac{G^2m^2}{r^2}
  \left({972}
+{166}\eta\right)\right.\nonumber\\
&&+\frac{Gm}{r}{\dot r}^2\left({170}\eta-{2866}\right)+{\dot
r}^2v^2\left(
  {3519}-{5004}\eta\right)\nonumber\\
  &&  \left.+\frac{G m}{r}v^2\left({3504}-140\eta\right)
  +v^4\left({1810}\eta-{1207}\right)\right]\nonumber\\
  &&+{(\mathbf{\hat{n}\times v})\mathbf{\cdot
}\mathbf{\Delta}}\frac{\delta
  m}{m} \left[{\dot r}^4\left(
{2676}\eta-\frac{7941}{4}\right) + \frac{G^2m^2}{r^2}\left(
  126\eta-{109}\right)\right.\nonumber\\
  &&+\frac{Gm}{r}{\dot r}^2\left({1031}\eta-\frac{6613}{2}\right)+{\dot r}^2v^2\left(
  {2364} - {3621}\eta\right) \nonumber\\
  &&\left.+\frac{G m}{r} v^2\left(\frac{4785}{2}-455\eta\right)
  +v^4\left({1040}\eta-\frac{2603}{4} \right)\right] \bigg\} \,.
\end{eqnarray}


\begin{thebibliography}{99}
\bibitem{advligo}{Advanced LIGO anticipated sensitivity curves, LIGO Document T0900288-v3,
{\tt
https://dcc.ligo.org/cgi-bin/private/DocDB/ShowDocument?docid=2974}}

\bibitem{advvirgo}{Advanced Virgo, project homepage, INFN. URL (accessed 16 July 2011).
External Link {\tt http://www.virgo.infn.it/advirgo/}.}

\bibitem{matchedfilter} B. Allen, Phys. Rev. D, \textbf{71}, 062001 (2005).

\bibitem{Blanchet06} L. Blanchet, Living Rev. Relativity \textbf{9}, 4. (2006). \\
{\tt http://www.livingreviews.org/lrr-2006-4}

\bibitem{Kidder} L. E. Kidder, Phys. Rev. D \textbf{52}, 821 (1995).

\bibitem{W93} \inserted{A. G. Wiseman, Phys. Rev. D \textbf{48}, 4757 (1993).}

\bibitem{spawaves}{B. S. Sathyaprakash and S. V. Dhurandhar, Phys. Rev. D 44, 43819 (1991).}

\bibitem{IyerWill} B. R. Iyer and C. M. Will, Phys. Rev. D \textbf{52}, 6882 (1995).

\bibitem{yunes} N. Yunes, K. G. Arun, E. Berti, and C. M. Will, Phys. Rev. D \textbf{80}, 084001 (2009).

\bibitem{wen} L. Wen, Astrophys. J. \textbf{598}, 419 (2003).

\bibitem{kocsis} R. M. O'Leary, B. Kocsis, and A. Loeb, Mon. Not. R. Astron. Soc. \textbf{395}, 2127 (2009).

\bibitem{janna} J. Levin, S. T. McWilliams, and H. Contreras,
Class. Quantum Grav.\textbf{28}, 175001 (2011).

\bibitem{exc} D. A. Brown and P. J. Zimmerman, Phys. Rev. D \textbf{81}, 024007 (2010).

\bibitem{Blanchet04} L. Blanchet, T. Damour, and G. E.-Far\`{e}se, Phys. Rev. D
\textbf{69}, 124007 (2004).

\bibitem{DJS01} T. Damour, P. Jaranowski, and G. Sch\"afer, Phys. Lett. B 513, 147 (2001).

\bibitem{Blanchet08} \inserted{K. G. Arun, L. Blanchet, B. R. Iyer, and M. S. S. Qusailah, Phys.
Rev. D \textbf{77}, 064034 (2008).}

\bibitem{WW} C. M. Will and A. G. Wiseman, Phys. Rev. D
\textbf{54}, 4813 (1996).

\bibitem{Maggiore} M. Maggiore, \textit{Gravitational waves}, Oxford University Press (2008).

\bibitem{LSCCommon} D. A. Brown et.al., LIGO-T070072-00-Z, arXiv:0709.0093 [gr-qc].

\bibitem{TOO} \inserted{H. Tagoshi, A. Ohashi, and B. Owen, Phys. Rev. D \textbf{63}, 044006
(2001).}

\bibitem{FBB} \inserted{G. Faye L. Blanchet, and A. Buonanno, Phys. Rev. D \textbf{74}, 104033
(2006).}

\bibitem{FC} L. S. Finn and D. F. Chernoff, Phys. Rev. D
\textbf{47}, 2198 (1993).

\bibitem{cbwavesdownload}{The CBwaves simulation software, {\tt http://grid.kfki.hu/project/virgo/cbwaves/}}

\bibitem{condor}{The Condor job scheduler, http://www.cs.wisc.edu/condor/}

\bibitem{etsens} C. K. Mishra, K. G. Arun, B. R. Iyer, and B. S. Sathyaprakash,
Phys. Rev. D \textbf{82}, 064010 (2010).

\bibitem{SPA} J. Mathews and R. L. Walker, \textit{Mathematical Methods of Physics} W. A. Benjamin, New York, (1970).

\bibitem{BT1} K. S. Thorne, Astrophys. J. \textbf{158}, 997 (1969).

\bibitem{BT2} W. L. Burke, J. Math. Phys. (N.Y.) \textbf{12}, 401 (1971).

\bibitem{DDPLA} T. Damour and N. Deruelle, Phys. Lett. A \textbf{87}, 81 (1981).

\bibitem{MoraWill} T. Mora and C. M. Will, Phys. Rev. D \textbf{69}, 104021 (2004).

\bibitem{ZengWill}  J. Zeng and C. M. Will, Gen. Rel. Grav. \textbf{39}, 1661 (2007).

\bibitem{gpv3} L. \'A. Gergely, Z. I. Perj\'es, and M. Vas\'uth, Phys. Rev. D \textbf{58}, 124001 (1998).

\bibitem{MGS} R.-M. Memmesheimer, A. Gopakumar, and G. Sch\"afer, Phys. Rev. D \textbf{70}, 104011 (2004).

\bibitem{spacing} B. J. Owen and B. S. Sathyaprakash, Phys. Rev. D \textbf{60}, 022002 (1999).

\bibitem{ffactor} T. A. Apostolatos, Phys. Rev. D54, 2421 (1996).

\bibitem{BCV} A. Buonanno, Y. Chen, and M. Vallisneri, Phys. Rev. D \textbf{67}, 104025 (2003).

\bibitem{SSC1} F. A. E. Pirani, Acta Phys. Polon. \textbf{15}, 389 (1956).

\bibitem{SSC2} T. D. Newton and E. P. Wigner, Rev. Mod. Phys. \textbf{21}, 400 (1949).

\bibitem{SSC3} E. Corinaldesi and A. Papapetrou, Proc. Roy. Soc. A \textbf{209}, 259 (1951).

\bibitem{WangWill} \inserted{H. Wang and C. M. Will, Phys. Rev. D \textbf{75}, 064017
(2007).}

\bibitem{WillPNSO} \inserted{C. M. Will, Phys. Rev. D \textbf{71}, 084027 (2005).}

\bibitem{GopaIyer} \inserted{A. Gopakumar and B. R. Iyer, Phys. Rev. D \textbf{56},
7708 (1997).}

\bibitem{Arun08} \inserted{K. G. Arun, L. Blanchet, B. R. Iyer, and M. S. S. Qusailah,
Phys. Rev. D \textbf{77}, 064035 (2008).}

\bibitem{FBB2} \inserted{L. Blanchet, A. Buonanno, and G. Faye, Phys. Rev. D \textbf{74},
104034 (2006).}

\end{thebibliography}
\end{document}